\documentclass[11pt,a4paper]{article}
\usepackage{lmodern}
\usepackage[T1]{fontenc}
\usepackage{microtype}
\usepackage[left=2.8cm,right=2.8cm,top=2.8cm,bottom=2.8cm]{geometry}
\usepackage{graphicx}
\usepackage{amsmath,amssymb,mathtools,enumerate}
\usepackage[bookmarks=true]{hyperref}
\usepackage{cite}
\usepackage{xcolor}
\usepackage{colonequals}

\numberwithin{equation}{section}
\usepackage[font=small,labelfont=bf,width=0.85\textwidth]{caption}

\providecommand{\hypersetup}[1]{}
\providecommand{\hyperref}[2][]{#2}
\providecommand{\texorpdfstring}[2]{#1}
\providecommand{\pdfbookmark}[3][]{}

\hypersetup{plainpages=false}
\hypersetup{pdfpagemode=UseOutlines}
\hypersetup{bookmarksnumbered=true}
\hypersetup{bookmarksopen=true}
\hypersetup{pdfstartview=FitH}
\hypersetup{colorlinks=false}
\hypersetup{citebordercolor={.6 .9 .6}}
\hypersetup{urlbordercolor={.7 .8 1}}
\hypersetup{linkbordercolor={1 .7 .7}}
\hypersetup{pdfborder={0 0 .5}}

\newcommand{\namedref}[2]{\hyperref[#2]{#1~\ref*{#2}}}

\makeatletter
\def\mr@ignsp#1 {\ifx\:#1\@empty\else #1\expandafter\mr@ignsp\fi}%
\newcommand{\multiref}[1]{\begingroup
\xdef\mr@no@sparg{\expandafter\mr@ignsp#1 \: }%
\def\mr@comma{}%
\@for\mr@refs:=\mr@no@sparg\do{\mr@comma\def\mr@comma{,}\ref{\mr@refs}}%
\endgroup}
\makeatother
\renewcommand{\eqref}[1]{(\multiref{#1})}

\makeatletter
\let\@myabstract\@empty
\let\@keywords\@empty
\let\@subject\@empty
\providecommand{\affiliation}[1]{\gdef\@affiliation{#1}}
\providecommand{\myabstract}[1]{\gdef\@myabstract{#1}}
\providecommand{\keywords}[1]{\gdef\@keywords{#1}}
\providecommand{\subject}[1]{\gdef\@subject{#1}}
\def\thetitle{\@title}
\def\theauthor{\@author}
\def\theaffiliation{\@affiliation}
\def\theabstract{\@myabstract}
\def\thesubject{\@subject}
\def\thedate{\@date}
\def\thekeywords{\@keywords}
\makeatother
\def\fillpdfdata{
\hypersetup{pdftitle={\thetitle}}%
\hypersetup{pdfsubject={\thesubject}}%
\hypersetup{pdfkeywords={\thekeywords}}%
}

\let\oldbfseries=\bfseries
\let\oldmdseries=\mdseries
\let\oldnormalfont=\normalfont
\renewcommand{\bfseries}{\oldbfseries\boldmath}
\renewcommand{\mdseries}{\oldmdseries\unboldmath}
\renewcommand{\normalfont}{\oldnormalfont\unboldmath}

\makeatletter
\newlength{\apb@width}
\newcommand{\autoparbox}[2][c]{\settowidth{\apb@width}{#2}\parbox[#1]{\apb@width}{#2}}

\makeatother

\newcommand{\beqa}{\begin{eqnarray}}
\newcommand{\eeqa}{\end{eqnarray}}
\newcommand{\beq}{\begin{equation}}
\newcommand{\eeq}{\end{equation}}
\newcommand{\lambdan}{\lambda}

\renewcommand{\l}{\lambda}

\newcommand{\D}{\Delta}

\newcommand{\BCn}{\textit{BC}$_N$}
\newcommand{\BCt}{\textit{BC}$_2$}
\newcommand{\BC}{\textit{BC}} 
\newcommand{\wall}{\omega}

\newmuskip\pFqskip
\mathchardef\pFcomma=\mathcode`,

\newcommand*\pFq[5]{%
   \begingroup
   \begingroup\lccode`~=`,
     \lowercase{\endgroup\def~}{\pFcomma\mkern\pFqskip}%
   \mathcode`,=\string"8000
   {}_{#1}F_{#2}\biggl(\genfrac..{0pt}{}{#3}{#4};#5\biggr)%
   \endgroup
}

\DeclareMathOperator*{\Res}{\text{Res}}

\begin{document}
\addtolength{\baselineskip}{2pt}


\title{Integrability of Conformal Blocks I: \\[3mm] Calogero-Sutherland Scattering Theory}

\myabstract{Conformal blocks are the central ingredient of the conformal bootstrap
programme. We elaborate on our recent observation that uncovered a relation with
wave functions of an integrable Calogero-Sutherland Hamiltonian in order to develop
a systematic theory of conformal blocks. Our main goal here is to review central
ingredients of the Heckman-Opdam theory for scattering states of Calogero-Sutherland
models with special emphasis to the relation with scalar 4-point blocks. We will
also discuss a number of direct consequences for conformal blocks, including a new
series expansion for blocks of arbitrary complex spin and a complete analysis of
their poles and residues. Applications to the Froissart-Gribov formula for conformal
field theory, as well as extensions to spinning blocks and defects are briefly discussed
before we conclude with an outlook on forthcoming work concerning algebraic
consequences of integrability.}



\noindent
\mbox{}\hfill DESY-17-178\\
\mbox{}\hfill WIS/05/17-Nov-DPPA

\author{%
	Mikhail Isachenkov\texorpdfstring{${}^a$}{}
	and Volker Schomerus\texorpdfstring{${}^b$}{}%
}

\hypersetup{pdfauthor={\theauthor}}

\vfill

\begin{center}
{\Large\textbf{\mathversion{bold}\thetitle}\par}
\vspace{1cm}

\textsc{\theauthor}
\vspace{5mm}

\textit{%
${}^a$Department of Particle Physics and Astrophysics, Weizmann Institute of Science, Rehovot 76100, Israel\\
${}^b$DESY Theory Group, DESY Hamburg, Notkestra{\ss}e 85, D-22607 Hamburg, Germany}\\
\vspace{3mm}
{\ttfamily
\href{mailto:mikhail.isachenkov@weizmann.ac.il}{mikhail.isachenkov@weizmann.ac.il}\\
\href{mailto:volker.schomerus@desy.de}{volker.schomerus@desy.de}}
\par\vspace{1cm}

\textbf{Abstract}\vspace{5mm}

\begin{minipage}{12.7cm}
\theabstract
\end{minipage}

\vspace{1cm}

\end{center}

\fillpdfdata

\hrule
\providecommand{\microtypesetup}[1]{}
\microtypesetup{protrusion=false}
\tableofcontents
\microtypesetup{protrusion=true}

\vspace{3ex}\hrule

\vfill

\newpage


\section{Introduction}
\label{sec:Intro}

The conformal bootstrap programme \cite{Mack:1969rr, Polyakov:1970xd, Ferrara:1973vz, Ferrara:1973yt, Polyakov:1974gs, Mack:1975jr}
was initially designed as an analytical approach to the non-perturbative dynamics
of critical systems. It relies on the careful separation of kinematical (or group
theoretic) input from dynamical data. In spite of significant early efforts to
develop the necessary background in representation theory of the conformal group,
see \cite{Dobrev:1977qv} and references therein, concrete implementations of the bootstrap
programme suffered from the fact that much of the relevant mathematics could not be
developed at the time. It took more than a decade before the real impact of the
conformal bootstrap was first demonstrated in the context of 2-dimensional conformal
field theories where the conformal symmetry in enhanced to the infinite dimensional
Virasoro algebra \cite{Belavin:1984vu}. Up until a few years ago, it was commonly assumed that
such a success of the bootstrap programme was only possible in 2-dimensional systems.
But since the recent numerical incarnation of the conformal bootstrap programme has
delivered data e.g.\ on scaling weights and operator products in the $d=3$ dimensional
Ising model with unprecedented precision, see \cite{Rattazzi:2008pe,ElShowk:2012ht,
El-Showk:2014dwa, Kos:2014bka, Simmons-Duffin:2015qma, Kos:2016ysd, Dymarsky:2017yzx} and \cite{Kos:2013tga, Kos:2015mba, Iliesiu:2015qra, Iliesiu:2017nrv, Rattazzi:2010yc, Poland:2011ey, Chester:2016wrc, Chester:2016ref} for some similar results in other theories, the conformal bootstrap has
attracted new attention.

The key kinematical data in the conformal bootstrap are the conformal blocks along
with the so-called crossing kernel. Explicit analytical results on conformal blocks
in $d > 2$ dimensional were scarce until the work of Dolan and Osborn \cite{Dolan:2000ut, Dolan:2003hv,
Dolan:2011dv} on scalar conformal blocks. In a few cases, such as for four scalar external
fields in even dimensions, Dolan and Osborn were able to construct blocks explicitly
in terms of ordinary single variable hypergeometric functions. Extensions to generic
dimensions and external field with spin or general defect blocks proved more difficult,
even though some remarkable progress has been achieved during the last few years, e.g.
through the use of differential operators, the concept and construction of seed
blocks etc., see e.g.\ \cite{Costa:2011mg, Costa:2011dw, SimmonsDuffin:2012uy, Rejon-Barrera:2015bpa,
Costa:2014rya, Penedones:2015aga, Echeverri:2015rwa,  Iliesiu:2015akf,  Costa:2016xah, Costa:2016hju,
Echeverri:2016dun,  Karateev:2017jgd, Kravchuk:2016qvl, Kravchuk:2017dzd} and references therein.
In most cases, however, a
construction of blocks in terms of ordinary hypergeometric functions could not be
found. In order to evaluate such more general blocks, Zamolodchikov-like recurrence
relations have become the most efficient tool. While these may suffice to provide the
required input for the numerical bootstrap, it is fair to say that a systematic and
universal theory of conformal blocks has not been developed to date. It is our main
goal to fill this important gap.

In some sense, the mathematical foundations for a modern and systematic theory
of conformal blocks, including those for external fields with spin, were actually
laid at about the same time at which the bootstrap programme was formulated, though
very much disguised at first. It gradually emerged from the systematic study of
solvable Schr\"odinger problems starting with the work by Calogero, Moser and
Sutherland \cite{Calogero:1970nt,Sutherland:1971ks,Moser:1975qp}. The quantum
mechanical models that were proposed in these
papers describe a 1-dimensional multi-particle system whose members are subject to
an external potential and exhibit pairwise interaction. It turned out that for
appropriate choices of the potentials and interactions, such models can be integrable.
One distinguishes two important series of such theories known as $A$ and $BC$ models.
While the former describe particles that move on the entire real line, particles are
restricted to the half-line in the case of $BC$-type models.

The study of eigenfunctions for Calogero-Sutherland (CS) models advanced rapidly after
a very influential series of papers by Heckman and Opdam that was initiated in
\cite{Heckman1987,Opdam1988} (inspired by earlier papers of
Koornwinder \cite{Koornwinder1974series})
and provided the basis for much of the modern theory of multivariable hypergeometric
functions. Subsequently, many different approaches were developed that emphasize
algebraic aspects (most notably due to Cherednik \cite{CherednikBook}), combinatorial identities \cite{MacdonaldBook} or relations to matrix models. The general
techniques that were developed in this context are fairly universal. On the other hand,
explicit formulas were often only worked out for $A$ models, with $BC$ trailing a bit behind.

These two different strands of seemingly unrelated developments were brought together by our
recent observation \cite{Isachenkov:2016gim} that conformal blocks of scalar 4-point functions
in a $d$-dimensional conformal field theory can be mapped to eigenfunctions of a 2-particle
hyperbolic Calogero-Sutherland Hamiltonian. Thereby, the modern theory of multivariable
hypergeometric functions and its integrable foundation enters the court of the conformal
bootstrap. This is the first of a series of papers in which we describe, and at various
places advance, the mathematical theory of Calogero-Sutherland models and develop the
applications to conformal blocks. Here we shall focus mainly on the classical
Heckman-Opdam theory for scattering states of Calogero-Sutherland models. Algebraic
consequences of integrability as well as advanced analytical features are subject of
a subsequent paper \cite{alg-structures}, see also concluding section \ref{sec:Conclusions} for a detailed outline.

The plan of this paper is as follows. In the \hyperref[sec:CSproblem]{next section} we introduce the relevant
hyperbolic Calogero-Sutherland models for \BCn\ root systems. After setting up some
notations, we spell out the \hyperref[subsec:CSpotential]{Hamiltonian} and describe  its \hyperref[subsec:AffineWeyl]{symmetries}, the associated
\hyperref[subsec:AffineWeyl]{fundamental domain} and different \hyperref[subsec:CoordinatesCS]{coordinate choices}. The we turn to the scattering
theory. In section \ref{sec:Wavefunctions} we introduce the notion of \hyperref[subsec:HarishChandra]{Harish-Chandra wave functions} and
discuss their analytic properties both in \hyperref[subsec:Monodromy]{coordinate space} and in the \hyperref[subsec:HChPoles]{space of
eigenvalues}. The former are controlled by a special class of \hyperref[subsec:Monodromy]{representations of
an affine braid group} which is discussed in detail. This will allow us to construct
the true wave functions of Calogero-Sutherland models as special linear combinations
of Harish-Chandra functions. We also describe the position of \hyperref[subsec:HChPoles]{poles of Harish-Chandra
functions} in the space of eigenvalues and provide explicit formulas for their residues,
at least for $N=2$. The latter have not appeared in the mathematical literature before
and they are obtained from a new series expansion for Harish-Chandra functions that we
derive in appendix \ref{app:zxexpansions}. The aim of section \ref{sec:Blocks} is to embed \hyperref[subsec:Casimir]{scalar conformal
blocks} into the general theory of Calogero-Sutherland wave functions. Our discussion includes
details on the choice of boundary conditions. This will enable us to discuss a number
of direct applications to scalar blocks in section \ref{sec:Applications}. These include a 
full classification of \hyperref[subsec:Poles]{poles of conformal blocks} for arbitrary (complex) values of the spin 
variable $l$. We also calculate the corresponding residues. When the spin variable is 
specialized to an integer (and the dimension $d$ is integer), our results for poles and residues
agree with those that were obtained using representation theory of the conformal Lie algebra
\cite{Penedones:2015aga}. The extension to complex spins may be seen as the main new advance 
of this work in the context of conformal field theory. Blocks with non-integer spin play an 
important role e.g. in the recent conformal Froissart-Gribov formula of Caron-Huot 
\cite{Caron-Huot:2017vep}, see also \cite{Simmons-Duffin:2017nub}.
We will sketch how such \hyperref[subsec:GribovFroissart]{inversion formulas} arise within the theory of Calogero-Sutherland
models and employ the algebraic structure of the monodromy representation to explain a
crucial numerical `coincidence' in Caron-Huot's derivation of the Froissart-Gribov formula.
The paper concludes with a extensive outlook, in particular to the second part where we will
discuss and exploit the rich algebraic structure of Calogero-Sutherland models. While
some of the more advanced results we describe are geared to the root system \BCt\ which
appears in the context of scalar 4-point functions, most of the more general discussion
is presented for general $N$. Larger values $N > 2$ turn out to be relevant
\cite{Schomerus:2017talk}
in the context of defect blocks \cite{Gadde:2016fbj,Fukuda:2017cup}.

\section{The \BCn\ Calogero-Sutherland problem}
\label{sec:CSproblem}

In this section we shall review the setup of the Calogero-Sutherland problem for the
\textit{BC}$_N$ root system. After a short discussion the simplest example, the famous
P\"oschl-Teller potential, we discuss the general setup. In our description we will put
some emphasis on the symmetries of the Calogero-Sutherland potential, possible domains
for the associated Schr\"odinger problem and the singularities at the boundary of these
domains.

\subsection{The P\"oschl-Teller Hamiltonian}
\label{subsec:PTHamiltonian}

The simplest example of what is now known as Calogero-Sutherland model goes back
to the work of P\"oschl and Teller in \cite{Poschl:1933zz}. The so-called modified P\"oschl-Teller
Hamiltonian takes the form
\begin{equation}\label{PT}
H^\textrm{PT}_{(a,b)} = - \partial_u^2 + V^{\text{PT}}_{(a,b)}(u) =
- \partial_u^2 -\frac{ab}{\sinh^2\frac{u}{2}} + \frac{(a+b)^2-\frac14}{\sinh^2 u}\ .
\end{equation}
and defines a 1-dimensional Schr\"odinger problem with a potential that depends
on two continuous parameters $a$ and $b$. P\"oschl and Teller noticed that the
corresponding eigenvalue problem can be mapped to the hypergeometric differential
equation and constructed the eigenfunctions in terms of hypergeometric functions.
All this is fairly standard, but there are a few things we would like to emphasize
in this example that will become important for extensions to the multi-particle
generalizations.
\smallskip

\begin{figure}[htb]
\centering\includegraphics[scale=.4]{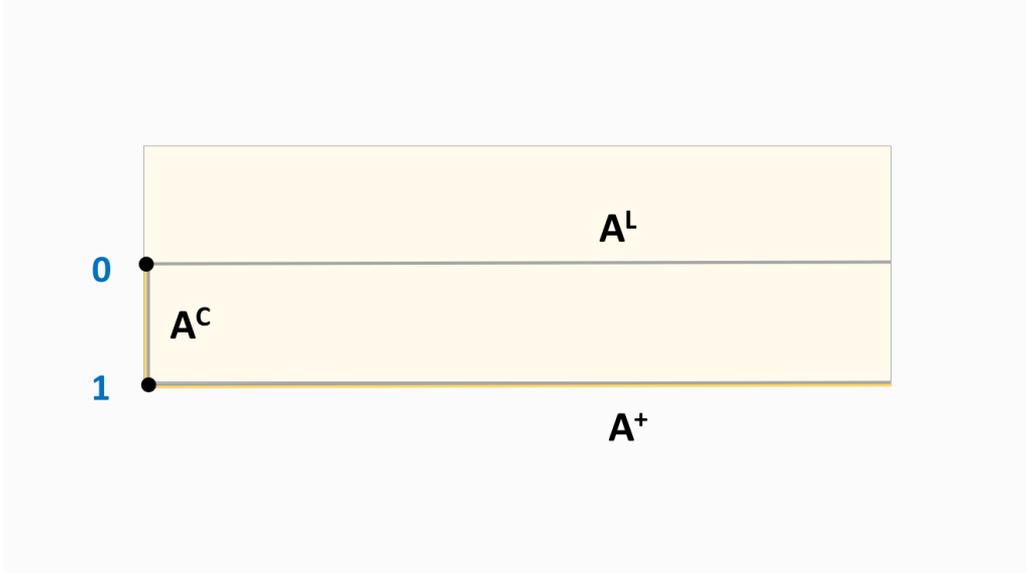}
\caption{The fundamental domain $D$ of the P\"oschl-Teller problem is a semi-infinite
cylinder whose end is squashed to an interval. It is obtained from the semi-infinite
strip by gluing the horizontal lines on top and bottom. Within this domain, the
potential diverges at two points $u_0 = i\pi$ and $u_1 = 0$ (black dots). Different
Schr\"odinger problems can be obtained, depending on the choice of a real subset $A$.
The figure shows three possible choices, $A^+, A^L= \tilde A^+$ and $A^C$. The
eigenvalue equations on $A^+$ and $A^L$ are related by a shift $\varrho$ that involves
the reflection of the coupling constant $b$ and they both lead to a continuous spectrum.
The 1-dimensional Schr\"odinger problem on the compact interval $A^C$ possesses a
discrete spectrum.}
\label{fig:BC1}
\end{figure}

For the moment we will consider $u$ as a complex variable. In the complex
$u$-plane, the P\"oschl-Teller potential possesses some symmetries. On the one
hand it is symmetric with respect to shifts $\tau:u \rightarrow u + 2\pi i$ of $u$
in the imaginary direction. These give rise to an action of $\mathbb{Z}$ on the
complex plane. In addition, the potential is also reflection symmetric, i.e. it is invariant
under the $\mathbb{Z}_2$ reflection $w:u \rightarrow -u$. Together, these two
transformations generate the symmetry group $\mathcal{W} = \mathbb{Z}_2 \ltimes
\mathbb{Z}$ of the P\"oschl-Teller potential. The fundamental domain
$$ D = \mathbb{C}/\mathcal{W} $$
for the action of $\mathcal{W}$ in the complex $u$-plane is shown in Figure
\ref{fig:BC1}.
After the appropriate identifications of boundary points it has the form of a
semi-infinite pillow, i.e. a semi-infinite cylinder whose end is squashed to
an interval. The two corners of this semi-infinite pillow correspond to the
two points $u_1 = 0$ and $u_0 = i\pi$ at which the P\"oschl-Teller potential
diverges. At the same time, these points are fixed under the action of a
non-trivial subgroup of the symmetry $\mathcal{W}$ on the complex $u$-plane.\footnote{When $b=0$, the fundamental domain is further reduced to
$ D' = \mathbb{C}/\left(\mathcal{W} \rtimes \mathbb{Z}_2 \right)$. In this case,
the complex torus has non-trivial (so called) center \cite{Heckman1987}, which is related to existence of a
{\it minuscule weight} for the reduced root system $C_1$. The denominator $\mathcal{W}
\rtimes \mathbb{Z}_2$ is called an extended affine Weyl group, and the additional
$\mathbb{Z}_2$ accounts for a permutation of an affine and a non-affine simple
root. The reduced fundamental domain $D'$ has $\Delta \Im u = \pi$, instead of
$\Delta \Im u = 2\pi$. When $b\neq 0$, there is a remnant of this symmetry which
we call $\varrho$, see below.}

The group $\mathbb{W}$ exhausts all the symmetries of the P\"oschl-Teller
potential that are associated with a transformation of the variable $u$
alone. But there exists one additional symmetry that involves a shift
in $u$ combined with a reflection of the coupling $b$.\footnote{This symmetry
is a twist of an ordinary translation symmetry of the $C_1$ P\"oschl-Teller problem,
where is acts on coordinates only since the parameter $b$ vanishes. It may be
regarded as a translation of the coordinate $u$ by a $2\pi i$ times a minuscule
coweight of the root system $C_1$, see below.} More precisely, it acts as
\begin{equation} \label{eq:varrho}
\varrho: u \rightarrow  u + i \pi \quad , \quad
b \rightarrow  - b
\end{equation}
while leaving the couplings $a$ invariant, i.e. $\tilde a = a$. In fact,
the P\"oschl-Teller potential is invariant under these replacements
\begin{eqnarray*}
\varrho: V^\textrm{PT}_{(a,b)}(u) \rightarrow
V^\textrm{PT}_{(a,-b)}(u+ i\pi) & = & -\frac{ab}{\cosh^2\frac{u}{2}}
+\frac{(a-b)^2-\frac{1}{4}}{\sinh^2u}
\\[2mm]
& = & - \frac{ab}{\sinh^2\frac{u}{2}} + \frac{(a+b)^2-\frac{1}{4}}{\sinh^2u}
= V^\textrm{PT}_{(a,b)}(u)\ .
\end{eqnarray*}
Note that the symmetry maps the two singular points $u_0 = 0$ and $u_1 =
i\pi$ in the fundamental domain $D$ onto each other.
\smallskip

After these comments on the symmetries of the P\"oschl-Teller potential let us
now discuss the possible setups for the Schr\"odinger problem. These correspond
to different 1-dimensional subsets $A \subset D$ on which the P\"oschl-Teller
potential is real. There are essentially three such choices which we shall
denote by $A^+, \tilde A^+$ and $A^C$. We define the set $A^+$ as
\begin{equation} \label{eq:Ap}
A^+ = \{\,  u \in \mathbb{R}\, |\,  u > 0\, \}\  .
\end{equation}
For this set, the Schr\"odinger equation reads
\begin{equation} \label{eq:PTeqAp}
\left[- \frac{d^2}{du^2} -\frac{ab}{\sinh^2\frac{u}{2}} +
\frac{(a+b)^2-\frac14}{\sinh^2u}\right] \psi(u) = \varepsilon \psi(u)\ .
\end{equation}
Note that the potential creates a wall at $u=0$ which shields the positive
half-line $u >0$ from the negative one. In this case, there is a continuum of
states with energy $\varepsilon > 0$. We will discuss these in the \hyperref[sec:Wavefunctions]{next
section}. A second possible choice is
\begin{equation}\label{eq:tAp}
 \tilde A^+ = \{\, \tilde u + i \pi\, | \tilde u \in \mathbb{R}_{>0}\,  \}\ .
 \end{equation}
On this set, the corresponding Schr\"odinger equation is the same as on
$A^+$, except that the parameter $b$ is sent to $-b$. Our previous comments
on the structure of the spectrum apply to this case as well.

There exists a qualitatively quite different choice for a domain on which
the Calogero-Sutherland Hamiltonian is real. It is given by
\begin{equation}\label{eq:AC}
 A^C = \{\, u = i \varphi\, | \, \varphi \in [0,\pi] \, \}
 \end{equation}
where the superscript $C$ stands for compact. On $A^C$ the Schr\"odinger
equation takes the form
\begin{equation}\label{eq:PTeqAC}
\left[- \frac{d^2}{d\varphi^2}
-\frac{ab}{\sin^2\frac{\varphi}{2}}  +
\frac{(a+b)^2-\frac14}{\sin^2 \varphi}\right] \psi(\varphi)
= \varepsilon \psi(\varphi)\ .
\end{equation}
Once again, there are infinite walls at $\varphi = 0, \pi$. This setup describes
a particle in a 1-dimensional box with infinitely high walls on both sides. In this
case the spectrum is discrete. We will recall the precise form of the wave functions
in the \hyperref[sec:Wavefunctions]{next section}.

\subsection{The Calogero-Sutherland potential}
\label{subsec:CSpotential}

\begin{figure}[htb]
\centering\includegraphics[scale=.4]{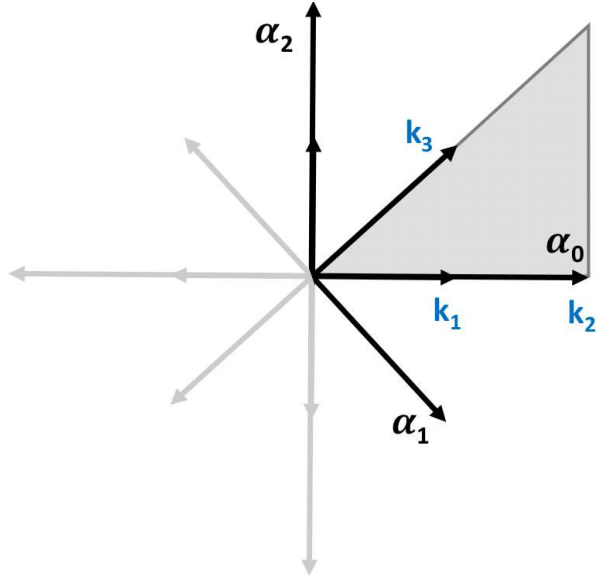}
\caption{The \textit{BC}$_2$ root system. The set $\Sigma^+$ of positive roots in
shown in bold black arrows. For $N>1$ there are three different types of roots that
belong to three different orbits of the Weyl group $W_N$. These carry multiplicity $k_{\alpha}$.
Our basis elements $\alpha_1$ and $\alpha_2$ are also depicted along with the highest
root $\alpha_0 = 2\alpha_1 + \alpha_2 = 2 e_1$. The linear spaces that are fixed
under the Weyl reflections $w_1$ and $w_2$ that are associated with $\alpha_1$
and $\alpha_2$ form the boundary of the Weyl chamber (shown in grey).}
\label{fig:roots}
\end{figure}

Before we can spell out the Calogero-Sutherland Hamiltonian, we need a bit of
notation. In general integrable Calogero-Sutherland Hamiltonians are associated
with a root system $\Sigma$. Here we shall focus on \BCn\ root systems, see Figure
\ref{fig:roots},
whose positive roots are given
\begin{equation}
\Sigma^+ = \{e_i, 2 e_i, e_i\pm e_j|1 \leq i,j
\leq N; i < j \}\ .
\end{equation}
We have used $e_i, 1 \leq i \leq N$ to denote a basis of $\mathbb{R}^N$ that is
orthonormal with respect to the canonical scalar product $\langle \cdot, \cdot
\rangle$. Note that this root system is not reduced, i.e.\ it contains roots
that are related by a factor of two.\footnote{The integrality condition in the
definition of a root system demands that a projection of a root onto any other
root is a half-integer multiple of the latter. One can easily see that this
can indeed happen if some roots possess a collinear partner differing by a
factor of two, but not more.} From time to time we will have to remove
the shortest positive roots, i.e. the roots $e_i$. The remaining set of positive
roots is denoted by $\Sigma^+_0$. Of course, $\Sigma^+_0$ are simply the positive
roots of the Lie algebra $C_N$. Let us also select a basis of $\Sigma^+_0$
consisting of
$$\alpha_N = 2e_N \quad , \quad \alpha_i = e_{i}-e_{i+1} $$
for $i=1, \dots, N-1$. Indeed, all elements of $\Sigma^+_0$ may be obtained as
linear combinations of $\alpha_i, i=1, \dots,N$ with non-negative integer
coefficients. The unique highest root is given by $\alpha_0 = 2e_1$. For any
root $\alpha$, we define
$$   \alpha^\vee = \frac{2\alpha}{\langle \alpha,\alpha\rangle}\ . $$
This concludes our short description of the root system and the special roots
that will play an important role below.
The potential of the associated Calogero-Sutherland model takes the form
\begin{equation}\label{CSgen}
V^{\text{CS}}(u_i) = \sum_{\alpha\in \Sigma^+}
\frac{k_\alpha(k_\alpha+ 2k_{2\alpha}-1)\langle \alpha,\alpha\rangle}
{4\sinh^2 \frac{\langle \alpha,u \rangle}{2} }\ .
\end{equation}
It involves the parameters $k_\alpha$, often referred to as multiplicities, that
are assumed to be invariant under the action of the Weyl group $W= W_N$ of $\Sigma$,
i.e.\ $k_{w\alpha} = k_\alpha$ for $w \in W$. Since the {\it BC}$_N, N > 1,$ root lattice
decomposes into three orbits under the action of the Weyl group, see Figure \ref{fig:roots},
the potential $V$ contains three independent parameters, which we parametrize as
\begin{equation}
k_3= k_{e_i \pm e_j} = \frac{\epsilon}{2} \quad ,\quad
k_2= k_{2e_i} = a+b+\frac12 \quad , \quad k_1= k_{e_i} = -2b \
\end{equation}
in terms of the three parameters $a,b$ and $\epsilon$. The reason for this choice
of parameters will become clear in the \hyperref[sec:Blocks]{fourth section}. In addition, we agree
that $k_{\beta}=0$ if $\beta \not \in \Sigma$. Finally, we have also introduced
$u = \sum u_i e_i$. It is easy to see that the formula \eqref{CSgen} reduces to
the P\"oschl-Teller potential upon setting $N=1$. Note that the root system for
\textit{BC}$_1$ consists of two orbits under the action of the Weyl group $W=W_1
= \mathbb{Z}_2$. Hence, the potential only contains two parameters, $a$ and $b$.
Let us also note that the case $b=0$ is somewhat special since there are no
contributions from the short roots $e_i$ in the potential. This means that the
underlying root system is $C_N$ rather than \textit{BC}$_N$.
\medskip

\noindent
{\bf Example:} The case we are most interested in appears for $N=2$. The
corresponding Calogero-Sutherland potential now contains all three parameters,
\begin{equation}
V_{(a,b,\epsilon)}^{\textrm{CS}}(u_1,u_2)  =
V_{(a,b)}^{\text{PT}}(u_1) +
V_{(a,b)}^{\text{PT}}(u_2) + \frac{\epsilon(\epsilon-2)}
{8\sinh^2\frac{u_1-u_2}{2}} + \frac{\epsilon(\epsilon-2)}
{8\sinh^2 \frac{u_1+u_2}{2}} \ . \label{CSpot}
\end{equation}
One may think of this potential as describing two P\"oschl-Teller particles
in the half-line that are interacting with an interaction strength depending
on $\epsilon$. Alternatively, one can think of a single particle that moves
in an external potential on a 2-dimensional domain $D_2$, see below.

\subsection{Affine Weyl group and the domain $D_N$}
\label{subsec:AffineWeyl}

The analysis of symmetries of the Calogero-Sutherland potential proceeds pretty much
in the same way as for the P\"oschl-Teller problem. Once again, we will think of $u_i$
as complex variables so that the potential is a function on $\mathbb{C}^N$. Obviously,
it is invariant under the $N$ independent discrete shifts
\begin{equation}
 \tau_i: u_j \longrightarrow u_j + 2\pi i \delta_{i,j} \ .
\end{equation}
These generate the abelian group $\mathbb{Z}^N$. In addition, the potential is also
left invariant by the action of the Weyl group $W_N$ of the \textit{BC}$_N$ root
system. This Weyl group can be generated by the $N$ Weyl reflections that are
associated with our basis $\alpha_i, i=1, \dots, N$. We shall denote these
Weyl reflections by $w_i = w(\alpha_i)$. It is not difficult to work out all
relations among these generators. They are given by
\begin{eqnarray}
w_i^2 = 1 , \quad , \quad w_{N-1} w_N w_{N-1} w_N = w_{N} w_{N-1} w_N w_{N-1}
\quad , \quad w_i w_{i+1} w_i = w_{i+1} w_i w_{i+1}
\end{eqnarray}
for $i = 1, \dots, N-2$. All other pairs of generators simply commute with each
other, i.e.
$$ w_i w_j = w_j w_ i \quad \textrm{for} \quad |i-j|\geq 2 \ . $$
The Weyl group $W_N$ acts on the translations $\tau_i$ by permutation and
inversion. More precisely one has the following set of non-trivial
relations
\begin{eqnarray}
w_N \tau^{-1}_N w_N =  \tau_N  \quad , \quad
w_i \tau_{i+1} w_i = \tau_{i}
\end{eqnarray}
for $i=1, \dots, N-1$. Here we used a multiplicative notation for the generators
$\tau_i$ of the abelian group $\mathbb{Z}^N$, i.e.\ we denote the shift of $u_i$
by $-2\pi i$ as $\tau_i^{-1}$ rather than $-\tau_i$. Together, the elements
$\tau_i$ and $w_i$ with $i=1, \dots, N$ generate the so-called \textit{affine
Weyl group}
\begin{equation} \label{eq:AWG}
\mathcal{W}_N \ = \ W_N \ltimes \mathbb{Z}^{N}\ .
\end{equation}
The affine Weyl group $\mathcal{W}_N$ describes all symmetries of the
Calogero-Sutherland potential \eqref{CSgen} that act on the coordinates
$u_i$ alone. It generalizes the group $\mathcal{W}_1 = \mathcal{W}$ we
had introduced in our discussion of the P\"oschl-Teller potential to the
case with $N \geq 2$.

Here we have described the affine Weyl group in terms of $2N$ generators
$\tau_i$ and $w_i$ with $i=1, \dots, N$. There exists a second description
in terms of $N+1$ generators $w_i$ with $i=0, \dots, N$. While $w_i$ with
$i\neq 0$ are the same Weyl reflections $w_i = w(\alpha_i)$ we used before,
the new generator $w_0$ is given by
\begin{equation}
w^{-1}_0 = w(\alpha_0) \tau_1 = w(2e_1)\tau_1  =  (w_1 \cdots w_{N-1} w_N)
(w_{N-1} \cdots w_1) \tau_1\ .
\end{equation}
One may check by explicit computation that this new element $w_0$ satisfies
the following relations with $w_i, i=1, \dots, N$
\begin{equation}
w_0^2 = 1 \quad , \quad  w_0 w_1 w_0 w_1 = w_1 w_0 w_1 w_0 \quad , \quad
w_0 w_i = w_i w_0 \
\end{equation}
for $i=2, \dots, N$. Note that the relation between $w_0$ and $w_1$ is identical
to the one between $w_{N-1}$ and $w_{N}$. Obviously, one can reconstruct the
generator $\tau_1$ from the element $w_0$ and the Weyl reflections $w_i, i=1,
\dots, N$. The other elements $\tau_i, i > 1$ are then obtained by conjugation
with $w_1 \cdots  w_{i-1}$.
\smallskip

As in the case of the P\"oschl-Teller potential there exists one additional
shift symmetry that requires a combined action on the coordinates and the
coupling constants. It is given by
\begin{equation}
\varrho: u_j \rightarrow u_j + i \pi \quad , \quad
b \rightarrow  - b
\end{equation}
while leaving the other two couplings $a$ and $\epsilon$ invariant. Note
that $\varrho$ involves a simultaneous action on all $N$ coordinates.
Following the same steps as in the section \ref{subsec:PTHamiltonian} one finds that
\begin{equation}
\varrho: V^\textrm{CS}_{(a,b,\epsilon)}(u) \rightarrow
V^\textrm{CS}_{(a,- b,\epsilon)}( u+i\pi) =
V^\textrm{CS}_{(a,b,\epsilon)}(u)\ .
\end{equation}
Let us stress that this symmetry is not part of the affine Weyl group
which acts only on coordinates.\footnote{As in the P\"oschl-Teller case, this
symmetry is a twist of an ordinary translation symmetry of the $C_N$ P\"oschl-Teller
problem for which it acts on the coordinates $u$ only. The latter may be considered as a
translation of $u$ by $2\pi i $ times a minuscule coweight of the reduced root
system $C_N$.}
\medskip

We are now well prepared to discuss the domain(s) on which we will
consider the Calogero-Sutherland system. We start with a set of $N$
complex coordinates $(u_i) \in \mathbb{C}^N$ and implement the
identification furnished by the the $N$ discrete shifts $\tau_i$.
This leaves us with an $N$-dimensional complex manifold
$$ \mathbb{T}_N = \mathbb{C}^N/\mathbb{Z}^N \ . $$
The manifold $\mathbb{T}_N$ contains an $N$-dimensional real
submanifold $A_N \subset \mathbb{T}_N$ that is parametrized by
$u_i \in \mathbb{R}$, modulo identification with $\tau_i$. We
can shift $A_N$ with $\varrho$ to obtain $\tilde A_N = \varrho
(A_N)$. The latter is parametrized by $u_i = \tilde u_i + i \pi$
with real $\tilde u_i \in \mathbb{R}$.

The Calogero-Sutherland potential diverges along the following walls of
real codimension two
$$ \wall_{\alpha} = \{\,  u \in \mathbb{T}_N \, | \, \langle \alpha ,
u\rangle = 0 \, \text{mod}\ 2\pi i \} \subset A_N \quad
\textrm{for} \quad \alpha \in \Sigma^+\  $$
that are in one-to-one correspondence with the positive roots of
our \textit{BC}$_N$ roots system. Note that for the long roots
$2e_j$, the set $\wall_{2e_j}$ possesses two disconnected components,
one of which coincides with the wall $\wall_{e_j}$ for the corresponding
short root. The walls associated with the roots $e_i\pm e_j$ contain
a single connected component. By construction, the walls $\wall_\alpha$
are invariant under the action of the reflection $w(\alpha)\in W_N$.

The walls we have just described are subspaces along which the
Calogero-Sutherland potential diverges so that the corresponding
Schr\"odinger problem can be restricted to various subsets within
the quotient space
\begin{equation}
 D_N = \mathbb{T}_N/W_N = \mathbb{C}^N / \mathcal{W}_N \ .
\end{equation}
which describes a fundamental domain\footnote{As in the P\"oschl-Teller case,
when $b=0$ the fundamental domain becomes $ D' = \mathbb{C}/\left(\mathcal{W}_N
\rtimes \mathbb{Z}_2 \right)$. Again, the root system becomes of reduced, $C_N$ type. The denominator $\mathcal{W}_N
\rtimes \mathbb{Z}_2$ is an extended affine Weyl group. The nontrivial element
in the additional $\mathbb{Z}_2$ accounts for a permutation of an affine and a
non-affine simple root that preserves the Weyl alcove. When $b\neq 0$, there is
a remnant of this symmetry which we again call $\varrho$.}
for the action of the Weyl group $W_N$ on $\mathbb{T}_N$.
Representatives of the quotient space in $\mathbb{T}_N$ intersect with only
$N+1$ walls $\wall_i, i=0, \dots, N$ where $\wall_i = \wall_{\alpha_i}$ for
$i=1, \dots, N-1,$ and $\wall_0,\wall_N$ are the two disconnected components of
$\wall_{\alpha_N}$ with $\wall_N = \wall_{e_N}$.

Once again, the fundamental domain $D_N$ for the Calogero-Sutherland problem
possesses several subsets along which the the potential is real. The most
important is the Weyl chamber
\begin{equation}
A_N^+ = \mathrm{WC}_N = \{u \in \mathbb{R}^N| \langle \alpha, u\rangle > 0 \
\textrm{ for all}  \ \alpha \in \Sigma^+_0 \} \subset D_N\ .
\end{equation}
In this case, the eigenfunctions possess $N$ continuous variables. As in
the case of the P\"oschl-Teller problem, we can also consider the shifted
Weyl chamber $\tilde A_N^+ = \varrho(A_N^+)$. It gives rise to a similar
set of wave functions except that the coupling $b$ must be replaced by
$\tilde b = - b$. Another extreme possibility is the case
\begin{equation}
A^C_N = \{\, u_i = i \varphi_i\, | \, \varphi_i \in [0,\pi]\, \}
\end{equation}
for which the spectrum of the Calogero-Sutherland model is discrete. But in
the multivariable case there are many other possibilities. We will discuss
a few of them for $N=2$.

\begin{figure}[htb]
\centering
\includegraphics[scale=.5]{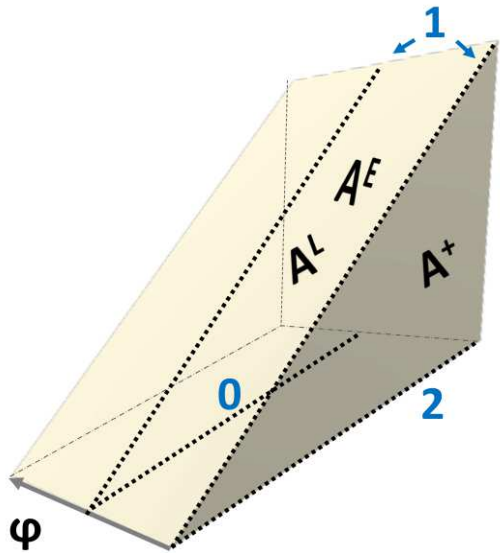}
\caption{The three-dimensional slice of the fundamental domain $D_2$ for the
\textit{BC}$_2$ Calogero-Sutherland model in $u$-space. Front and the back
side of the wedge should be identified. The fixed points (walls) under the
action of $w_2$ and $w_1$ are shown as bold dashed lines. Fixed points of
$w_2$ fall into two disconnected components which carry the labels $0$ and
$2$. The fixed points of $w_1$ on the other hand form a connected set that
intersects our 3-dimensional slice in two lines labeled by $1$. The shaded
area in front is the Weyl chamber $A_2^+$. It is bounded by the walls $\wall_1$
and $\wall_2$. The shifted Weyl chamber $A_2^L = \tilde A_2^+$ is bounded by the
walls $\wall_0$ and $\wall_1$. The subset $A^E_2$ is the 2-dimensional semi-infinite
strip of width $\pi$ on top of the wedge. It is bounded by pieces of wall
$\wall_1$ only, except in the corners.}
\label{fig:D2}
\end{figure}
\medskip

\noindent
{\bf Example:} Let us give some additional details for $N=2$. In this case, the Weyl
group $W_2$ consists of eight elements. It can be generated from $w_2 = w(2e_2)$ and
$w_1= w(e_1-e_2)$ subject to the relations
\begin{equation} \label{eq:W2}
w_1^2 = 1 = w_2^2 \quad , \quad w_1 w_2 w_1 w_2 = w_2 w_1 w_2 w_1 \ .
\end{equation}
The fundamental domain $D_2$ for the action of the Weyl group $W_2$ on $\mathbb{T}_2$,
or rather a 3-dimensional subspace thereof that satisfies $\Im (u_1 +u_2) = 0$, is
shown in Figure \ref{fig:D2}.

Once again, we can consider the Schr\"odinger equation for the Calogero-Sutherland
potential on various real subsets ${\mathcal A}_2$. The most standard choice in
the mathematical literature is
\begin{equation} \label{eq:A2p}
 A^+_2 = \{ (u_1,u_2) | \langle \alpha,u\rangle > 0 \ \textrm{for}
\ \alpha \in \Sigma^+\}\ .
\end{equation}
This is simply a Weyl chamber for the {\it BC}$_2$ root system. As for $N=1$ the
Calogero-Sutherland potential diverges along the walls of the chamber, see our
discussion above.

There are two additional choices we want to discuss here because of their relevance
for conformal field theory, see section \ref{sec:Blocks} below. The first one is given by
\begin{equation} \label{eq:A2L}
A^L_2 = \tilde A^+_2 = \{ (u_1,u_2) = (\, \tilde u_1 + i \pi,\tilde u_2 - i\pi)\, | \,
\tilde u_i > 0\,  ;\, , \tilde u_1 > \tilde u_2\}\ .
\end{equation}
It may be obtained from $A^+_2$ by application of $\varrho$ combined with a translation
by $-2\pi i e_2$. The associated Schr\"odinger problem has the same form as on $A^+_2$,
except that the parameters are changed, see our discussion above
\begin{equation}
V^\textrm{CS}_{(a,b,\epsilon)}(u_1,u_2) =
V^\textrm{CS}_{(a,-b,\epsilon)}(\tilde u_1,\tilde u_2)
\end{equation}
Another relevant possible subset that leads to real potential is the one in which
the two coordinates $u_1$ and $u_2= u_1^\ast $ are complex conjugates of each other,
$$ {A}^E_2 = \{\,  (u_1,u_2)\, |\,  u_1 =  u_2^\ast, \Im u_1 \in [0,\pi]\}\ . $$
We shall decompose $u_1 = u + i \varphi$ into its real and imaginary part. In this
case we are dealing with a particle that moves on a 2-dimensional semi-infinite
strip given by $u \geq 0$ and $\varphi \in [0,\pi]$. For large $u$ the potential
becomes
$$ V_{(a,b,\epsilon)}^{\textrm{CS}}(u,\varphi) \sim - \frac{\epsilon(\epsilon-2)}
{8\sin^2\varphi} \  \mbox{ for } \ u \rightarrow \infty\ .$$
So, we see that in this asymptotic regime, wave functions are given by a
product of a P\"oschl-Teller bound state and a plane wave in the $u$-direction.

\subsection{Coordinates in the CS problem}
\label{subsec:CoordinatesCS}

\begin{figure}[htb]
\centering
\includegraphics[scale=.4]{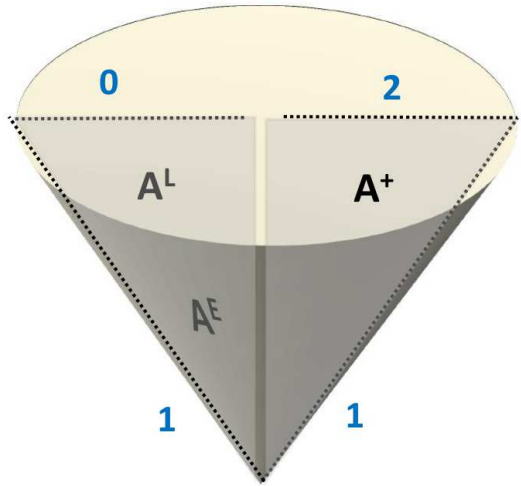}
\caption{The fundamental domain for the \textit{BC}$_2$ Calogero-Sutherland model in
$x^{-1}$-space. The shaded areas in the radial direction are the Weyl chamber $A_2^+$
and the shifted domain $A^L_2$. On the boundary of the cone, $A^+_2$ is bounded by the
walls $\wall_1$ and $\wall_2$ while $A^L_2$ is bounded by $\wall_1$ and $\wall_0$. The subset $A^E_2$
is mapped to half of the the mantle.}
\label{fig:D2x}
\end{figure}

Let us conclude this section with a short comment on coordinates. So far,
we have described the Calogero-Sutherland problem in terms of coordinates
$u_i$ in which the kinetic term is simply the standard flat space Laplacian.
Since the $u_i$ are coordinates on $\mathbb{T}_N$ it is tempting to apply
the exponential map that sends  $\mathbb{T}_N$ to $\mathbb{C}^N$.\footnote{To
be more precise, this map is injective, with the image $\left(\mathbb{C}^{\times}
\right)^N$, and thus defines a partial compactification of the complex torus. By
the action of the Weyl group, it extends to a toroidal compactification of the
torus corresponding to its decomposition into Weyl chambers. This gives the toric
variety of $x$-coordinates, see \cite{HeckmanBook} (page 55), \cite{MR495499, OdaBook}
for details.} Indeed, we shall often use the coordinates
\begin{equation}\label{eq:xfromu}  x_i = e^{u_i}
\end{equation}
instead of $u_i$. This has the advantage that the identification $u_i \equiv
u_i + 2\pi i$ is manifest. Upon application of the exponential map (and inversion),
the 3-dimensional slice of the fundamental domain $D_2$ that is shown in
Figure \ref{fig:D2} gets mapped to a cone, see Figure \ref{fig:D2x}. The map
sends the Weyl chamber $A_2^+$ and the space $A^L_2= \tilde A^+_2$ to one
half of a section through the cone each, while the set $A_2^E$ becomes half
of the mantle of the cone. Figure \ref{fig:D2x} also keeps track of the
location of the singularities.
While the coordinates $x_i$ make the identification with $l_i : u_i \rightarrow u_i
+ 2\pi i$ manifest so that they are proper coordinates on $\mathbb{T}_N$, the identification
from the action of the Weyl group $W_N$ is not built into these coordinates. It is
often good to do a little better and to use coordinates that are invariant at least
under the action of the Weyl reflections $w(e_i), i=1, \dots N$. Any function of
$x_i + x_i^{-1}$ would do the job, but we shall adopt a very specific one, namely
\footnote{A bit more precisely, for $\BC_2$ we will use the branches $z_1 =
\frac{e^{-i\pi}}{\sinh^2\frac{u_1}{2}}$ $z_2 =  \frac{e^{+i\pi}}{\sinh^2\frac{u_2}{2}}$
in order to be consistent with definition \eqref{eq:A2L}.}
\begin{equation} \label{eq:zfromux}
z_i = - \frac{1}{\sinh^2\frac{u_i}{2}} = \frac{4}{2-x_i-x^{-1}_i} \ .
\end{equation}
These coordinates send the domain $D_N$ to configurations $(z_i) \in \mathbb{C}^N$
that are symmetric under the action of the permutation group $S_N \subset W_N$.
The latter is generated by the Weyl reflections $w_i=w(e_i \pm e_{i+1})$ for $i=
1, \dots, N-1$.

\section{Wave functions of the Calogero-Sutherland model}
\label{sec:Wavefunctions}

Having set up the eigenvalue equation we want to study we now turn to a discussion
of the solutions. As a \hyperref[subsec:WavefunctionsPT]{warmup}, we briefly look at the example of the P\"oschl-Teller
problem for which the study of wave functions involves some fairly basic facts from
the theory of Gauss' hypergeometric functions. Then we turn to \hyperref[subsec:HarishChandra]{general $N$} and
discuss a basis of scattering states that are known as \hyperref[subsec:HarishChandra]{Harish-Chandra functions}.
We will discuss their definition and a new \textit{series expansion}
formula for $N=2$ along with a few direct consequences. As a main
application, in the \hyperref[subsec:HChPoles]{third subsection} we provide a complete analysis of \textit{poles and residues}
of Harish Chandra functions for $N=2$.
The \hyperref[subsec:Monodromy]{fourth subsection} finally,
is devoted to the construction of physical wave functions. Through a general discussion
of the \textit{monodromy representation} for the Calogero-Sutherland eigenvalue problem
we are led to consider special linear combinations of Harish-Chandra functions that
possess good analytic properties along the walls of the scattering problem. The optimal
choice corresponds to the so-called Heckman-Opdam multivariable hypergeometric functions
and, for $N=2$, a Euclidean analogue thereof.

\subsection{Wave functions of the P\"oschl-Teller problem}
\label{subsec:WavefunctionsPT}

Here we will mostly study the P\"oschl-Teller Hamiltonian on the domain $A^+$ that
was introduced in the \hyperref[subsec:PTHamiltonian]{previous section}. The corresponding eigenvalue equation was
stated in eq.\ \eqref{eq:PTeqAp}. We shall make the following Ansatz for the wave
function
\begin{equation} \label{eq:Hypgauge}
 \Psi(k;u) = \Theta(k;u) \Phi(\lambda,k;u) = 4^{a+\frac12} \sinh^{a-b+\frac12} \frac{u}{2}
\cosh^{a+b+\frac12} \frac{u}{2} \Phi(\lambda,k;u).
\end{equation}
As before, $k = (k_1,k_2) = (-2b,a+b+1/2)$ denotes the parameters of the
P\"oschl-Teller potential and $\lambda$ is the momentum. In our conventions it is
related to the energy eigenvalue $\epsilon$ in eq.\ \eqref{eq:Hypgauge} through
$\lambda^2 = - \epsilon$ so that $\lambda$ is purely imaginary for positive
energy solutions.

Since the P\"oschl-Teller potential tends to zero at $u = \infty$, the eigenfunctions
of the P\"oschl-Teller Hamiltonian are superpositions of an outgoing and an incoming
plane wave in this asymptotic regime. We can choose a basis for which one of the two
wave functions of energy $\epsilon$ is purely outgoing while the other is purely
incoming, i.e.\
\begin{equation} \label{CPWasym}
 \Psi(\pm\lambdan,k;u) \sim e^{\pm \lambda u} + \dots  \quad \textrm{for} \quad u
\rightarrow \infty \ .
\end{equation}
Wave functions with these asymptotic properties are also referred to as Harish-Chandra
functions. These two wave functions with eigenvalue $\varepsilon = - \lambda^2$ can easily
be constructed in terms of Gauss' hypergeometric functions as\footnote{Here and in the
following, we choose the principal branch for $(-z)^{A}$ and insist on $|\Im u| <
\pi$ in the $u$-plane.}
\begin{equation}\label{BC1-zexpansion}
\Phi(\pm\lambdan,k;z) = \left(\frac{1}{4}\frac{z}{z-1}\right)^{a+1/2\mp\lambda}
\pFq{2}{1}{1/2+a\mp\lambdan \pFcomma 1/2-b\mp\lambdan}{1\mp2\lambdan}{\frac{z}{z-1}},
\end{equation}
where $z$ and $u$ are related through equation \eqref{eq:zfromux}. For large values
of $u$, the argument of the hypergeometric functions approaches zero and the asymptotics
of the prefactor combine with that of the gauge transformation $\Theta$ to give the
desired asymptotics \eqref{CPWasym}. Since the standard series expansion for $ _2F_1
(\alpha,\beta;\gamma|y)$ converges on the real $y$-line for $-1 < y < 1 $ and our
parameter $z$ takes values $z \in (-\infty,0)$ on $A^+$, the usual series expansion
of the function \eqref{BC1-zexpansion} converges in the entire Weyl chamber.

In the context of Calogero-Sutherland models, one usually requires the series expansions
in the variable $x^{-1} = \exp (-u)$ to be convergent throughout the Weyl chamber, see below.
We shall refer to such expansions as $u$-expansions, even though they are really expansions
in $\exp(- u)$. To obtain such a $u$-expansion for the case at hand, one simply expresses
$z$ through $u$ in the series expansion of \eqref{BC1-zexpansion}\footnote{To avoid
additional subtleties, we keep the momenta generic here, namely we assume
that $\pm a+1/2\mp \lambda, \pm b+1/2\mp\lambda \neq 0,-1,-2,\dots$. Special cases can
be obtained by carefully taking limits, see comments in section \ref{sec:Wavefunctions} and appendix \ref{app:Integerspin}.} and then
expands $(1-e^{-u})^{-2(a+1/2\mp \lambda+k)}$ in powers of $\exp(-u)$. This second
expansion is  absolutely convergent for $u>0$. We arrive at
\begin{align}\label{BC1-uexpansion}
\Phi(\pm\lambdan;k;u)=\sum_{p=0}^{\infty}\frac{e^{(\pm\lambdan-1/2-a-p)u}}{p!}(2a+1 \mp2\lambda)_p\,
\pFq{3}{2}{-p\pFcomma 1/2+b\mp\lambda \pFcomma 1 +2a \mp 2\lambda+p}{1\mp2\lambda \pFcomma 1
+a \mp\lambda}{1}.
\end{align}
In accord with the general analysis (see \hyperref[subsec:HarishChandra]{next subsection}), this
expansion is convergent on the entire
domain $A^+$. It can be analytically continued to the strip $\{u | \Re u>0, |\Im u|<\pi \}$, which
is a tube-like neighborhood of $A^+$. If $b=0$ in \eqref{BC1-uexpansion}, one can use Watson's
summation for $_3F_2$ \cite{AAR} to sum the $u$-expansion formula \eqref{BC1-uexpansion} into
\begin{align*}
\Phi(\pm\lambdan;a,b=0;u)=e^{\left(\pm\lambda-\frac{1}{2}-a\right)u}\pFq{2}{1}{1/2+a\pFcomma 1/2+a \mp
\lambda}{1\mp\lambda }{e^{-2u}}\ .
\end{align*}
The resulting expression is well known from the theory of Calogero-Sutherland wave functions for
the reduced root system $C_1 \simeq B_1 \simeq A_1$.

After this short detour on series expansions for the in- and outgoing wave functions
we come back to the problem of constructing physical wave functions for the P\"oschl-Teller
problem. Clearly, the two wave functions we considered so far are badly behaved when we
approach the wall at $u=0$. In fact, $u=0$ is a branch point. But there exists a unique
linear combination of these two wave functions that is analytic at $u = 0$. It is given by
\begin{eqnarray}
\Phi^{W}(\lambdan,k;z) & = & c(\lambdan,k) \Phi(\lambda,k;z) + c(-\lambda,k)
\Phi(-\lambdan,k;z)\nonumber \\[2mm] \label{eq:PhiW}
& = & \
\pFq{2}{1}{a+1/2+\lambdan \pFcomma a+1/2-\lambdan}{1+a-b}{\frac{1}{z}} \ \ .
\end{eqnarray}
with coefficients given by
$$ c(\lambda,k) = 4^{-\lambdan+a+1/2} \frac{\Gamma(a-b+1)\Gamma(2\lambdan)}
{\Gamma(1/2+\lambdan+a)\Gamma(1/2+\lambdan-b)} \ . $$
These values then allow to apply Kummer's identity in order to pass from the first
to the second line in eq.\ \eqref{eq:PhiW}. Obviously, the branch point at $u=0$,
or equivalently at $z=-\infty$, has been removed now. After multiplication  with
the factor $\Theta$, $\Phi^W$ gives what we would usually consider the physical
solution of the hyperbolic P\"oschl-Teller system. The eigenfunctions
$\Psi^W(\lambda,k;z)$ with $\lambda = i p, p \geq 0,$ form a complete orthonormal
basis of eigenfunctions for the hyperbolic P\"oschl-Teller problem on $A^+$
\cite{Koornwinder-Jacobi}, with (appropriately normalized) measure $d\mu \sim
|\Theta(k;u)|^2 du$.

The branch points of $\Phi(\pm \lambdan,k;z)$ at $u=0$ prevent us from continuing
the purely in- and outgoing wave functions beyond $A^+$ into the compact domain
$A^C$ that was defined in eq.\ \eqref{eq:AC}. On the other hand, the function
$\Phi^W$ can be continued into $A^C$. For generic choices of $\lambda$ the
resulting wave function of the trigonometric P\"oschl-Teller problem
\eqref{eq:PTeqAC} possesses a branch point at $u=i\pi$. This branch point
can only be avoided for a discrete set of $\lambda$. In this way one obtains
the usual eigenfunctions $\psi_n$ of the trigonometric P\"oschl-Teller problem
\begin{equation} \label{eq:ACBC1}
\psi_n(a,b;\varphi) \sim \sin^{a-b+\frac12} \frac{\varphi}{2} \cos^{a+b+\frac12}
\frac{\varphi}{2} \
\pFq{2}{1}{-n \pFcomma 2a+1+n}{1+a-b}{\sin^2\frac{\varphi}{2}}
\end{equation}
for $\pm\lambda = n+a+1/2$ where the variable $\varphi$ is restricted to the interval
$\varphi \in [0,\pi]$ so that it parametrizes $A^C$. Note that for $n=0$, i.e. for the
ground state, the hypergeometric function contributes a trivial factor. Hence, the wave
function of the ground state in the compact domain coincides with the function $\Theta$
we introduced in eq.\ \eqref{eq:Hypgauge}.

Let us finally also discuss the wave functions of the P\"oschl-Teller problem
on the shifted domain $\tilde A^+ = A^L$ that we defined in eq.\ \eqref{eq:tAp}.
In this case, we select the Harish-Chandra functions such that the eigenfunctions
$\Psi$ possess the standard asymptotics in the real coordinate $\tilde u$ on
$\tilde A^+ = A^L$, i.e.
\begin{equation} \label{tildeCPWasym}
 \tilde \Psi(\pm\lambdan,k;u) \sim e^{\pm \lambda \tilde u} + \dots =
 e^{\pm i \pi \lambda}   e^{\pm \lambda u} + \dots  \quad \textrm{for}
 \quad \tilde u = u-i\pi \rightarrow \infty \ .
\end{equation}
Of course this implies that the Harish-Chandra functions $\tilde \Phi$ are related
to $\Phi$ by a $\lambda$-dependent gauge transformation
\begin{equation} \label{eq:tPhiPhi}
 \tilde \Phi(\pm \lambda,k;z) = e^{ i \pi (\pm\lambda-a-1/2)} \Phi(\lambda,k;z) \  .
\end{equation}
Solutions of our Calogero-Sutherland problem with these asymptotics take the
form
\begin{equation}\label{tildePhi-zexpansion}
\tilde \Phi(\pm\lambdan,k;z) = \left(\frac{z}{4}\right)^{a+1/2\mp\lambda}
\
\pFq{2}{1}{a\mp\lambdan+1/2\pFcomma b\mp\lambdan+1/2}{1\mp 2\lambdan}{z}.
\end{equation}
The relation \eqref{eq:tPhiPhi} with the standard Harish-Chandra functions $\Phi$ is then a
consequence of the Pfaff transformation for Gauss hypergeometric function \cite{AAR}. In
writing our formula for $\tilde \Phi$ we agree to use the principal branch for $z^A$
so that in $z$-plane the twisted Harish-Chandra function has cuts along $(-\infty,0)\cup
(1,+\infty)$ for generic values of parameters. Note that on $\tilde A^+ = A^L$ the
variable $z$ takes values in $z = [0,1[$, so that this usual series expansion of the
functions $\tilde \Phi$ is convergent on the entire shifted Weyl chamber. As above,
this function can be analytically continued to a semistrip $\{u| \Re u>0, 0<\Im u<2\pi\}$.
Once again we need to form a special linear combination of these twisted Harish-Chandra
functions $\tilde \Phi$ to obtain the physical wave function which is regular at $u =
i \pi$ or equivalently $z=1$. It is given by
\begin{eqnarray}
\tilde \Phi^{W}(\lambdan,k;z) & = & \tilde c(\lambdan,k) \tilde\Phi(\lambda,k;z)
+ \tilde c(-\lambda,k) \tilde\Phi(-\lambdan,k;z)\nonumber \\[2mm] \label{eq:tildePhiW}
& = & z^{a+\frac12-\lambda}
\
\pFq{2}{1}{a+1/2-\lambdan\pFcomma b+1/2-\lambdan}{1+a+b}{1-z} \ \ .
\end{eqnarray}
with coefficients
$$
\tilde c(\lambda,k) = 4^{-\lambdan+a+1/2} \frac{\Gamma(a+b+1)\Gamma(2\lambdan)}
{\Gamma(1/2+\lambdan+a)\Gamma(1/2+\lambdan+b)} =: c_0(\lambda,k)\ .
$$
We observe that $\tilde c$ can be obtained from $c$ through the inversion $b \rightarrow
-b$ of the parameter $b$. This replacement agrees with the action of the shift $\varrho$
on the coupling constants in the P\"oschl-Teller problem, see eq.\ \eqref{eq:varrho}. The
functions $\tilde \Psi^W(\lambda,k;z)$ for $\lambda = ip, p \geq 0,$ provide us with a
complete orthonormal basis of eigenfunctions for the P\"oschl-Teller problem on $\tilde A^+$.
\medskip

There is a different, more algebraic way to express these results on the construction
of physical wave functions for the Calogero-Sutherland problems through representations
of the fundamental group $\pi_1(D)$. Recall from the \hyperref[subsec:PTHamiltonian]{previous section}, that the domain
$D$ is a semi-infinite pillow. Hence, its fundamental group is freely generated by two
elements $g_0$ and $g_1$. These are described by loops around the singular points $u_0
= i\pi$ and $u_1 = 0$. The Harish-Chandra functions $\Phi(\pm \lambda,k;z)$ carry a
2-dimensional monodromy representation $M$ of this fundamental group. We can easily
infer the representation matrices from standard properties of hypergeometric functions,
along the lines of our discussion above. For $g_1$ one finds that
\begin{equation}
 M(g_1) = M_1 =  C^{-1} \left( \begin{array}{cc} 1 & 0 \\ 0 & e^{2\pi i(a-b)}
 \end{array}\right) C ,
\end{equation}
where the matrix $C$ is given by
$$
C(\lambda,k) = \left( \begin{array}{cc} c(\lambda,k) &  c(-\lambda,k)\\
                      c(\lambda,k') & c(-\lambda,k')
                      \end{array} \right)
$$
and $k' = (k'_1,k'_2) = (2b,-a-b+1/2)$ is obtained by reversing the sign of both
$a$ and $b$. Our discussion of the regular solution at $u_0$ shows that $M(g_0) =
M_0$ can be computed in the same way after we perform a gauge transformation from
$\Phi$ to $\tilde \Phi$, i.e.
\begin{equation}
 M(g_0) = M_0 = \Omega^{-1} \tilde M_1 \Omega  \\[2mm]
\end{equation}
where $\tilde M_1$ is obtained from $M_1$ by the substitution $b\rightarrow -b$ and
the matrix $\Omega$ encodes the gauge transformation \eqref{eq:tPhiPhi}. It reads
$$
\Omega(\lambda,k) = \left( \begin{array}{cc} e^{i\pi (\lambda-a-1/2)}  & 0 \\
                      0 & e^{i\pi(-\lambda-a-1/2)} \end{array} \right)\ .
$$
We observe that both monodromy matrices possess one eigenvector with unit eigenvalue. This
signals the existence of a special linear combination of Harish-Chandra functions
which has trivial monodromy around $u_0$ and $u_1$, respectively. Explicitly, these
eigenvectors were given by the functions $\Phi^W$ and $\tilde \Phi^W$. Finally, we
also want to stress that the product $ M_0 M_1$ of the two monodromies describes the
monodromy of the Harish-Chandra functions at $u = \infty$. The latter is fully
determined by asymptotic behavior of Harish-Chandra functions, i.e.
\begin{equation}
M_\infty = M_0 M_1  = e^{2\pi i (a+\frac12)} \left( \begin{array}{cc}
e^{-2\pi i \lambda} & 0  \\ 0 & e^{2\pi i \lambda}  \end{array}\right).
\end{equation}
We will now
explain that all this carries over to $N \geq 2$. In particular, the monodromy group
and its representation on Harish-Chandra functions is explicitly known, see the
section \ref{subsec:Monodromy} below.

\subsection{Harish-Chandra series expansions}
\label{subsec:HarishChandra}

Let us now discuss eigenstates of the hyperbolic Calogero-Sutherland Hamiltonian for
$N >1 $. Our discussion will start with the Weyl chamber $A^+_N$ in which all
the $u_i$ are real and positive. To construct the solutions we are interested in, we
note that in the region of large $u$ where we are far away from all walls of the
Weyl chamber, the Calogero-Sutherland potential goes to zero and hence, in this regime,
any wave function is a superposition of plane waves. Before we give precise definitions
let us split off a simple factor from the eigenfunctions $\Psi$ and introduce a new
function $\Phi$ through
\begin{equation} \label{eq:Theta}
 \Psi(k;u) = \Theta(k;u) \Phi(\lambda,k;u) = \prod_{\alpha \in \Sigma^+}
\left( 2 \sinh \frac{\langle\alpha,u\rangle}{2}\right)^{k_\alpha} \
\Phi(\lambda,k;u).
\end{equation}
The factor $\Theta(k;u)$ is split off for convenience, see below. In the case of $N=1$
it reduces to the one we worked with in the \hyperref[subsec:WavefunctionsPT]{previous subsection} in order
to map the eigenvalue equation for the P\"oschl-Teller Hamiltonian to the hypergeometric differential
equation in the variable $1/z$. We will often refer to the factor $\Theta$ as a \textit{gauge
transformation} and to $\Phi$ as a Calogero-Sutherland wave functions in hypergeometric
gauge. Let us note in passing that the function $\Theta(k,u)$ possesses the following
asymptotics for large $u$,
\begin{equation}
\Theta(k;u) \sim e^{\langle \rho(k), u\rangle} + \dots \quad \textrm{ with } \quad \rho(k) =
\frac12\sum_{\alpha \in \Sigma^+} \alpha k_\alpha\ .
\end{equation}
So-called Harish-Chandra wave functions $\Phi(\lambdan,k;z)$ are $W_N$ symmetric solutions
of the Calogero-Sutherland Hamiltonian for which $\Phi$ possesses the following simple
asymptotic behavior
\begin{equation}\label{HCasym}
\Phi(\lambdan,k;u)\  \sim \ e^{\langle \lambdan - \rho(k), u \rangle}
+ \dots \ \mbox{ for } \ u \rightarrow \infty\ \mathrm{in} \ A^+_N = \textit{WC}_N
\end{equation}
where $\lambdan = \sum \lambdan_i e_i$ is the vector of momenta and $u \rightarrow
\infty$ in $A^+_N$ means that all components become large while preserving the order
$u_N < u_{N-1} < \dots < u_1$. Recall that $W_N$ symmetry means that $\Phi$ depends
on the $u_i$ through $z_i$ and that it is symmetric under all permutations of the
$z_i$. The  condition \eqref{HCasym} selects a unique solution of the scattering
problem describing a single plane wave. It is analytic in the Weyl chamber
$A^+_N$. The corresponding eigenvalue of the Calogero-Sutherland Hamiltonian is
given by
$$ \varepsilon = \varepsilon(\lambda) = - \sum \lambda_i^2 \ . $$
When we required the Harish-Chandra functions to be symmetric, we used the action
of the Weyl group $W_N$ on the coordinate space $\mathbb{T}_N$. On the other hand,
the Weyl group also acts in a natural way on the asymptotic data $\lambda$ of the
Harish-Chandra functions by sending any choice of $\lambda$ through a sequence of
Weyl reflections to $w \lambda, w \in W_N$. Since the eigenvalue $\varepsilon$ is
invariant under all the reflections, our Harish-Chandra functions come in families.
For generic choices of $\lambda$, one obtains $|W_N| = N! 2^N$ solutions
$\Phi(w\lambda,k;z)$ which all possess the same eigenvalue of the Hamiltonian.

At least for sufficiently generic values of the momenta,\footnote{A precise formulation
of the condition is given below through the inequalities \eqref{HCh-rec-condition} and 
the subsequent discussion.} Harish-Chandra
functions can be constructed as a series expansion in the variables $x_i = \exp u_i$
\begin{align}\label{HCh-Heckman-def}
\Phi(\lambdan,k;u) = \sum_{\mu\in Q_+} \Gamma_{\mu}
(\lambda,k)e^{\langle \lambda-\rho(k)-\mu,u\rangle}, \quad \quad  \Gamma_{0}(\lambda,k)=1,
\end{align}
where we adopt $|\Im u_i|<\pi$ for $i=1, \dots,N$ on the principal branch of \textit{BC}$_N$
Harish-Chandra functions and we sum over elements $\mu$ of the $\mathbb{Z}_{\geq 0}$-cone $Q_+$
over the positive roots, i.e. the set
$$ Q_+ = \{ \mu = \sum_{i=1}^N n_i \alpha_i \, | \, n_i \geq 0 \ \textrm{for} \ i=1,
\dots, N\, \}\ . $$
For later discussions we note that $Q_+$ comes equipped with a partial order $\preceq$ where
$\mu \preceq \nu$ iff $\nu - \mu \in Q_+$.

It is not difficult to derive recursion relations of the expansion coefficients
$\Gamma_{\mu}(\lambda,k)$ directly from the Calogero-Sutherland eigenvalue problem, see
e.g. \cite{HeckmanBook} \footnote{The recursion derived in \cite{Hogervorst:2013sma} is
closely related to this one, specialized for $\BC_2$. However, our expansion here is in
monomials, not in Gegenbauer polynomials, although it is not difficult to go between
the two.}
\begin{align}\label{Heckman-recursion}
\left\langle 2\lambda-\mu, \mu \right\rangle  \Gamma_{\mu}(\lambda,k)=
2\sum_{\alpha\in \Sigma^+} k_{\alpha} \sum_{j\geq 1}
\left\langle \lambda-\rho(k)-\mu+j \alpha, \alpha \right\rangle  \Gamma_{\mu-j\alpha}(\lambda,k).
\end{align}
This can be solved uniquely, if
\begin{align}\label{HCh-rec-condition}
2\langle \lambda,\mu \rangle -\langle \mu,\mu \rangle \neq 0 \quad \textrm{ for all } \ \mu\in Q_+\ .
\end{align}
The resulting series are known to converge\footnote{The series converges absolutely and uniformly on compacta in
the set of complexified momenta and multiplicities times $A^+$, as long as they are chosen to avoid the hyperplanes
\eqref{HCh-rec-condition}.} within the Weyl chamber \cite{HeckmanBook}, as is fairly obvious from the physics they
describe. Even when one of the conditions \eqref{HCh-rec-condition} is violated it is possible to obtain a complete
basis of series solutions $\Phi(w\lambda,k ; u), w\in W_N$. There is a subset of such cases where things are a bit
subtle, namely when $\lambda$ is chosen such that $\langle \lambda,\alpha^{\vee} \rangle$ is integer which implies 
that one of the conditions in eq.\ \eqref{HCh-rec-condition} is violated. For such values of non-generic momenta 
$\lambda$ some of the series solutions are logarithmic. This is analogous to usual properties of the Gauss
hypergeometric function when the difference of exponents becomes integer \cite{AAR}. To obtain their expansions,
one needs to see which fundamental solutions coincide on the corresponding locus and take the limit of their 
rescaled difference as a new fundamental solution.
\medskip

\noindent
{\bf Example:} Let us look at the Harish-Chandra functions for $N=2$ in some detail. Although,
we could literally repeat the entire $N=1$ discussion here, we leave much of it for appendix
\ref{app:zxexpansions}.\footnote{In appendix \ref{app:zxexpansions}, we derive a $z$-expansion for twisted $BC_2$ Harish-Chandra function which is related to the present one by formula \eqref{twistedHCh-BC2}}. In particular, this appendix contains explicit expressions for the
expansion coefficients $\Gamma_\mu$ in the $u$-expansion \eqref{HCh-Heckman-def}, see eq.\
\eqref{BC2-uexpansion}. We do not want to repeat these here and will rather discuss a
somewhat intermediate form of a $z_i$-expansion from which we shall derive many interesting
properties properties of Harish-Chandra functions in the remainder of this and in the \hyperref[subsec:HChPoles]{next subsection}.
It is given by
\begin{align}\label{BC2-zexpansion}
&\Phi(\lambda_1,\lambda_2;k_i;z_1,z_2)=\frac{1}{4^{2a+1+\epsilon/2-\lambda_1-\lambda_2}}
\sum_{n,m=0}^{\infty}\frac{\left(1/2+a-\lambda_1,1/2-b-\lambda_1,\frac{\epsilon}{2}-\lambda_1+\lambda_2\right)_{n}}
{\left(1-2\lambda_1,1-\lambda_1+\lambda_2\right)_n}\nonumber\\[2mm]
&\hspace*{1cm} \times\frac{\left(1/2+a-\lambda_2,1/2-b-\lambda_2, \epsilon/2+\lambda_1-\lambda_2\right)_m}
{\left(1-2\lambda_2,1+\lambda_1-\lambda_2\right)_m} \frac{\left( 1-\epsilon/2-\lambda_1+\lambda_2\right)_{n-m}}
{\left(-\lambda_1+\lambda_2\right)_{n-m}}\nonumber\\[2mm]
&\hspace*{1.5cm} \times \pFq{4}{3}{-n \pFcomma  -m \pFcomma 1-\epsilon/2 \pFcomma  1-\epsilon/2-\lambda_1-\lambda_2  }
{1-\epsilon/2 +\lambda_1-\lambda_2 -n \pFcomma 1-\epsilon/2 -\lambda_1+\lambda_2 -m \pFcomma 1-\lambda_1-\lambda_2 }{1}
\\[3mm]
& \times\frac{1}{n!m!}\, \left(\frac{z_1}{z_1-1}\right)^{\frac{1+\epsilon}{2}+a-\lambda_1+n}\left(\frac{z_2}{z_2-1}\right)^{\frac{1}{2}+a-\lambda_2+m} \,
\pFq{2}{1}{\epsilon/2-\lambda_1+\lambda_2-m+n \pFcomma \epsilon/2}{1-\lambda_1+\lambda_2-m+n}{\frac{z_1}{z_2}\frac{z_2-1}{z_1-1}}.\nonumber
\end{align}
As in the discussion for $N=1$ we choose the principal branch for $(-z_i)^{A}$, so that $z_i/(z_i-1)=\sinh ^{-2}(u_i/2)$. It is obvious that
this function has the correct asymptotic behaviour. Let us stress that, unlike a somewhat similar
expansion for conformal blocks that appears in \cite{Dolan:2003hv}, our expansion for Harish-Chandra
functions is also valid for non-integer spins $l$. We can use it to derive a corresponding expansion for conformal
blocks once we have explained how to construct blocks from Harish-Chandra functions in the
\hyperref[sec:Blocks]{next section}. The derivation of eq.\ \eqref{BC2-zexpansion}, its features and equivalent
expansions are described in appendix \ref{app:zxexpansions}.
We note that the $z$-expansion \eqref{BC2-zexpansion} is convergent for arguments in the region
\begin{align}
\Re z_i < \frac{1}{2}, i=1,2, \quad \,\, \left|\frac{z_1}{z_1-1}\right|<\left|\frac{z_2}{z_1-1}\right|,
\end{align}
which includes the entire Weyl chamber, similarly to the $BC_1$ case.
Here we assume that the parameters are generic.

As an immediate application of the series expansion \eqref{BC2-zexpansion} we can evaluate Harish-Chandra
functions for some special values of the multiplicities $k_i$. For $d = \epsilon + 2 = 2$, for example, the
multiplicity $k_3 = 0$ vanishes so that one of the upper parameters in the balanced $_4F_3(1)$ inside the
sum coincides with one of the lower ones. The sum involving the resulting balanced $_3F_2(1)$ is summable
via Saalsch\"utz identity \cite{BaileyBook} and we obtain
\begin{align}
\Phi(\lambda_i;k_1,k_2,k_3 = 0;z_i)&= \prod_{i=1,2} \left(\frac{1}{4}\frac{z_i}{z_i-1}
\right)^{1/2+a-\lambda_i}\pFq{2}{1}{1/2+a-\lambda_i \pFcomma 1/2-b-\lambda_i}{1-2\lambda_i}{\frac{z_i}{z_i-1}}.
\nonumber
\end{align}
The result is a product of Harish-Chandra functions for the P\"oschl-Teller problem $N=1$, see eq.\ \eqref{BC1-zexpansion}.
The case of $d=4$ is even simpler to evaluate. Indeed, for this value of $d$, the parameter $k_3 = (d-2)/2 = 1$
and hence one of the upper parameters in the balanced $_4F_3(1)$ tends to zero, so that we obtain
\begin{align}
\Phi(\lambda_1,\lambda_2;k_1,k_2,k_3=1;z_1,z_2)&= \nonumber \\[2mm] & \hspace*{-3cm} =
\frac{1}{4}\frac{z_1 z_2}{z_1-z_2}\, \prod_{i=1,2} \left(\frac{1}{4}\frac{z_i}{z_i-1}\right)^{a+1/2-\lambda_i}
\pFq{2}{1}{1/2+a-\lambda_i \pFcomma 1/2-b-\lambda_i}{1-2\lambda_i}{\frac{z_i}{z_i-1}}. \nonumber
\end{align}
Before we derive further properties of Harish-Chandra functions for $N=2$ from the series expansion
\eqref{BC2-zexpansion}, we want to review a few more general properties that hold for any $N$.
\medskip

The last two formulas for $N=2$ that we derived from the $z$-expansion \eqref{BC2-zexpansion} possess a
nice generalization to arbitrary values of $N$. For $k_3 = 0,1$ and generic (non-resonant) values of the
eigenvalues $\lambda$ is known that \cite{Shimeno2008}
\begin{align}\label{Shimeno-factor}
\Phi(\lambda_i;k_i;z_i)&=\Delta_m^{-k_3}\  \prod_{i=1}^N \left(\frac{1}{4}
\frac{z_i}{z_i-1}\right)^{a+1/2-\lambda_i}
\pFq{2}{1}{1/2+a-\lambda_i \pFcomma 1/2-b-\lambda_i}{1-2\lambda_i}{\frac{z_i}{z_i-1}}, \nonumber
\end{align}
where
$$\Delta_{m}= \prod_{\alpha\in\Sigma^+, \textit{ middle}}
\left(e^{\langle\alpha,u\rangle}-e^{-\langle\alpha,u\rangle}\right)$$
is a Weyl denominator for middle roots.
More generally, it is known \cite{OpdamDunkl,Shimeno2008} that Harish-Chandra functions for any positive
integer value of the multiplicity $k_3$ are multilinear combinations of P\"oschl-Teller wave functions.
The most elegant derivation of such expressions, one that is also completely universal in $N$, involves
$N(N-1)/2$ differential or difference operators which shift $k_3$ by one unit. For scalar 4-point blocks
similar constructions are known from the work of Dolan-Osborn \cite{Dolan:2003hv, Dolan:2011dv}. We will
show how explicit closed formulas for these operators follow from the integrable structure of
Calogero-Sutherland models in a forthcoming publication \cite{alg-structures}.
\medskip

\subsection{Poles and residues of Harish-Chandra functions}
\label{subsec:HChPoles}

Our short excursion to explicit expressions for Harish-Chandra functions which exist for
special values of the multiplicities only should not mislead the reader to think that
Harish-Chandra functions can only be understood for very special cases. In fact, it is
one of the central virtues of Heckman-Opdam theory that one is able to say so much about
Harish-Chandra functions often without knowing their explicit series expansions or
integral formulas. As an example for one of the many further properties that is well
understood beyond the simple case of $N=1$ let us mention that $\exp(\langle-\lambda+\rho(k),u\rangle)
\Phi(\lambda;k;u)$ are entire functions of the multiplicities $k_i$ and meromorphic function
of asymptotic data $\lambda_i$, for any choice of $u$ in the fundamental domain. They are known
to possess simple poles whenever the set of $\lambda_i$ satisfies one of the following conditions
\begin{equation} \label{eq:poles}
 \langle \lambda_\ast, \alpha^\vee \rangle =  s
\quad  \textrm{for}  \quad s = 1,2, \dots \quad , \quad \alpha \in \Sigma^+\ .
\end{equation}
Let us note that given $\lambda_\ast$ satisfying this condition, this $\lambda_\ast$
violates the condition \eqref{HCh-rec-condition}. In fact, the quadratic expression in eq.\
\eqref{HCh-rec-condition} vanishes at least for $\mu = s \alpha^\vee$. The converse is not
true, i.e. there exist many values of $\lambda$ that violate the inequalities
\eqref{HCh-rec-condition} but do satisfy the condition \eqref{eq:poles}. At such values
of $\lambda$ Heckman and Opdam showed that the singularities are only apparent. For the
true poles at $\lambda_\ast = \lambda_{\alpha,n}$, the residues are given by
(see e.g. \cite{OpdamDunkl})
\begin{equation} \label{eq:residues}
\text{Res}_{(\alpha,s)} \Phi(\lambda,k;z) \sim \Phi(w(\alpha)\lambda_{\alpha,s},k;z)\ .
\end{equation}
where $\sim$ indicates that the relation with the Harish-Chandra function $\Phi(w(\alpha))$
holds only up to an constant factor. The latter is not known in general, but we shall
explain later how it can be determined and provide explicit expressions for $N=2$ with
the help of the series expansion \eqref{BC2-zexpansion}.

Relation \eqref{eq:residues} is actually a little more subtle than it may appear at first.
According to a theorem by Heckman (\cite{HeckmanBook}, Cor. 4.2.4), eq.\ \eqref{eq:residues}
holds true as it stands whenever the quadratic equation $\langle 2\lambda_0-\mu, \mu\rangle
= 0$ has exactly one non-zero solution $\mu_0\in Q_+$. In this case, one has
\begin{align}\label{H-theorem}
\left\{\langle 2\lambda-\mu_0, \mu_0\rangle \Phi(\lambda;k;u)\right\}\biggl |_{\lambda_0}=\left\{ (2\lambda-\mu_0,\mu_0) \Gamma_{\mu_0}(\lambda,k) \right\}\biggl |_{\lambda_0}\ \Phi(\lambda_0-\mu_0;
k;u) \ .
\end{align}
Here, the Harish-Chandra function on the right hand side appears from summing over all
terms in the original series with $\mu \succeq \mu_0$ with respect to the partial ordering
of $Q_+$ we introduced above. If the quadratic equation has several solutions in $Q_+$, on
the other hand, relation \eqref{eq:residues} holds only for a properly defined Harish-Chandra
series on the right hand side. One can do this by approaching the desired limit point $\lambda_\ast$
along a sequence of irrational values $\lambda$ for which $\langle 2\lambda-\mu, \mu\rangle = 0$ has a
unique solution. Then, provided one knows the series expansion of Harish-Chandra function for
such $\lambda$, it is possible to calculate its residues by Heckman's theorem.
\medskip

\noindent
{\bf Example:} Let us continue our tradition to describe a few more details in the
case of $N=2$. As we discussed before, the corresponding Weyl group $W_2$ is eight
dimensional. It is generated by the two reflections $w_1$ and $w_2$. All eight
elements are listed in table \ref{tab:table2} along with their action $\lambda
\rightarrow w \lambda$ on the asymptotic data $\lambda$ of the Harish-Chandra
functions. In the case at hand, the shift by $\rho(k)$ reads
$$\rho(k) = 1/2 (k_1+2k_2 + 2 k_3, k_1+2 k_2) = (a+1/2 + \epsilon/2, a+1/2)\ . $$
The final column of table \ref{tab:table2} will be explained in section \ref{sec:Blocks}. It
describes the asymptotic data in a different parametrization.

\begin{table}[h!]
  \centering
  \caption{Elements of the Weyl group $W_2$ for the \textit{BC}$_2$ root system (first column)
  along with their action on the asymptotic data $\lambda = (\lambda_1,\lambda_2)$ (second
  column). The third column is identical to the second, but uses a different parametrization
  of the asymptotic data that we will discuss in section \ref{sec:Blocks}.}
  \label{tab:table2}
  \begin{tabular}{l|c|c}
  $w\in W_2$ & $w(\lambdan_1,\lambda_2)$ & $w(\Delta,l)$ \\[2mm]
    \hline & & \\[-3mm]
    $e$  & $\left(\lambdan_1,\,\, \lambdan_2\right)$ & $\left(\Delta,l \right)$ \\[1mm]
    $w_1$ & $\left(\lambdan_2,\,\, \lambdan_1\right)$ & $\left(\Delta,2-l-d\right)$  \\[1mm] \hline & & \\[-3mm]
    $w_2$  & $\left(\lambdan_1,\,\, -\lambdan_2\right)$ & $\left(1-l,1-\Delta\right)$ \\[1mm]
    $w_1 w_2$  & $\left(-\lambdan_2,\,\, \lambdan_1\right)$ & $\left(1-l,\Delta-d+1\right)$  \\[1mm]\hline & & \\[-3mm]
    $w_2 w_1$  & $\left(\lambdan_2,\,\, -\lambdan_1\right)$ &  $\left(l+d-1,1-\Delta\right)$ \\[1mm]
    $w_1w_2w_1$  & $\left(-\lambdan_1,\,\, \lambdan_2\right)$ &   $\left(l+d-1,\Delta + 1 -d\right)$ \\[1mm]\hline & & \\[-3mm]
    $w_2 w_1 w_2 $  & $\left(-\lambdan_2,\,\, -\lambdan_1\right)$ & $\left(d-\Delta,l\right)$  \\[1mm]
    $w_1w_2w_1w_2$  & $\left(-\lambdan_1,\,\, -\lambdan_2\right)$ &  $\left(d-\Delta,2-l-d\right)$ \\
  \end{tabular}
\end{table}

If we denote an element of $\mu \in Q_+$ as $\mu = n \alpha_1 + m \alpha_2 = n e_1 + (m-n) e_2 =(
n,m-n)$ for $n,m\in \mathbb{Z}_{\geq 0}$, the locally finite set of hyperplanes on which the
definition of the Harish-Chandra series requires an appropriate limiting procedure reads
\begin{align}\label{BC2-diofantine-gen}
2n \lambda_1+2(m-n)\lambda_2=n^2+(m-n)^2.
\end{align}
As we stated before, only a small subset of these hyperplanes give actual poles. Namely,
according to equation \eqref{eq:poles} the Harish-Chandra functions possess four series of poles in
the asymptotic data $\lambda$. In fact, while the set $\Sigma^+$ of positive roots contains six elements,
the solutions of eq.\ \eqref{eq:poles} for the longest roots $2e_i$ form a subspace of the solutions for
the short roots $e_i$. The relevant four roots are listed table \ref{tab:table3} along with the
corresponding Weyl reflection written in terms of our fundamental generators $w_1$ and $w_2$. For
each of these four cases, the solutions of eq.\ \eqref{eq:poles} are listed in the third column. They
depend on a non-negative integer and one free parameter $\lambda$. The residues of our Harish-Chandra
functions for these values of $\lambda_\ast$ are again given by Harish-Chandra functions with different
asymptotic data. The latter is listed in the forth column of table \ref{tab:table3}. The information
of the third and forth column is repeated in the last two columns in a different parametrization of
the asymptotic data that we explain in section \ref{sec:Blocks}.

\begin{table}
  \centering
  \caption{Pole positions (first column) and asymptotic data of their residues (second column) of the
  Harish-Chandra functions $\Phi(\lambda_1,\lambda_2,k;z)$ for $N=2$. The third and forth column contain
  the same information in a different parametrization of asymptotic data, see section \ref{sec:Blocks}.}
  \label{tab:table3}
  \begin{tabular}{l|c||c|c||c|c}
  $\alpha$ & $w(\alpha)$ & $(\lambda_1\lambda_2)_\ast$ & $w(\alpha)(\lambda_1,\lambda_2)_\ast$ &
  $(\Delta,l)_\ast $ &  $w(\alpha)(\Delta,l)_\ast$ \\[2mm]
    \hline  & & & & & \\[-3mm]
  $e_1$     & $w_1 w_2 w_1$ & $(s/2,\lambda)$       & $(-s/2,\lambda)$      & $(l+d-s-1,l)$      & $(l+d-1,l-s)$ \\[1mm]
  $e_2$     & $ w_2 $       & $(\lambda,s/2)$       & $(\lambda,-s/2)$      &   $(1-s-l,l)$        & $(1-l,l+s)$\\[1mm]
  $e_1+e_2$ & $w_2w_1w_2$   & $(-\lambda+s,\lambda)$ & $(-\lambda,\lambda-s)$ & $(d/2-s,l)$        & $(d/2+s,l)$ \\[1mm]
  $e_1-e_2$ & $w_1$         & $(\lambda+s,\lambda)$  & $(\lambda,\lambda+s)$  & $(\Delta,1-d/2+s)$ & $(\Delta,1-s-d/2)$\\
  \end{tabular}
\end{table}

In order to obtain exact formulas for the residues, including the numerical coefficients, we shall employ the
series expansion \eqref{BC2-zexpansion}. The general idea is simple to state. As we know, the residue should be
proportional to a Weyl-reflected Harish-Chandra function, so one just needs to locate the terms of the series
expansion that diverge as we send $\lambda$ close to a corresponding pole position. Within the set of these
divergent terms one needs to identify the one that gives the leading power of $\Phi(w(\alpha)\lambda_{\alpha,s},
k;z)$ and read off the coefficient. Heckman's theorem quoted above guarantees that we can interchange the
summation and the limit. In principle, the described steps can be carried out both for the usual $u$-expansion
(see appendix \ref{app:zxexpansions}) and the $z$-expansion we spelled out in equation \eqref{BC2-zexpansion}.
Here we shall sketch the derivation from the latter and spell out the explicit residues for all four families
of poles.

According to our table, the residue for the series $\lambda_1=s/2+\xi$ is proportional to $\Phi(-\frac{s}{2},
\lambda_2;k;z(u))$. It is easy to see that the leading divergent term in the $z_i$-expansion \eqref{BC2-zexpansion}
arises from the summands with $n=s$, $m=0$, $p=0$. Here and in the following, the summation index $p$ refers to the
series expansion of the hypergeometric function $ _2F_1$ in the last line of equation \eqref{BC2-zexpansion}. Hence,
the two hypergeometric functions in the summands of eq.\ \eqref{BC2-zexpansion} can be replaced by $F \sim 1$ in
all divergent terms. Keeping in mind that $1/\left(1-s-2\xi\right)_s \sim (-1)^s/2\xi(s-1)!$, we obtain
\begin{align*}
&\Phi\left(\lambda,k;z(u)\right)\biggl|_{\lambda_1 \sim \frac{s}{2}+\xi}= \\[2mm]
&\frac{1}{\xi} \left(\frac{1}{2}\frac{4^s}{s!(s-1)!}\frac{\left(a+\frac{1-s}{2},b+\frac{1-s}{2}, \frac{\epsilon}{2}+
\lambda_2-\frac{s}{2},1-\frac{\epsilon}{2}+\lambda_2-\frac{s}{2}\right)_s}{\left(\lambda_2-\frac{s}{2}, 1+
\lambda_2-\frac{s}{2}\right)_s}x_1^{-a-\frac{1+\epsilon}{2}-\frac{s}{2}}x_2^{-a-\frac{1}{2}+\lambda_2}+\dots\right),
\end{align*}
so that the residues are given by
\begin{align}\label{res1}
\Res_{\lambda_1=s/2}\Phi\left(\lambda,k;z(u)\right)&=\frac{1}{2}\frac{4^s}{s!(s-1)!}\frac{\left(a+\frac{1-s}{2},b+\frac{1-s}{2}, \frac{\epsilon}{2}+\lambda_2-\frac{s}{2},1-\frac{\epsilon}{2}+\lambda_2-\frac{s}{2}\right)_s}{\left(\lambda_2-\frac{s}{2}, 1+
\lambda_2-\frac{s}{2}\right)_s}\\[2mm]    & \hspace*{7cm} \times \,\, \Phi(-\frac{s}{2}, \lambda_2;k;z(u)).\nonumber
\end{align}
The evaluation of the residue for the family $\lambda_2=s/2+\xi$ is even simpler. Since it is proportional to
$\Phi(\lambda_1,-\frac{s}{2};k;z(u))$, it must arise from the terms $n=0$, $m=s$, $p=0$ in eq.\ \eqref{BC2-zexpansion}.
Once again it is easily seen that the two inner hypergeometric functions that appear in the summands do not contribute
so that the residue reads
\begin{align}\label{res2}
\Res_{\lambda_2=s/2}\Phi\left(\lambda,k;z(u)\right)=\frac{1}{2}\frac{4^s}{s!(s-1)!}\left(a+\frac{1-s}{2},
b+\frac{1-s}{2}\right)_s\ \Phi(\lambda_1,-\frac{s}{2};k;z(u)).
\end{align}
For the third family of poles at $(\lambda_1,\lambda_2)_*=(-\lambda+s,\lambda)$ the residue is proportional to
$\Phi(-\lambda,\lambda-s;k;z(u))$. The relevant terms in the sum arise from $n=m=s$, $p=0$. For these values of
indices, the inner $_2F_1$ still does not contribute, but the $_4F_3$ does, as none of the upper parameters is
zero anymore. Furthermore, in the present case the Pochhammer symbol that produces the desired pole is among the
lower ones in the series expansion of the inner $_4F_3$ function. To take it out, we use a Whipple transformation
for balanced $_4F_3$ \cite{BaileyBook}
\begin{align*}
&\pFq{4}{3}{-n \pFcomma  -m \pFcomma 1-\epsilon/2 \pFcomma  1-\epsilon/2-\lambda_1-\lambda_2  }
{1-\epsilon/2 +\lambda_1-\lambda_2 -n \pFcomma 1-\epsilon/2 -\lambda_1+\lambda_2 -m \pFcomma 1-\lambda_1-\lambda_2 }{1}\\[2mm]
& \hspace*{2cm} =\frac{\left(1-\epsilon/2-\lambda_1+\lambda_2, 1-\lambda_1-\lambda_2+m\right)_n}{\left(1-\lambda_1-\lambda_2,
1-\epsilon/2-\lambda_1+\lambda_2-m\right)_n} \times \\[2mm]
& \hspace*{3cm} \times \pFq{4}{3}{-n \pFcomma  -m \pFcomma \lambda_1-\lambda_2 -n \pFcomma  2\lambda_1 -n  }{\epsilon/2 +
\lambda_1-\lambda_2 -n \pFcomma 1-\epsilon/2 +\lambda_1-\lambda_2 -n \pFcomma \lambda_1+\lambda_2 -m -n }{1}.
\end{align*}
Substituting $(\lambda_1, \lambda_2)_*$, taking $n=m=s$ and expanding the balanced $_4F_3$ on the right hand side of
the last formula in $\xi$, we see that its leading term is not singular now. Moreover, one upper and one lower parameter
in it cancel against each other and we are left with a balanced $_3F_2$ function that can be summed by a usual
Saalsch\"utz formula \cite{BaileyBook}:
\begin{align*}
\pFq{4}{3}{-s \pFcomma  -s \pFcomma -2\lambda +\xi \pFcomma  -2\lambda+s+2\xi  }{\epsilon/2 -2\lambda+\xi \pFcomma
1-\epsilon/2 -2\lambda+\xi \pFcomma -s+\xi }{1} \sim \frac{\left(\frac{\epsilon}{2}, 1-\frac{\epsilon}{2}\right)_s}
{\left(\frac{\epsilon}{2}-2\lambda, 1-\frac{\epsilon}{2}-2\lambda\right)_s}  +\dots
\end{align*}
where the dots denote subleading orders in $\xi$. All in all, the residue is thus again given just by a product of
Pochhammer symbols
\begin{align}\label{res3}
&\Res_{\lambda_1=-\lambda+s,\lambda_2=\lambda}\Phi\left(\lambda,k;z(u)\right)=\frac{4^{2s}}{s!(s-1)!}\\[2mm]
&\times\frac{\left(\frac{1}{2}+a+\lambda-s,\frac{1}{2}+b+\lambda-s,\frac{1}{2}+a-\lambda,\frac{1}{2}+b-\lambda,
\frac{\epsilon}{2},1-\frac{\epsilon}{2}\right)_s}{\left(-2\lambda, 1-2\lambda \right)_{2s}}\times \,\,
\Phi(-\lambda,\lambda-s;k;z(u)).\nonumber
\end{align}
The last sequence of poles at $(\lambda_1,\lambda_2)_*=(\lambda+s,\lambda)$ with residues proportional to
$\Phi(\lambda,\lambda+s;k;z(u))$, is again simple. In this case the leading terms arise from to $n=m=0$, $p=s$,
so the inner $_4F_3$ does not contribute, whereas the $_2F_1$ does. Therefore, we readily obtain
\begin{align}\label{res4}
&\Res_{\lambda_1=\lambda+s,\lambda_2=\lambda}\Phi\left(\lambda,k;z(u)\right)=\frac{1}{s!(s-1)!}
\left( \frac{\epsilon}{2},1-\frac{\epsilon}{2}\right)_s\times \,\, \Phi(\lambda,\lambda+s;k;z(u)).
\end{align}
Notice that all the residues are symmetric under $a,b,\leftrightarrow -a,-b$, $\epsilon \leftrightarrow 2-\epsilon$,
as they should be in consistency with the symmetries of $\BC_2$ Harish-Chandra function described above. Let us also
note that, for generic multiplicities $k$ and eigenvalues $\lambda$, all four families contribute an infinite
sequences of poles.

\subsection{Monodromy group and representation}
\label{subsec:Monodromy}

As in the case of the P\"oschl-Teller problem, imposing good features of the wave functions at the walls
of $A_N^+$ requires to consider certain linear combinations of Harish-Chandra functions. For the
\textit{BC}$_N$ Calogero-Sutherland model, the Weyl chamber $A^+_N$ possesses $N$ walls $\wall_i, i
= 1,\dots, N,$ which are in one-to-one correspondence to the generators $w_i$ of the Weyl group.
As we have discussed above, there is one more wall $\wall_0$, but it does not bound the domain $A^+_N$
and so is not of concern for us, at least for most of this section. Since there are $|W_N|$
Harish-Chandra functions and hence $|W_N|$ coefficients to fix in their linear combinations, one
might naively expect that analyticity at the $N$ walls $\wall_i, i=1, \dots, N$ would leave some
coefficients undetermined. It turns out, however, that there exists a unique linear combination
(up to a constant factor) that is analytic at all $N$ walls on the boundary of the Weyl chamber.

The prescription to build such analytic wave functions is not difficult to state. Suppose
we want the wave function $\Phi$ to be analytic at some subset $\wall_{i_1}, \dots, \wall_{i_r}$
consisting of $r \leq N$ of the $N$ walls that bound $A^+_N$, i.e \ $i_\nu \neq 0$. To
each of these walls there is a generator $w_{i_\nu}$ of the Weyl group and so our set of
$r$ walls is associated with a subgroup $V \subset W_N$ of the Weyl group that is generated
by $w_{i_1}, \dots, w_{i_r}$. Given this subgroup we now define the following superposition
of Harish-Chandra functions
\begin{equation}\label{parabolic-sum}
\Phi^V(\lambdan,k;z) = \sum_{w \in V} c(w\lambdan,k) \Phi(w\lambdan,k;z)
\end{equation}
where the so-called Harish-Chandra c-function\footnote{This function is an analytic continuation
of the so-called Gindikin-Karpelevic c-functions \cite{Gindikin1962} in harmonic analysis.} reads
\begin{eqnarray} \label{eq:cfunction}
c(\lambdan,k) & = & \frac{\gamma(\lambda,k)}{\gamma(\rho(k),k)} \quad , \quad
\gamma(\lambdan,k) = \prod_{\alpha\in \Sigma^+}  \gamma_{\alpha}(\lambdan,k) \quad , \\[2mm]
 & & \gamma_{\alpha}(\lambdan,k)  =
\frac{\Gamma\left(\frac12 k_{\alpha/2}+ \langle \lambdan, \alpha^\vee\rangle \right)}
{\Gamma\left( \frac12 k_{\alpha/2}+k_\alpha + \langle \lambdan, \alpha^\vee\rangle\right)}.
\end{eqnarray}
Any wave function of this form turns out be be regular at the walls $\wall_{i_1}, \dots,
\wall_{i_r}$. There are two extreme cases of this construction. If we just demand regularity
at a single wall $\wall_i$, then the subgroup $V$ consists of two elements, the identity and
the reflection $w_i$. Hence, from the $W_N$ orbit of $|W_N|=N! 2^N $ Harish-Chandra functions
$\Phi(w\lambda,k;z)$ we obtain $N! 2^{N-1}$ linear combinations that are regular at $\wall_i$.
If, on the other hand, we want a wave function that is regular at all the walls $\wall_i, i=1,
\dots, N$ that bound $A^+_N$, then the subgroup $V = W_N$ coincides with the Weyl group
and we end up with a unique linear combination. Let us note that while this function
$\Phi^W$ is analytic in a neighborhood of $A^+_N$, it fails to be analytic at the wall
$\wall_0$. The function $F^+=\Phi^W$ is also known as Heckman-Opdam hypergeometric function.
More generally, linear combinations of the form \eqref{parabolic-sum} are referred to
$\Theta$-hypergeometric functions, see e.g. \cite{Pasquale2004} and references therein.

One can also build a function $\tilde\Phi^W$ that is analytic at $\wall_0, \dots,
\wall_{N-1}$, but has monodromy around $\wall_N$. We will give an exact formula in the
case of $N=2$ below. The corresponding wave functions $\Psi^W(\lambda,k;z)$ and $\tilde
\Psi^W(\lambda,k;z)$ with $\lambda_j = i p_j, p_j \geq 0,$ are the physical wave functions
of the Calogero-Sutherland problem on $A^+_N$ and $\tilde A^+_N = A^L$, respectively. Hence,
they provide an orthogonal basis within the space of functions on the $A^+_N$ and $A^L$.\cite{HeckmanBook, opdam1995, OpdamDunkl} In hypergeometric gauge the Hermitian scalar product on
$A_N^+$ reads\footnote{This formula defines an inner product only if certain conditions
on the multiplicities $k_\alpha$ are satisfied, e.g. if $\Re (k_{\alpha}) > 0$ for all
$\alpha \in \Sigma^+$.}
\begin{align}
(\phi_1,\phi_2)=\int_{A_N^+} \phi_1(u)\overline{\phi_2(u)}\left|\Theta(k;u)\right|^2 du.
\end{align}
where $du$ is a suitably normalized measure on $A_N^+$ \cite{opdam1995}. The gauge
transformed Calogero-Sutherland Hamiltonian $ \Theta(k;u)^{-1} H^\textrm{CS}\Theta(k;u)$ is
self-adjoint with respect to this inner product.
\medskip

As in the case of $N=1$ there exists a nice reformulation in terms of the representation
theory of the fundamental group $\mathcal{M}_N = \pi_1(D_N)$. The latter is generated by $N+1$
elements $g_i, i=0, \dots, N$ which are associated with closed loops around the $N+1$ walls
$\wall_i, i=0, \dots, N$ of the Calogero-Sutherland potential \cite{VIETDUNG1983425, HeckmanBook}
(see also \cite{Brieskorn1971}). Note that the real codimension of these walls in complex space
is two. Since the walls are associated with generators $w_i$ of the affine Weyl group $\mathcal{W}_N$,
so are the generators $g_i \in \mathcal{M}_N$. One may show \cite{VIETDUNG1983425} that the generators
$g_i, i = 0,\dots, N$ of the fundamental group $\mathcal{M}_N = \pi(D_N)$ satisfy the following set
of relations
\begin{eqnarray}
 \  g_i g_j & = & g_j g_i \quad
\textrm{for} \ |i-j|\geq 2  \, , \\[2mm]
g_i g_{i+1} g_i  & = & g_{i+1} g_i g_{i+1} \quad \textrm{for} \ i=1, \dots\, N-2 \, , \\[2mm]
g_{0} g_{1} g_{0} g_{1} = g_{1} g_{0} g_{1} g_{0}
& , & g_{N-1} g_{N} g_{N-1} g_{N} = g_{N} g_{N-1} g_{N} g_{N-1} \  .
\end{eqnarray}
These are very similar to the defining relations of the affine Weyl group, except that
in the fundamental group $g_i^2 \neq 1$. So, the relation between the monodromy group
of $D_N$ and the affine Weyl group mimics the relation between the braid group and the
permutation group. Therefore, the fundamental group $\mathcal{M}_N$ is also referred to
as (Artin) affine braid group. Note that the relation in the second row is the usual braid or
Yang-Baxter equation. The relations in the third line, which involve the elements $g_0$
and $g_N$, on the other hand, resemble forth order reflection equations in which two
factors on each side arise from the reflection at a boundary while the other two are
associated with scattering in the bulk.

For the affine Weyl group we actually discussed two different presentations, one in terms
of $N+1$ generators $w_i, i=0, \dots, N$ and a second one that involves $2N$ generators
$\tau_i$ and $w_i$ with $i=1, \dots, N$. In complete analogy to the construction of the
generators $\tau_i \in W_N$, we can build \cite{van1983homotopy, ExtArtin1983, HeckmanBook}
(see also \cite{Deligne1972}) a set of commuting generators $l_i \in \mathcal{M}_N$
through
\begin{equation}
\ell_i =  g_i^{-1} \cdots g_{N-1}^{-1} (g_N^{-1} \cdots g_0^{-1}) g_1 \cdots g_{i-1}.
\end{equation}
The reader is invited to verify that $\ell_i \ell_j = \ell_j \ell_i$. Hence the group
elements $\ell_i$ generate an $N$-dimensional lattice $\mathbb{Z}^N \cong \Gamma \subset
\mathcal{M}_N$. The generators $g_i, i=1, \dots, N$ act on this
lattice as
\begin{eqnarray}
g_i \ell_{i} g_i  & = &  \ell_{i+1} \ , \quad 1 \leq i < N \\[2mm]
[g_i,\ell_j] & = & 0  \ , \quad |i-j| \geq 2 \ \mbox{ or } \ (i,j)=(N,N-1)\ .
\end{eqnarray}
Our conventions on directions of the basic loops match \cite{HeckmanBook} (section 4.3)
if we identify our generators $\ell_i$ with the generators $l_i$ used by Heckman as
$\ell_i = \prod_{k=i}^N l^{-1}_k$.

For generic values of $\lambda$, the space of Harish-Chandra wave functions
$\Phi(w\lambda,k;z)$ forms a $|W_N|$-dimensional representation of the monodromy
group $\mathcal{M}_N$. This representation is explicitly known due to the work of
Heckman and Opdam. With respect to the action of $M_i$, $i=1, \dots, N$ the space
of Harish-Chandra functions splits into $|W_N|/2$ 2-dimensional subspaces each of
which is spanned by a pair of Harish-Chandra functions $\Phi(\lambda,k;z)$ and
$\Phi(w_i\lambda,k;z)$. On these subspaces, the action is given by
\begin{eqnarray}
M(g_i) & = & M_i = C_i^{-1} \left( \begin{array}{cc} 1 & 0 \\ 0 & e^{\pi i(1+\epsilon)}
\end{array}\right) C_i \quad \textrm{for} \quad i=1, \dots, N-1, \label{eq:Mi}\\[2mm]
M(g_N) & = &  M_N = C_N^{-1} \left( \begin{array}{cc} 1 & 0 \\ 0 & e^{2\pi i(a-b)}
\end{array}\right) C_N\ . \label{eq:MN}
\end{eqnarray}
The matrices $C_i, i=1, \dots, N,$ are defined as
$$
C_i(\lambda,k) = \left( \begin{array}{cc} c_i(\lambda,k) &  c_i(w_i\lambda,k)\\
c_i(\lambda,k'') & c_i(w_i\lambda,k'') \end{array} \right),
$$
where $k''=(k_1'',k_2'',k_3'')=(-k_1,1-k_2,1-k_3)$ denotes an involution of the
multiplicities (see also below) and
\begin{eqnarray}
c_i(\lambda,k) & = & 4^{-\lambda_i+\lambda_{i+1} +\epsilon/2}
   \frac{\Gamma(\frac12 + \frac{\epsilon}{2})\Gamma(2\lambda_i - 2\lambda_{i+1})}
   {\Gamma(\frac{\epsilon}{2}+\lambda_i-\lambda_{i+1})
   \Gamma(\frac12 + \lambda_i -\lambda_{i+1})} \quad \textrm{ and } \\[2mm]
c_N(\lambda,k) & = & 4^{-\lambdan_N+a+1/2} \frac{\Gamma(a-b+1)\Gamma(2\lambdan_N)}
{\Gamma(1/2+\lambdan_N-b)\Gamma(1/2+\lambdan_N+a)}
\end{eqnarray}
is the same function that appeared in our discussion of the P\"oschl-Teller problem.
The monodromy representation $M_0 = M(g_0)$ of the element $g_0$ is once again obtained
from $M_N$ by the simple replacement $b\rightarrow - b$ along with a gauge transformation
of the form
$$
\Omega(\lambda,k) = \textit{diag\/}_{w\in W_N}\left(e^{i\pi
\langle w\lambda-\rho ,\theta\rangle}\right)   \   $$
where $\theta = \sum_i e_i$ is the sum of short roots.\footnote{The mathematical origin of
this $\theta$ is that it is twice the minuscule co-weight of root system $C_N$. In this way,
it is related to a symmetry that interchanges the $0$-th and $N$-th affine root, see previous
discussion.} So, in formulas one has
\begin{equation}\label{eq:M0}
M(g_0) = M_0 =    \Omega^{-1} \tilde M_N \Omega \ .
 \end{equation}
Here $\tilde M_N$ denotes the matrix in eq.\ \eqref{eq:MN} with $b$ replaced by $-b$.
Note that the monodromy matrices $M_i, i=1, \dots, N-1,$ for the curves around the
walls $\wall_1, \dots, \wall_{N-1}$ do not depend on $b$ and are left invariant by
the gauge transformation, i.e. $\Omega^{-1}  \tilde M_i \Omega = M_i$. One may check
that the monodromy matrices $M_r, r=0, \dots, N$ satisfy all the defining relations
of the fundamental group $\mathcal{M}_N$. In addition they possess the following
Hecke property
\begin{eqnarray} \label{eq:Hecke}
(M_r-1)(M_r-\gamma_r) & = &  0 \quad , \quad \mathrm{where}\  \\[2mm]
\gamma_0 = e^{2\pi i (a+b)} \ , \ \gamma_i & = & e^{\pi i(1+\epsilon)} \ , \
\gamma_N = e^{2\pi i(a-b)} \ \nonumber
\end{eqnarray}
for $r = 0, \dots, N$ and $i=1, \dots, N-1$. The Hecke property of the monodromy
matrix is obvious from the eqs.\ \eqref{eq:Mi}, \eqref{eq:MN} and \eqref{eq:M0}. The
formulas for the monodromy representation we have displayed here imply that the functions
$\Phi^V$ we defined above indeed have trivial monodromy with respect to $g_{i_1},\dots,
g_{i_r}$.

One might wonder why these formulas are so similar to the ones in the P\"oschl-Teller
problem. The reason is however not so difficult to grasp. If we start our curve near
asymptotic infinity we can reach each of the walls separately, i.e.\ we can move near
any of the wall $\wall_i$ while staying far away from the others $\wall_j, j\neq i$. In this
way, the monodromy problem for a given wall can really be solved within the theory
of single variable hypergeometric functions. The walls $\wall_0$ and $\wall_N$ are in fact
entirely equivalent to the walls in the P\"oschl-Teller problem while the intermediate
ones possess a different dependence on the coupling constants. The latter may be
understood through a replacement $b\rightarrow 0$ and $a \rightarrow(\epsilon-1)/2$
in the expression for $M_N$.

We have also promised some more comments on the origin of the multiplicities that occur
in the second row of the $C$ matrices. These may be traced back to symmetries of the
Harish-Chandra functions. Indeed, it is well known that \cite{Opdam1993}
\begin{align}\label{HCh-involution}
\Phi(\lambda,k'';u) = \Theta_0(-(2k)'';u) \,  \Phi(\lambda,k;u) \quad \textrm{where} \quad
\Theta_0(k;u)=\prod_{\Sigma^{+}_0}\left( 2 \sinh \frac{\langle\alpha,u\rangle}{2}\right)^{k_\alpha}
\end{align}
and $\Sigma^{+}_0$ denotes the set of middle and long positive roots, i.e. in the product
that defines $\Theta_0$ we do not include factors for the short roots. In terms of our
parameters $(a,b,\epsilon)$ the involution reads $(a'',b'',\epsilon'')=(-a,-b,2-\epsilon)$.

Before we return to our discussion of the special case of $N=2$, we want to conclude the
general discussion on the monodromy representation of the affine braid group by stating a general result from \cite{HeckmanBook} (proposition 4.3.10) that provides a
criterion for the irreducibility of the $|W_N|$-dimensional representation in the
space of Harish-Chandra functions. According to this theory, the representation we have
described is irreducible, provided that the momenta $\lambda$ and the multiplicities $k$
are sufficiently generic, i.e. that
\begin{equation} \label{eq:irred}
\langle \lambda , \alpha^\vee\rangle + k_\alpha + k_{2\alpha} \not\in \mathbb{Z}
\end{equation}
for all roots $\alpha$ of the $\BC_N$ root system. When one of these conditions is violated,
the monodromy representation may contain non-trivial subrepresentations. We will see one
such example below.
\medskip

\noindent
{\bf Example:} Let us discuss some more details for the case of $N=2$ that is
most relevant in the context of conformal field theory. We have noted already that
the corresponding Weyl group possesses $|W_2| = 8$ elements. The eight Harish-Chandra
functions $\Phi(w\lambdan, k;z), w \in W_2$ give rise to an eight-dimensional monodromy
representation of the fundamental group $\mathcal{M}_2$. The latter is generated by three
elements $g_0,g_1,g_2$  subject to the following relations
$$ g_0 g_2 = g_2 g_0 \quad , \quad g_0 g_1 g_0 g_1 = g_1 g_0 g_1g_0
\quad g_1 g_2 g_1 g_2 = g_2 g_1 g_2 g_1\ . $$
These are associated with the three walls in Figure \ref{fig:D2}. Among the linear
combinations of Harish-Chandra functions, there are a few that we would like to
highlight. The first one is associated with the subgroup $V \subset W_2$ that is
generated by $w_1$. According to the general theorem we quoted above, the
following $w_1$ symmetric combination of Harish-Chandra functions
$$ \Phi^s (\lambda,k;z) = c_1(\lambda_1,\lambda_2,k) \Phi(\lambda_1,\lambda_2,k;z)
+ c_1(\lambda_2,\lambda_1,k) \Phi(\lambda_2,\lambda_1,k;z)\ .
$$
is regular at the wall $\wall_1$, see Figure \ref{fig:D2}. Note that $w_1$ acts on momenta of
Harish-Chandra functions by exchanging $\lambda_1$ and $\lambda_2$. The functions
$\Phi^s$ coincide with the functions $\Phi^V$ with $V = \{e,w_1\}\subset W_2$ we
introduced above up to a constant ($z$-independent) factor. Since $\Phi^s$ is
analytic at the wall $\wall_1$, it can be continued from $A^+_2$ to $A^E_2$ and
further into $A^L_2 = \tilde A^+_2$. On the other hand it fails to be analytic
at the walls $\wall_0$ and $\wall_2$. In order to obtain a regular wave function
at the wall $\wall_2$ and hence in a neighborhood of the domain $A^+_2$, we must
add a whole 8-dimensional orbit of Harish-Chandra functions under the action of
the Weyl group on the asymptotic data $\lambda$. The coefficients  are given by
the Harish-Chandra c-function that we defined in eq.\ \eqref{eq:cfunction}.

If instead we want a wave function that is regular at the walls $\wall_1$ and
$\wall_0$ and hence in a neighborhood of $\tilde A^+_2$ we must sum the twisted
Harish-Chandra functions\footnote{We assume $0<\Im u_1 < 2\pi$, $-2\pi < \Im u_2 <0$ here.}
\begin{align}\label{twistedHCh-BC2}
\tilde \Phi(\lambda_1,\lambda_2;a,b,\epsilon;u_1(z),u_2(z)) &= e^{i\pi(- \epsilon/2 +
\lambda_1-\lambda_2)} \Phi(\lambda_1,\lambda_2;a,b,\epsilon; u_1(z),  u_2(z))\\
&=\Phi(\lambda_1,\lambda_2;a,-b,\epsilon;\tilde u_1(z), \tilde u_2(z))\
\end{align}
over an entire orbit of the Weyl group $W_2$ with coefficients given by a twisted
version $\tilde c$ of the Harish-Chandra c-functions \eqref{eq:cfunction} ( i.e. in which
$b$ is replaced by $-b$ ). Since the element $w_2$ of the Weyl group does not commute
with the gauge transformation $\Omega$, the resulting function $\tilde \Phi^W$ is
not a constant multiple of $\Phi^W$. By construction, the Heckman-Opdam hypergeometric
functions $F^+ = \Phi^W$ and $\tilde F^+ = F^L = \tilde \Phi^W$ trivialize two
monodromies each, e.g.
$$ M_1 F^+ = F^+ \quad , \quad M_2 F^+ = F^+ \  $$
and similarly for $\tilde F^+ = F^L$ but with $M_0$ instead of $M_2$. This concludes
our discussion of wave functions on $A^+_2$ and the Lorentzian domain $A^L$.
\medskip

The hypergeometric functions we constructed in the previous example provide physical
wave functions for the domain $A^+_2$ and the Lorentzian domain $\tilde A^+_2 = A^L$,
respectively. Their construction is well known in the mathematical literature. For
our purposes below we are also interested in the physical wave functions for the
Euclidean domain $A^E$. As far as we know, there exists no general theory for these
functions, but for the specific example of $N=2$ such wave functions have been
discussed in the context of conformal field theory \cite{Costa:2012cb, Caron-Huot:2017vep}.
The former paper also contains a partial characterization in terms of monodromies. Recall
that the Euclidean domain $A^E$ is bounded by the wall $\wall_1$. To be slightly more
precise, the strip is bounded by two semi-infinite lines satisfying $u_1 = u_2$ (up to
shifts by $2\pi i$) which arise from the wall $\wall_1$ and a compact interval along
$u_1 = i \varphi = - u_2$ that is associated with the image of the wall $\wall_1$ under
the Weyl reflection with $w_2$. Consequently, one can characterize the physical wave
functions $F^E$ for the Euclidean strip $A^E$ through the following three monodromy
conditions
\begin{eqnarray}
 M_1 F^E(\lambda_1,\lambda_2,k;z_1,z_2) & = &
 F^E(\lambda_1,\lambda_2,k;z_1,z_2)\nonumber \\[2mm]
 M_2^{-1} M_1 M_2F^E(\lambda_1,\lambda_2,k;z_1,z_2) & = &
 F^E(\lambda_1,\lambda_2,k;z_1,z_2)\ , \label{eq:charF} \\[2mm]
 \tilde M_2^{-1} M_1 \tilde M_2F^E(\lambda_1,\lambda_2,k;z_1,z_2)
 & = & F^E(\lambda_1,\lambda_2,k;z_1,z_2)\ .\nonumber
 \end{eqnarray}
It is easy to see that these three conditions cannot be solved simultaneously unless
the $\lambda_1-\lambda_2 - k_3$ is a non-negative integer. From now on until the end of this section we will assume that the parameter $\epsilon$
is generic, i.e. $\epsilon/2$ is not an integer. In this case, the unique
(up to normalization) solution is given by
\begin{equation}\label{eq:FEdef}
F^E_{\lambda_1,\lambda_2}(z_1,z_2) \sim \Phi(\lambda_1,\lambda_2,k;z_1,z_2) +
N(\lambda_1,\lambda_2,k) \Phi(-\lambda_2,-\lambda_1,k;z_1,z_2)
\end{equation}
with
\begin{eqnarray}
N(\lambda_1,\lambda_2,k) & = & \frac{c_1(-\lambda_2,\lambda_1,k) M_2(\lambda_1,\lambda_2)_{12}}
{c_1(\lambda_1,-\lambda_2) M_2(-\lambda_2,-\lambda_1)_{12}}  \\[2mm]
& & \hspace*{-3cm} = 4^{2\lambda_1+ 2\lambda_2} \frac{\Gamma(- \lambda_1-\lambda_2) \Gamma(\frac{\epsilon}{2} +
\lambda_1+ \lambda_2)}{\Gamma(\frac{\epsilon}{2} - \lambda_1-\lambda_2) \Gamma(\lambda_1 + \lambda_2)}
\ \frac{\Gamma(-2\lambda_2)\Gamma(1-2\lambda_2) \Gamma(\frac12 + \lambda_1 \pm a) \Gamma(\frac12 + \lambda_1 \pm b)}
{\Gamma(2\lambda_1)\Gamma(1+2\lambda_1) \Gamma(\frac12 - \lambda_2 \pm a) \Gamma(\frac12 - \lambda_2 \pm b)}
\nonumber
\end{eqnarray}
where $\Gamma(x\pm y)$ is the conventional shorthand for $\Gamma(x\pm y) = \Gamma(x+y)
\Gamma(x-y)$ and $M_2(\lambda_1,\lambda_2)_{12}$ denotes a matrix element of the monodromy
matrix \eqref{eq:MN} for $N=2$. While the first two monodromy conditions in the list
\eqref{eq:charF} possess four linearly independent solutions, the third condition puts the
wave function on the semi-infinite strip and hence imposes strong additional constraints. As in our
discussion of the $\BC_1$ theory, regular wave functions can only exist for a discrete
set of values $n = \lambda_1-\lambda_2 - \epsilon/2 = 0,1,2, \dots$. Since the strip is
infinitely extended in one direction, the sum $\lambda_1 + \lambda_2$ can assume any
imaginary value. When $u_1 + u_2$ becomes infinite, the wave functions $F^E$ factorize
into a product of Poeschl-Teller wave function of the form \eqref{eq:ACBC1}\footnote{With
parameters $a,b$ replaced by $a \rightarrow (\epsilon-1)/2$ and $b \rightarrow 0$ to account
for the strength of the potential at the wall $\wall_1$.}  and a plane wave in the coordinate
$u_1 + u_2$. The latter possesses an incoming and an outgoing component.

Let us conclude this discussion of the Euclidean domain with a few general observations
concerning the structure of the monodromy representations that occur when we specialize
the spin $\lambda_1 - \lambda_2 -\epsilon/2$ to be integer. As we will argue now, the
corresponding eight-dimensional representations of the affine braid group are indecomposable
but no longer irreducible, i.e.\ they contain a non-trivial invariant subspace which turns
out to be four-dimensional. This is consistent with the criterion \eqref{eq:irred}. In fact,
for $\alpha = - e_1 + e_2$ it states
$$ - \lambda_1 + \lambda_2 + \epsilon/2 \not \in \mathbb{Z} $$
which is violated for the representations we are considering. It is indeed easy to see that
the four Harish-Chandra functions
$$  \Phi(\lambda_1,\lambda_2,k;z_1,z_2) \ , \ \Phi(-\lambda_2,\lambda_1,k;z_1,z_2)  \ , \
 \Phi(\lambda_1,-\lambda_2,k;z_1,z_2) \ , \ \Phi(-\lambda_2,-\lambda_1,k;z_1,z_2) $$
span a four-dimensional subrepresentation. In fact, while the first two functions are
invariant under the action of $M_1$, the remaining ones are just mapped onto each other.
The monodromy matrix $M_2$, on the other hand, mixes the first and the third function and
similarly the second and the fourth. All four functions are eigenfunctions of the generators
$\ell_i$ and hence no element within the affine braid group can ever get us out of the
space spanned by these four functions. On the other hand, the remaining four Harish-Chandra
functions from the eight-dimensional monodromy representation do not span an invariant
subspace, i.e.\ the monodromy representations are indecomposable when $\lambda_1-\lambda_2
- \epsilon/2$ is integer. Let us also note that the wave function $F^E$ we constructed
above, is contained in the four-dimensional invariant subspace.

\section{Wave functions and conformal blocks}
\label{sec:Blocks}

After this lightning review of Heckman and Opdam's work on the Calogero-Sutherland
problem we want to explain how all this is related to the theory of conformal blocks
for four external scalar fields. We shall begin with a brief reminder on the \hyperref[subsec:Casimir]{Casimir
equation} for conformal blocks before we explain the relation in the case of the \hyperref[subsec:BlocksPT]{$N=1$
theory} that applies to 4-point blocks of bulk fields in 2-dimensional theories
as well as to blocks of two scalar bulk fields in the presence of a boundary in
any dimension \cite{Liendo:2012hy}. Then we turn to \hyperref[subsec:BlocksHCh]{conformal 4-point blocks in any
dimension} and their relation to the \textit{BC}$_2$ Calogero-Sutherland model,
following \cite{Isachenkov:2016gim}. In particular we will construct higher dimensional
conformal blocks and their shadows in terms of the \textit{BC}$_2$ Harish-Chandra functions.

\subsection{Conformal blocks and the Casimir equation}
\label{subsec:Casimir}

It is time now to embed the theory of scalar conformal blocks into the framework
of the Calogero-Sutherland model that we have sketched in the previous \hyperref[sec:CSproblem]{two} \hyperref[sec:Wavefunctions]{sections}.
To begin with, let us recall the precise definition of scalar conformal 4-point blocks as
solutions of the so-called Casimir equation. \cite{Dolan:2003hv} We consider the
correlation function of four scalar conformal primary fields of weight $\Delta_i, i
= 1,\dots,4$ in a $d$-dimensional conformal field theory. Using conformal symmetry,
this 4-point correlator can be written as
\beqa
& & \hspace*{-3cm} \nonumber
\langle\phi_1(x_1)\phi_2(x_2)\phi_3(x_3)\phi_4(x_4)\rangle \\
& & \hspace*{-8mm}
= \frac{1}{x_{12}^{\frac12(\D_1+\D_2)}x_{34}^{\frac12(\D_3+\D_4)}}
\left(\frac{x_{14}}{x_{24}}\right)^a
\left(\frac{x_{14}}{x_{13}}\right)^b  G(z,\bar z)
\eeqa
with $x_{ij} = x_i - x_j$ and $2a=\D_2-\D_1, 2b= \D_3-\D_4$. The conformal
invariants $z, \bar z$ are introduced to parameterize the more familiar
cross ratios as
\beq \label{zixirel}
u=\frac{x^2_{12}x^2_{34}}{x^2_{13}x^2_{24}}  =  z\bar z   \quad , \quad
v=\frac{x^2_{14}x^2_{23}}{x^2_{13}x^2_{24}}  =  (1-z)(1-\bar z)  \ .
\eeq
For a Euclidean theory, $z,\bar z$ are complex conjugates. When we work
with Lorentzian signature, on the other hand, the variables $z$ and
$\bar z$ are real and taken from the interval $z,\bar z \in [0,1[$.
The function $G$ receives contributions from all the primary fields that can
appear in the operator product expansion of the fields $\phi_1$ and $\phi_2$
\beq
G(z,\bar z) = \sum_{\D,l} \lambdan^{12}_l(\D) \lambdan^{34}_l(\Delta)
G_{\D,l}(z,\bar z)\ .
\eeq
This expansion separates the dynamically determined coefficients $\lambdan$
of the operator product from the kinematic {\em conformal blocks}
$G_{\Delta,l}$. The latter are eigenfunctions of the quadratic
conformal Laplacian $D^2_\epsilon$,
\beq \label{CPWdeq}
 D^2_\epsilon G(z,\bar z) = \frac12 C_{\D,l} G(z,\bar z)
\eeq
with eigenvalues
\beq \label{CPWev}
 C_{\D,l} = \Delta(\Delta-d) +l (l+d-2) \ .
\eeq
The form of the conformal Laplacian has been worked out in
\cite{Dolan:2003hv},
\beq
 D^2_\epsilon := D^2 + \overline{D}^2 +
 \epsilon\left[ \frac{z \overline{z}}{\overline{z}-z}
 \left( \overline{\partial} - \partial\right)
 \label{Diffop}  +(z^{2}\partial-\overline{z}^{2}
 \overline{\partial})\right]
\eeq
where $\epsilon = d-2$ and
\beq
D^2 = z^{2}(1-z)\partial^{2} -(a+b+1)z^{2}\partial\ - ab z.
\eeq
$\overline{D}^2$ is defined similarly in terms of $\bar z$.
In order to fully determine the conformal blocks, we impose the
following boundary condition
\beq \label{CPWasymBlock}
G_{\D,l}(z,\bar z) \sim c_l (z\bar z)^{\frac12(\D-l)}
(z+\bar z)^l  + \dots
\eeq
when $z, \bar z$ are close to $z,\bar z =0$ and we choose
the normalization factor $c_l$ to be
\begin{align*}
c_l=\frac{\left(\epsilon/2\right)_l}{\left(\epsilon\right)_l}.
\end{align*}
The condition \eqref{CPWasymBlock} selects a unique solution of the
conformal Casimir equation and therefore completes our definition of
scalar 4-point blocks.\footnote{Our normalization conventions match those
of \cite{Dolan:2011dv} (up to a change $z\leftrightarrow \bar{z}$ which is
a symmetry of blocks). To match with \cite{Caron-Huot:2017vep}, one should
strip off the prefactor $c_l$.}

\subsection{Blocks and the P\"oschl-Teller problem}
\label{subsec:BlocksPT}

We can gain some first insight into the relation between conformal
blocks and Harish-Chandra functions by setting $d=2$. The conformal
Casimir equation then decomposes into two independent equations that
determine the dependence of blocks on $z$ and $\bar z$, respectively.
Focusing on the $z$-dependence leads to the following second order
differential equation
$$
D^2 G_h(z) = h(h-1)G_h(z)\ .
$$
Note that $D^2$ and hence the eigenfunctions $G_h$ depend on $a,b$ but
we do not display this dependence explicitly.  The same equation also
appears for the blocks of two scalar fields in the presence of a boundary
in $d$-dimensional conformal field theory, see \cite{Liendo:2012hy}. Following
\cite{Isachenkov:2016gim} we perform the gauge transformation
\beq \label{PTGT}
\Psi(u) := 2^{1-2h} \frac{(1-z)^{\frac{a+b}{2}+\frac{1}{4}}}{\sqrt{z}} \ G(z)
\eeq
where the coordinates $z$ and $u$ are related through eq.\ \eqref{eq:zfromux}
as before. Note that we consider the region in which $z \in [0,1]$ which means
that $u = \tilde u + i\pi$ with
$$ z = -\sinh^{-2} \frac{u}{2} = \cosh^{-2} \frac{\tilde u}{2}\ . $$
Inserting these relations it is easy to see that the function $\Psi$
is an eigenfunction of the P\"oschl-Teller Hamiltonian with potential
$V^\textrm{PT}_{(a,b)}(u)$ where the coupling constants $a,b$ are the same
as the parameters $a,b$ in the conformal Casimir equation, i.e.\ they are
related to the holomorphic weights $h_i$ of the external fields through $a=
h_2-h_1$ and $b = h_3-h_4$. The eigenvalue $\varepsilon = - \lambda^2$, on
the other hand, is determined by the holomorphic weight $h$ of the
intermediate field $\varepsilon = - (2h-1)^2$. Hence we read off that
$h = 1/2 - \lambda$.

If we compare the relation \eqref{PTGT} between conformal blocks $G$ and wave
functions $\Psi$ with the gauge transformation $\Theta$  between wave functions
and (twisted) Harish-Chandra functions we find that
$$ G(z) = \vartheta(z) \Phi(\lambda,k;z) := 2^{2a+1-2\lambda} z^{-a}
          \tilde \Phi(\lambda,k;z)\ .  $$
Note that the gauge transformation $\vartheta(z)$ that relates the twisted
Harish-Chandra functions to conformal blocks takes a very simple form. Similarly,
the second twisted Harish-Chandra function $\tilde \Phi(-\lambda,k;z)$ is
related to the so-called shadow block. Finally, there exists a special linear combination
of blocks and their shadows that is analytic along the wall that bounds the Lorentzian
domain $A^L_1$, as discussed in section \ref{subsec:WavefunctionsPT}. These form a complete basis in the space
of functions on the half-line.

\subsection{Blocks from Harish-Chandra wave functions}
\label{subsec:BlocksHCh}

As was observed in \cite{Isachenkov:2016gim}, the Casimir equation for
conformal blocks is equivalent to the Calogero-Sutherland model
for reflection group {\it BC}$_2$. The parameters $a,b$ and
$\epsilon$ that are determined by the conformal weights of the
external scalar and the dimension $d$ coincide with the parameters
$a,b$ and $\epsilon$ in the potential \eqref{CSpot}. To relate the
associated Schr\"odinger problem on the {\it BC}$_2$ Weyl chamber
with the eigenvalue equation \eqref{CPWdeq} for the conformal
Laplacian we employ
\beq \label{CSGT}
\Psi(u_1,u_2) := 2^{d-2\D} \prod_i
\frac{(1-z_i)^{\frac{a+b}{2}+\frac{1}{4}}}{z_i^{\frac12+\frac{\epsilon}{2}}}
|z_1-z_2|^{\frac{\epsilon}{2}} G(z_1,z_2),
\eeq
where $z_1 = z $ and $z_2 = \bar  z$. It is not difficult to verify that
this gauge transformation, along with the usual relation \eqref{eq:zfromux}
between $z_i$ and $u_i$, turns the conformal Laplacian into the
Calogero-Sutherland Hamiltonian for the potential \eqref{CSpot}, with the
eigenvalue
$$\varepsilon = - d(d-2)/4-(C_{\D,l}+1)/2\ . $$
In order to identify the conformal block with the precise eigenfunction of the
Calogero-Sutherland model, we must also compare the asymptotic behavior \eqref{CPWasymBlock}
with the asymptotics \eqref{HCasym} of twisted Harish-Chandra functions. Note that the limit
$u_1,u_2 \rightarrow \infty$ maps to the limit $z_1,z_2 \rightarrow 0$. In Harish-Chandra
theory we perform this limit in the Weyl chamber $A^L_2 = \tilde A^+_2$ where $\tilde u_1
> \tilde u_2$. This maps to real cross ratios $z_1,z_2$ with $0<z_1 < z_2<1$. Taking into
account the asymptotics of the gauge transformation \eqref{CSGT} and of the factor
$\Theta$, comparison gives
\begin{equation}\label{eq:lafromDl}
\lambdan_1 = \frac12 + \frac{\epsilon}{2} - \frac12 (\Delta -l)  \quad , \quad
\lambdan_2 = \frac12 - \frac12(\Delta + l)  \ .
\end{equation}
If we solve these for the conformal weight $\Delta$ and the spin $l$ of the exchanged
field, we obtain
\begin{equation} \label{eq:Dlfromla}
\Delta = \frac{d}{2} - \lambdan_1 - \lambdan_2 \quad , \quad
l = \lambdan_1 - \lambdan_2 - \frac{\epsilon}{2}.
\end{equation}
In order for the twisted Harish-Chandra wave function to possess oscillatory behavior
at infinity, the parameters $\lambda_i$ must be imaginary. The corresponding values
of $\Delta=d/2+ic$ are then associated with the principal continuous series representation of the
conformal group \cite{Dobrev:1977qv}.

Note that the asymptotic condition \eqref{CPWasymBlock} is symmetric with respect to the
exchange of $z = z_1$ and $\bar z = z_2$. Hence the block should not be identified
with a single twisted Harish-Chandra function $\tilde \Phi(\lambdan,k;u)$ but rather
with the superposition of $\tilde \Phi(\lambdan,k;u)$ and $\tilde \Phi(w_1\lambdan,k;u)$.
This is the symmetric superposition $\Phi^s = \Phi^V$ with $V = \{ e, w_1\}$ we have
briefly discussed at the end of the \hyperref[subsec:Monodromy]{previous section},
\begin{eqnarray}\label{eq:GDl}
G_{\Delta,l}(z,\bar z)&=&\frac{2^{4a+d-2\lambda_1-2\lambda_2}}{ (z\bar z)^{a}}
 \left[c_{\alpha_1}(\lambdan_1,\lambdan_2)\tilde\Phi(\lambdan_1,\lambdan_2,k;z_1,z_2) +
 c_{\alpha_1}(\lambdan_2,\lambdan_1)
 \tilde\Phi(\lambdan_2,\lambdan_1,k;z_1,z_2)\right] \nonumber \\[2mm]
\textrm{where} & & c_{\alpha_1}(\lambdan;k)=\frac{\gamma_{\alpha_1}(\lambda,k)}
{\gamma_{\alpha_1}(\rho(k),k)}  = \frac{\Gamma(\lambda_1-\lambda_2)\Gamma(\epsilon)}
{\Gamma(\lambda_1-\lambda_2+\epsilon/2)\Gamma(\epsilon/2)}.
\end{eqnarray}
Here the parameters $\lambdan$ on the right hand side are determined from the conformal weight
$\Delta$ and spin $l$ of the exchanged field through eqs.\ \eqref{eq:lafromDl}, the weights
$k_\alpha$ should be fixed from the conformal weights $\Delta_i$ of the external scalars and
the spacetime dimension $d$ through
$$ k_1 = \D_4-\D_3 \quad , \quad k_2 = \frac12 (\D_2 + \D_3 -\D_1 -\D_4 +1 )
\quad , \quad k_3 = \frac{d}{2}-1 $$
and the variables $u_i$ are related to the cross ratios by eq.\ \eqref{zixirel}. By construction,
the conformal block trivializes the monodromy $M_1$, i.e.\
$$ M_1\,  G_{\Delta,l}(z,\bar z) = G_{\Delta,l}(z,\bar z) \ . $$
This means that it is regular along the wall $\omega_1$ only. On the other hand, the block fails to
be regular along the walls $\omega_0$ and $\omega_2$. We will come back to this issue in the \hyperref[subsec:GribovFroissart]{next section}.
Let us also note that in the case of integer spin, both twisted Harish-Chandra functions contribute
to the block if and only if the dimension $d$ is even. Otherwise, the coefficient $c_{\alpha_1}
(\lambda_2,\lambda_1)$ in front of the second Harish-Chandra function vanishes due to a divergent
gamma factor in the denominator that does not cancel against a divergence in the numerator. Hence,
for integer $l$ and generic dimension $d$, the conformal block is given by a single Harish-Chandra
function.

As discussed in \cite{Isachenkov:2016gim}, for integer $l$\footnote{See appendix \ref{app:Integerspin} for some details of this limit.} our $z-$expansion \eqref{BC2-zexpansion} for
Harish-Chandra functions implies an expansion for conformal block that was first derived by Dolan and
Osborn in \cite{Dolan:2003hv}. Indeed, when $-\epsilon/2+\lambda_1-\lambda_2=l$ is a non-negative
integer, we can use Whipple's identity to rewrite the summands in eq.\ \eqref{BC2-zexpansion}
through
\begin{align*}
&\frac{\left(\frac{\epsilon}{2}-\lambda_{12}\right)_{n-m}}{\left(1-\lambda_{12}\right)_{n-m}}\,\,\pFq{4}{3}{\epsilon/2-\lambda_{12}-m+n \pFcomma 2\lambda_2-m
\pFcomma \epsilon/2 \pFcomma  -m }{\epsilon/2 -\lambda_{12} -m \pFcomma \epsilon/2 +\Lambda_{12} -m \pFcomma 1-\lambda_{12}-m+n }{1}=\\[2mm]
&\frac{\Gamma(\epsilon/2,-\epsilon/2+2\lambda_1,\epsilon/2+\lambda_1\pm\lambda_2)}{\Gamma(\epsilon,2\lambda_1,\lambda_1\pm\lambda_2)}
\frac{\left(\epsilon/2-\lambda_{12}\right)_{n-m}}{\left(1-\epsilon/2-\lambda_{12}\right)_{n-m}}\frac{\left(1-2\lambda_1\right)_{n}}
{\left(1+\epsilon/2-2\lambda_1\right)_{n}}\frac{\left(1-\Lambda_{12}\right)_{m}}{\left(1-\epsilon/2-\Lambda_{12}\right)_{m}}\\[2mm]
&\times \pFq{4}{3}{\epsilon/2-\lambda_{12}-m+n \pFcomma \epsilon/2-\lambda_1\pm \lambda_2  \pFcomma \epsilon/2 }{\epsilon/2 -\lambda_{12} -m
\pFcomma 1+\epsilon/2 -2\lambda_1+n \pFcomma \epsilon }{1}
\end{align*}
where $\lambda_{12} = \lambda_1-\lambda_2$ and $\Lambda_{12} = \lambda_1 + \lambda_2$. Once this is
inserted into eq.\ \eqref{BC2-zexpansion}, using the relation \eqref{twistedHCh-BC2} we recover the expansion from \cite{Dolan:2003hv}. On the
other hand, the expansion formula in \cite{Dolan:2003hv} fails to work for non-integer spin while our
generalized binomial formula for \textit{BC}$_2$ provides a valid solution with prescribed asymptotics.
In order to correct the expansion of Dolan and Osborn in case of non-integer spin $l$ one should add a
second term containing $_4F_3$ on the right hand side of the last identity, which then becomes a particular
instance of three-term relation for a balanced $_4F_3$. Moreover, the two symbols $_4F_3$ that appear
on the right hand side of this relation are non-terminating, yet still convergent due to balancedness.
The analytical continuation of the left hand side, on the other hand, is trivial, since this function
stays terminating, due to a negative integer among the upper parameters. Of course, instead of
correcting the expansion formula of Dolan and Osborn one can work with our expansion \eqref{BC2-zexpansion}
which is valid for integer as well as non-integer spin $l$.

\section{Some applications of scattering theory to conformal blocks}
\label{sec:Applications}

Now that we have explained how the conformal blocks are embedded into the Calogero-Sutherland
scattering theory, we can begin to apply some of the general results from Heckman-Opdam theory
to conformal field theory. The \hyperref[subsec:Poles]{first subsection} aims at understanding poles and residues of
conformal blocks and hence their construction through Zamolodchikov-like recurrence relations.
We will recover and extend existing results on scalar blocks in the conformal field theory literature, most
notably those in \cite{Penedones:2015aga, Kos:2013tga,Kos:2014bka}, from our results on poles and residues of
Harish-Chandra functions, see section \ref{subsec:HChPoles}. As an application of our discussion of the
monodromy representation in section \ref{subsec:Monodromy}, we will approach a recent technical observation
by Caron-Huot that was instrumental for deriving a \hyperref[subsec:GribovFroissart]{Gribov-Froissart-like formula} for
conformal correlators in \cite{Caron-Huot:2017vep}.

\subsection{Poles and residues of blocks}
\label{subsec:Poles}

The first application concerns the analysis of poles of conformal blocks and their residues.
Such relations play an important role for the explicit evaluation of blocks via Zamolodchikov-type
recursions in the numerical bootstrap. In this context they were first discussed in
\cite{Penedones:2015aga}\footnote{For scalar blocks, the residue formulas were first presented
in \cite{Kos:2013tga,Kos:2014bka}.}, see also \cite{Costa:2016xah, Karateev:2017jgd} for
subsequent development. The existing results on poles and their residues were derived
\footnote{The derivation of residues for scalar blocks presented in \cite{Penedones:2015aga}
(appendix B.1) is somewhat incomplete for Type III poles, but the results are correct.}
from representation theory of the conformal algebra. Here we will re-derive the main
results for scalar blocks directly from our understanding of the pole structure of
Harish-Chandra functions, see eqs.\ \eqref{eq:poles} and \eqref{eq:residues}, but with
an important improvement: As we stressed before, our series expansion and all its
consequences in sections \ref{subsec:HarishChandra}, \ref{subsec:HChPoles} are valid for arbitrary complex spin $l$. Hence we are able
to classify poles of scalar conformal blocks and compute their residues for any complex
spin $l$.

Even though the equations \eqref{res1, res2, res3, res4} were obtained for ordinary
Harish-Chandra functions, they also hold for twisted Harish-Chandra functions if we replace
$\Phi$ by $\tilde \Phi$, the parameter $b$ by $-b$ and the domain $A^+$ by $A^L$, because of
the simple relation \eqref{twistedHCh-BC2} between the two sets of functions. Consequently,
according to our general analysis in subsection \ref{subsec:HChPoles}, for $u$ in the
fundamental domain that contains $A^L$, complex values of the multiplicities $k$ and complex
momenta $\lambda\in  \mathbb{C}^2$, the (normalized) twisted Harish-Chandra function $\exp(\langle
-\lambda+\rho(k),u\rangle)\tilde\Phi(\lambda;k;u)$ is analytic within the (complex) fundamental
domain of $u$ variables, entire in $k$ and meromorphic along the hyperplanes defined by $\langle
\lambda, \alpha^{\vee}\rangle=s$ with $s=1,2, \dots $. Typically, the twisted Harish-Chandra
functions possess simple poles along these hyperplanes, except where two or more of them meet.
At such intersection points the pole orders add up.\footnote{Namely, for a subvariety where
$n$ such hyperplanes meet, multiplying our function by $n$ linear forms cancelling those
poles, we get an isolated singularity which is removable by Hartogs' theorem \cite{Hartogs1906,
OsgoodBook}.}We will see some examples in our discussion later on and also explain how to
extract the residues in such cases.

Let us recall that for type $\BC_2$ we distinguished four families of poles corresponding
to the roots $e_1,e_2,e_1+e_2,e_1-e_2$ (see table 3). For compactness we denote the residues
of twisted Harish-Chandra function at these simple poles as
\begin{align}\label{twistedBC2-residues}
&r_1\left(a,b,\epsilon; \frac{s}{2}, \lambda\right)=\frac{1}{2}\frac{4^s}{s!(s-1)!}\frac{\left(a+\frac{1-s}{2},-b+\frac{1-s}{2}, \frac{\epsilon}{2}+\lambda-\frac{s}{2},1-\frac{\epsilon}{2}+\lambda-\frac{s}{2}\right)_s}{\left(\lambda-\frac{s}{2}, 1+
\lambda-\frac{s}{2}\right)_s},\nonumber\\[2mm]
&r_2\left(a,b,\epsilon; \lambda , \frac{s}{2}\right)=\frac{1}{2}\frac{4^s}{s!(s-1)!}\left(a+\frac{1-s}{2},
-b+\frac{1-s}{2}\right)_s,\nonumber\\[2mm]
& r_{12}^+\left(a,b,\epsilon; -\lambda+s , \lambda\right)=\frac{4^{2s}}{s!(s-1)!}\\[2mm]
&\hspace*{3cm}\times\frac{\left(\frac{1}{2}+a+\lambda-s,\frac{1}{2}-b+\lambda-s,\frac{1}{2}+a-\lambda,\frac{1}{2}-b-\lambda,
\frac{\epsilon}{2},1-\frac{\epsilon}{2}\right)_s}{\left(-2\lambda, 1-2\lambda \right)_{2s}},\nonumber\\[2mm]
&r_{12}^-\left(a,b,\epsilon; \lambda+s , \lambda\right)
=\frac{1}{s!(s-1)!}
\left( \frac{\epsilon}{2},1-\frac{\epsilon}{2}\right)_s.\nonumber
\end{align}
We will often refer to these four series of poles as $e_1$, $e_2$, $e_1+e_2$ and $e_1-e_2$
series, corresponding to the simple roots in the $\BC_2$ root system they are associated with.
In our subsequent analysis of poles and residues of conformal blocks we will assume that the
dimension $d\ \geq 2$ is integer in order to reduce the number of cases we have to investigate
separately. Let us stress, however, that this restriction is merely of pedagogical origin. On
the other hand, we will keep the spin $l = \lambda_1-\lambda_2 - \epsilon/2$ arbitrary to exploit
at least some of the potential our approach has compared to the existing representation theoretic
derivations of residue formulas. Of course, our analysis with fixed dimension $d$ will also give
us access to integer values of $l$.\footnote{As we comment in appendix \ref{app:Integerspin}, a consistent
(usual) rule for obtaining special cases as limits of generic expressions is to first take limits for
multiplicity parameters such as the dimension or $\epsilon=d-2$ and then for combinations of the
momenta $\lambda_i$ such as the spin $l$. Performing the limits in opposite order gives different
results.}

In order to simplify notations, let us introduce a new symbol $\mathcal{G}$ for the following function
\begin{align}
\mathcal{G}_{\Delta,l}(\lambda_i; z_i)=G_{\Delta,l}(\lambda_i, z_i)\  \frac{\left(z_1 z_2\right)^a}{4^{2a+\epsilon/2+1-\lambda_1-\lambda_2}},
\end{align}
that is obtained by stripping off a simple factor from the block $G$. According to our formula \eqref{eq:GDl}
the new functions $\mathcal{G}$ are linear combinations of twisted Harish-Chandra functions of the form
\begin{align*}
\mathcal{G}_{\Delta,l}(\lambda_i; z_i)=c_{\alpha_1}(\lambda_1,\lambda_2)
\tilde\Phi\left(\lambda_1,\lambda_2;k;z_1,z_2\right)+c_{\alpha_1}(\lambda_2,\lambda_1)\tilde\Phi\left(\lambda_2,\lambda_1;k;z_1,z_2\right).
\end{align*}
In going from twisted Harish-Chandra functions to these blocks, the pole patterns can change since some
singularities of the two summands may cancel each other. Indeed, we now show that all residues related to
the root $\alpha_1=e_1-e_2$ cancel in the block. Although this follows from general theorems of
\cite{Pasquale2004}, it is also easy to see using our expressions for residues of Harish-Chandra
functions. Let us first check this statement in case $\epsilon$ is not a positive even integer.
Using eq.\ \eqref{twistedBC2-residues} we then find
\begin{align}
\Res_{\lambda_{12} =s} \mathcal{G}_{\Delta,l}&=\frac{\Gamma\left(s, \epsilon\right)}{\Gamma\left(s+\epsilon/2,\epsilon/2\right)}\Res_{\lambda_{12}=s}\tilde \Phi(\lambda_1,\lambda_2)+\Res_{\lambda_{12}=s}\frac{\Gamma\left(\lambda_{12}, \epsilon\right)}{\Gamma\left(-\lambda_{12}+
\epsilon/2,\epsilon/2\right)}\ \tilde\Phi(\lambda_2,\lambda_1)\nonumber\\[2mm]
&=\frac{\Gamma\left(\epsilon\right)}{\Gamma\left(\epsilon/2\right)^2}\frac{\left(1-\epsilon/2\right)_s}{s!}\
\left(1+(-1)\right)\tilde\Phi(\lambda_2,\lambda_1)=0.
\end{align}
In the first line we set $\lambda_{12} = \lambda_1-\lambda_2$ and we used the standard notation $\Gamma(A,B) =
\Gamma(A) \Gamma(B)$. Indeed, the residues from the first and second term cancel each other. Now let us see what
happens for $\epsilon \in 2\mathbb{Z}_{\geq 0}$.\footnote{For $\epsilon=0$ there is a singularity in the gamma
factors which is obviously removable and just gives the factor of $1/2$.} If $s<\epsilon/2$, our above calculation
is still valid. As soon as $s$ becomes equal to $\epsilon/2$ or exceeds this value, the residue of the first summand
vanishes due to eq.\ \eqref{twistedBC2-residues} while the residue of the second term vanishes because the ratio as
one of the two gamma functions in the second term becomes finite in the limit. The absence of poles in our $e_1-e_2$
series is of course consistent with \cite{Penedones:2015aga} where three series of poles were found to occur for a scalar block. More
precisely, our first series of poles that is associated with the root $e_1$ corresponds to Type II in
\cite{Penedones:2015aga} while the series that come with the roots $e_2$ and $e_1+e_2$ are Type I
and III, respectively.

In all three series, the poles are located along a discrete family of 1-dimensional lines of real momenta $\lambda_i
\in \mathbb{R}$. In the first family, $\lambda_1 = s/2$ is constant and $\lambda_2=\lambda$ is free while in the second family we have $\lambda_2 =
s/2$ and $\lambda_1 = \lambda$ is free. The third family that comes with the root $e_1 + e_2$ is described by the
equations $\lambda_1 + \lambda_2 = s$. So, obviously, we can have double and even triple intersections of these
lines. This motivates to distinguish the following five different cases, see also the discussion below,
\begin{itemize}
\item[A)] $\epsilon \in 2\mathbb{Z}_{\geq 0}$ and  $l \in \mathbb{Z}_{\geq 0}$
\item[B)] $\epsilon \in 2\mathbb{Z}_{\geq 0}+1$ and $l \in \mathbb{Z}_{\geq 0}+1/2$
\item[C)] $\epsilon \in 2\mathbb{Z}_{\geq 0}$ and $l \in \mathbb{Z}_{\geq 0}+1/2$
\item[D)] $\epsilon \in 2\mathbb{Z}_{\geq 0}+1$ and $l \in \mathbb{Z}_{\geq 0}$
\item[E)] $\epsilon \in \mathbb{Z}_{\geq 0}$ and $l\notin \mathbb{Z}_{\geq 0}/2$.
\end{itemize}
Only two of these cases, namely A and D, involve integer spins and so for these two families our
results match the outcome of the analysis in \cite{Penedones:2015aga}. Case E we will refer
to as the case of 'generic momenta'. It will turn out that cases A and C give rise to double
poles while even triple poles appear in case B. Cases D and E, on the other hand possess simple
poles only and hence they are the easiest to analyse.

\begin{figure}[htb]
\hspace*{-2.6cm}\includegraphics[scale=.6]{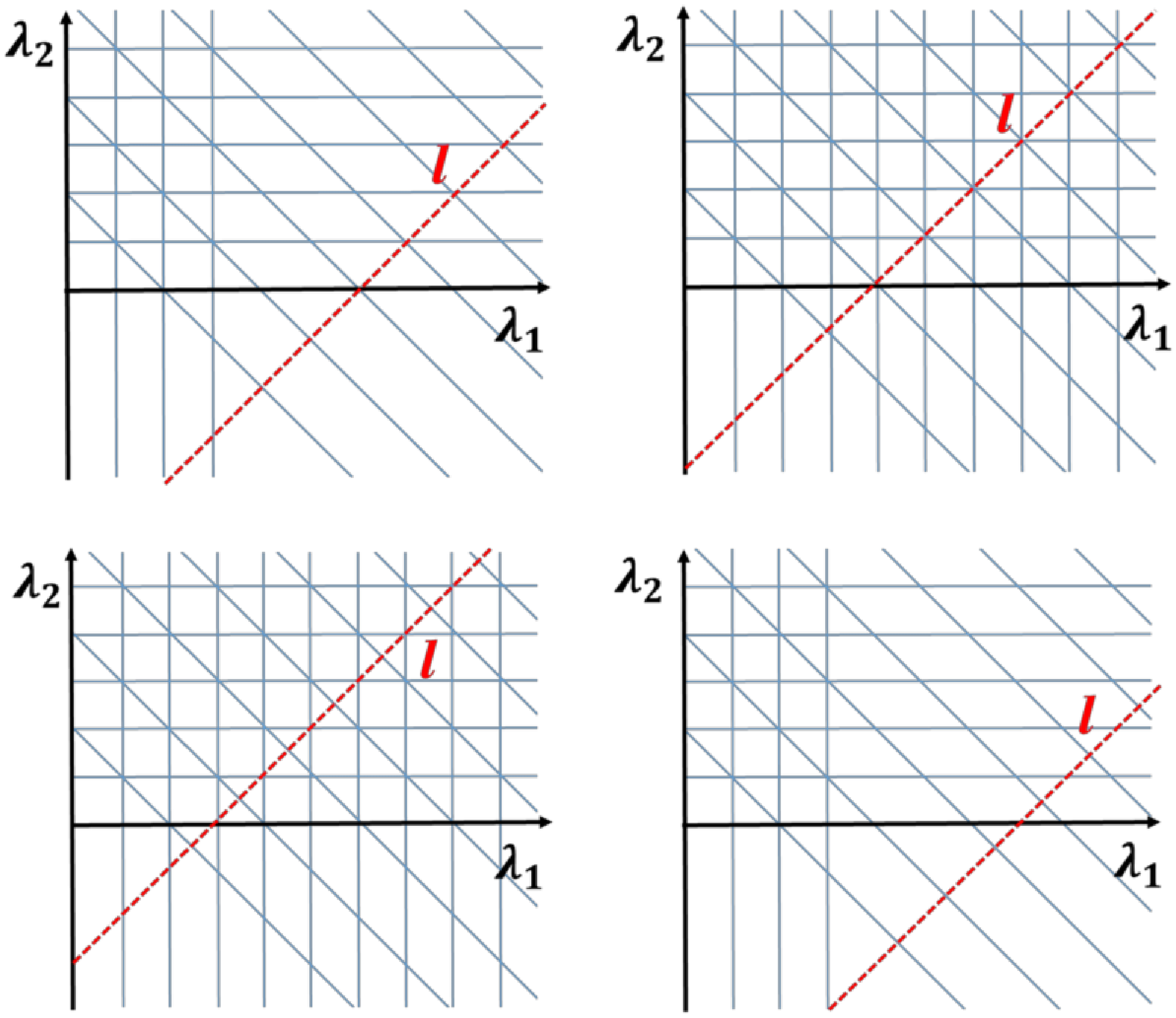}
\caption{Illustration of the poles positions in case A-D. The three sets of blue lines depict the position
of singularities for the three different series with $e_1$ singularities along the vertical lines, $e_2$ along
horizontal and $e_1+e_2$ along diagonal lines. Note that in cases A and D there is a finite number of $e_1$
singularities. The dotted red lines depict the choice of spin $l$, i.e. they run along the curves $\lambda_2
= \lambda_1 - l - \epsilon/2$. For a fixed choice of $l$ we see those as the singularities that lie on the red
lines. In several cases these involve collisions of two or even three poles. The corresponding conformal weight
at which the singularity occurs is determined as $\Delta = d/2 - \lambda_1 - \lambda_2$.}
\label{figure4}
\end{figure}

Before we now discuss the poles coming from the Harish-Chandra functions, we want to make one
important observation concerning the first $e_1$ series of poles. From the explicit formula for
the residue we listed above one can infer that, when the spin $l$ is integer, i.e. in cases A
and D, the residue vanishes for all but the first $l$ values of $s= 1, \dots, l$. Hence in this
case we only have a finite number of singular lines of type $e_1$. For the values $l$ assumes in
case B and C on the other hand, there is an infinite family of $e_1$ poles with non-vanishing
residues. We have depicted the singular lines (light blue) for the cases $A$-$D$ in Figure \ref{figure4}
in order to illustrate the following discussion of these cases.
\begin{itemize}
\item[A)] As we have just discussed, case A involves a finite number of $e_1$ singularities.
After we fix $l$ to be a positive integer (see red line in Figure \ref{figure4}), we obtain a
finite set of simple poles at $\Delta=l+\epsilon,l+\epsilon-1,\dots ,\epsilon+1$ from the $e_1$
series, a finite set of simple poles at $\Delta=\epsilon/2, \epsilon/2-1, \dots 1-l$ from the
$e_1 + e_2$ series and finally, an infinite number of double poles with $\Delta=-l,-l-1,\dots$
from the collision of the $e_2$ and the $e_1 + e_2$ series.
\item[B)] This case is similar to the previous, except that now we have infinitely many $e_1$
singularities. This implies that even triple pole collisions can occur. After we fix $l$ we
obtain (see upper right image in Figure \ref{figure4}) single poles for $\Delta=l+\epsilon,
l+\epsilon-1, \dots , \epsilon/2+1$, (from $e_1$ series) double poles for $\Delta=\epsilon/2,
\epsilon/2-1, \dots , 1-l$ (from the intersection of $e_1$ and $e_1+e_2$ series) and an
infinite number of triple poles for $\Delta=-l, -l-1, \dots$ from the triple intersections
of all three singularities.
\item[C)] As in the previous case there are infinitely many $e_1$ singularities, but now the
choice of $d$ and $l$ implies that we do not encounter triple poles. In fact, the red line in
the lower left image of Figure \ref{figure4} only cuts through intersections of the $e_1$
and the $e_2$ series. Consequently, after fixing $l$ we see single poles for $\Delta=l+\epsilon,
l+\epsilon-1, \dots , 1-l$ and $\Delta=\epsilon/2, \epsilon/2-1, \dots$ as well as double poles
for $\Delta=-l, -l-1, \dots$.
\item[D, E)] All three families of poles are simple.
\end{itemize}
This completes our analysis of poles of conformal blocks in complex momentum space. Let us note
that they all appear for real values of $\lambda_1,\lambda_2$ and outside the cone that is
allowed by unitarity bounds.

In our analysis of residues we will first state the complete results for the generic momenta, i.e. case E.
For the three different families of poles, the residues are given by
\begin{align}
\Res_{\lambda_1=s/2} \mathcal{G}_{\Delta,l}(z,\bar z)&= r_1\left(a,b,\epsilon; \frac{s}{2},\lambda_2\right)
\frac{\left(-\frac{s}{2}-\lambda_2\right)_s}{\left(\frac{\epsilon-s}{2}-\lambda_2\right)_s}\
\biggl [ c_{\alpha_1}\left(-\frac{s}{2},\lambda_2\right) \tilde \Phi\left(-\frac{s}{2},\lambda_2 \right)\\[2mm]
& \hspace*{1cm} + c_{\alpha_1}\left(\lambda_2, -\frac{s}{2}\right) \tilde \Phi\left(\lambda_2,-\frac{s}{2}\right) \
\frac{\sin \left(\pi \left(\frac{s}{2}-\lambda_2\right)\right)\sin \left(\pi \left(\frac{s-\epsilon}{2}+\lambda_2\right)\right)}
{\sin \left(\pi \left(\frac{s}{2}+\lambda_2\right)\right)\sin \left(\pi \left(\frac{s+\epsilon}{2}-\lambda_2\right)\right)}\biggl ],
\nonumber
\end{align}

\begin{align}
\Res_{\lambda_2=s/2} \mathcal{G}_{\Delta,l}(z,\bar z)&= r_2\left(a,b,\epsilon; \lambda_1, \frac{s}{2}\right)
\frac{\left(\frac{\epsilon-s}{2}+\lambda_1\right)_s}{\left(-\frac{s}{2}+\lambda_1\right)_s}\
\biggl [ c_{\alpha_1}\left(\lambda_1, -\frac{s}{2}\right) \tilde \Phi\left(\lambda_1, -\frac{s}{2} \right)\\[2mm]
&\hspace*{1cm}+ c_{\alpha_1}\left(-\frac{s}{2}, \lambda_1 \right) \tilde \Phi\left(-\frac{s}{2}, \lambda_1\right) \
\frac{\sin \left(\pi \left(\frac{s}{2}+\lambda_1\right)\right)\sin \left(\pi \left(\frac{s+\epsilon}{2}-\lambda_1\right)\right)}
{\sin \left(\pi \left(\frac{s}{2}-\lambda_1\right)\right)\sin \left(\pi \left(\frac{s-\epsilon}{2}+\lambda_1\right)\right)}\biggl ],
\nonumber
\end{align}

\begin{align}
\Res_{\lambda_1=-\lambda+s, \lambda_2=\lambda} \mathcal{G}_{\Delta,l}(z,\bar z)&= r_{12}^+\left(a,b,\epsilon; -\lambda+s, \lambda \right)\
\biggl [ c_{\alpha_1}\left(-\lambda, \lambda-s \right) \tilde \Phi\left(-\lambda, \lambda-s \right)\\[2mm]
&\hspace*{1cm} + c_{\alpha_1}\left(\lambda-s, -\lambda \right) \tilde \Phi\left(\lambda-s, -\lambda\right) \biggl ]
= r_{12}^+\, \mathcal{G}_{d/2+s,l}(z,\bar z).\nonumber
\end{align}

In the last family, the residue of the block at $(\Delta,l)_\ast = (d/2-s,l)$ turns out to be proportional to the
block with permuted and reflected momenta, i.e. with $(\Delta,l) = (d/2+s,l)$. For the first two families, however,
the residues of blocks are not proportional to a block anymore, unless the spin $l$ becomes integer in which case
$$
\Res_{\lambda_1=s/2} \mathcal{G}_{\Delta,l}(z,\bar z) \sim \mathcal{G}_{d-1+l,l-s}(z,\bar z) \quad , \quad
\Res_{\lambda_2=s/2} \mathcal{G}_{\Delta,l}(z,\bar z) \sim \mathcal{G}_{1-l, l+s}(z,\bar z)\ .
$$
These relations for conformal blocks with integer $l$ mimic those of the Harish-Chandra functions, see table
\ref{tab:table3}. Within our list of poles (cases A-D), there are two cases in which the residues of the first
two families of poles are proportional to a single block. Actually, this happens in case D since the coefficient
in front of the second Harish-Chandra function vanishes, see our discussion in the \hyperref[subsec:BlocksHCh]{previous section}, so that the
block is actually given by a single Harish-Chandra function. In addition, it can also happen in case A for poles of
the $e_1$ series and for those poles of $e_1+e_2$ series which do not collide to the ones from $e_2$ series. On the other hand, double poles arising from the
colliding series of $e_2$ and $e_1+e_2$ singularities need a special treatment. Thus, once again, we are in
agreement to the conformal group analysis of \cite{Penedones:2015aga}.

When double or triple pole collision appear in the cases A-C, the residues, of course, will look differently
from the expressions we gave above for the generic case. In order to evaluate these such residues for an $n$-th
order pole at some point $\Delta = \Delta^*$ we use the standard prescription
\begin{align*}
\Res_{\Delta = \Delta^*} \mathcal{G}_{\Delta,l}=\lim_{\Delta \to \Delta^*} \frac{\partial^n}{\partial \Delta^n}
\biggl [\left(\Delta-\Delta^*\right)^n \mathcal{G}_{\Delta, l}\biggl ].
\end{align*}
in our explicit expansion \eqref{tildeBC2-zexpansion} (or its $u$-counterpart). Computing the corresponding
residues for case A-C is in principle straightforward now, but we refrain from giving explicit expressions here.
Our analysis provides a new view on the 'irregularities' that were noticed in \cite{Penedones:2015aga, Costa:2016xah}, when
poles of higher orders appear or residues are not proportional to blocks. From our 'analytically continued'
point of view, such problems are resolved into two separate issues. First, the blocks are linear combinations of
twisted Harish-Chandra functions which for generic parameters leads to a linear combination of twisted Harish-Chandra
functions rather than a simgle block. Furthermore, there is an issue of non-generic momenta $\lambda$ for which care
should be taken in defining Harish-Chandra functions as described in sections \ref{subsec:HarishChandra}, \ref{subsec:HChPoles} and appendix \ref{app:Integerspin}.

Before we conclude this subsection, we want to discuss in a bit more detail in which respect our analysis advances the one in
\cite{Penedones:2015aga}, where the authors analysed conformal blocks as sums over states in certain parabolic Verma modules
of the conformal group. While Penedones et al. presented a nice general pattern of pole counting and calculating of residues of
conformal blocks labeled by representations of the conformal group, we instead focused on the case of scalar blocks while applying
the finer optics of Calogero-Sutherland scattering theory. Our analysis treats blocks as meromorphic functions in the labels of
the intermediate field, i.e. the momenta $(\lambda_1,\lambda_2)$ for scalar blocks, so that we can really explore a full
(complexified) parameter space in momenta $\lambda$.\footnote{The restrictions we imposed, namely that the dimension $d$ is
integer and that $\Re l$ is positive, are both easy to lift.} This also gives some new singularities, in particular the triple
poles for odd dimension $d$ and half-integer $l$. Blocks with non-integer spin $l$ have recently started to play an important
role e.g. in the Froissart-Gribov type formula of \cite{Caron-Huot:2017vep}, see also the \hyperref[subsec:GribovFroissart]{next subsection}. What we have seen here
is just a first illustration of a recurrent theme in Calogero-Sutherland theory. In fact, the integrable DAHA \cite{CherednikBook}
setting whose relation to conformal blocks we are developing is tailored to explorations of full complex parameter spaces. We
will dwell on algebraic structures  and their implications in a forthcoming paper \cite{alg-structures}. At the same time, it
seems to us that analytical continuation in dimension $d$ and representation labels would be difficult to achieve with the
representation theoretic methods of \cite{Penedones:2015aga}, or at least requires to implement significant new technology
such as the interpolating Karoubian categories for representation theories of conformal groups \cite{Deligne1990} to put it on a firm ground.

\subsection{On the derivation of a Gribov-Froissart formula for CFT}
\label{subsec:GribovFroissart}

In a recent paper \cite{Caron-Huot:2017vep}, Caron-Huot derived a Lorentzian inversion formula
that allows to obtain the operator product expansion coefficients from the discontinuity of the
four-point correlation function $\mathcal{G}(z,\bar z)$, see also \cite{Simmons-Duffin:2017nub}
for an alternative derivation. The formula actually extracts a function that is analytic in the
spin $l$. The mathematical counterpart/origin of formulas of Gribov-Froissart type goes under
the name of Ramanujan's master theorem, see \cite{RamanujanBook} (p.297) and \cite{Amdeberhan2012}
for a more recent review. Related inversion formulas for $\BC_N$ Calogero-Sutherland Hamiltonians
have been studied in the mathematical literature \cite{Olafsson:2013a}. To explain the basic setup
let us stress that the (twisted) Heckman-Opdam hypergeometric functions on the usual domain $A^+$
or the Lorentzian domain $A^L$ provide an orthonormal basis of wave functions that in many
respects is similar to the basis of exponentials $\exp(i k u)$ for functions on the real line.
Just as the latter give rise to the usual Fourier transform, Heckman-Opdam functions define an
interesting integral transform that has been investigated e.g. in \cite{HeckmanBook, OpdamDunkl,
opdam1995, Cherednik1997}.  For the special case of $\BC_1$, this transform, conventionally known
as Jacobi (function) transform \cite{Koornwinder-Jacobi}, has also been considered in the conformal bootstrap
literature where it was dubbed ``$\alpha$ space transform'', see \cite{Hogervorst:2017kbj,
Hogervorst:2017sfd}.

Questions that are natural to study for Fourier transforms are also natural to study
for the integral transform that is obtained from (twisted) Heckman-Opdam functions.
In particular, one may look at how the behaviour of a function is related to the asymptotic behavior of its transform in the parameters
$l$ and $\Delta$ for various choices of function spaces \cite{Schapira2008} on the Weyl
chamber or the Lorentzian domain. A large body of such results in mathematics go under
the name Tauberian theorems \cite{KorevaarBook}. In the physics literature, similar issues
were studied in \cite{Pappadopulo:2012jk, Qiao:2017xif} by application of
classical (Hardy-Littlewood and Wiener) Tauberian theorems for the Fourier transform,
see also \cite{Fitzpatrick:2012yx, Komargodski:2012ek, Fitzpatrick:2014vua} for related physical arguments. In conformal
field theory, these provide bounds on the asymptotic behaviour of the spectral density
that arise from the short distance behaviour of the correlator. Wiener's Tauberian theorem
tailored\footnote{A usual Wiener type Tauberian theorem states that the translations of
an integrable function span a dense space iff its Fourier transform is everywhere non-zero.
For Heckman-Opdam transforms the translation is replaced by a generalized translation that
is determined by the positive convolution structure.} to the $\BC_1$ case can be found in
\cite{Jianming1997}, while Tauberian theorems for special cases of $\BC_N$ Heckman-Opdam
functions coming from representation theory appear in \cite{2009arXiv0905.3018N}.\footnote{In particular, for the $\BC_1$ case, it would be interesting to see whether the
Jacobi transform Tauberian theorem of \cite{Jianming1997} can improve the conformal field
theory analysis of \cite{Qiao:2017xif} in order to constrain the subleading behaviour of
spectral density and, perhaps, to show that the spectrum of the exchanged operators
approaches the one of generalized free theory. It would also be nice to see if the
$\BC_2$ case can provide a rigorous justification for results in \cite{Fitzpatrick:2012yx,
Komargodski:2012ek} on the lightcone limit, as approached in \cite{Qiao:2017xif} through
the usual Wiener Tauberian theorem.}

A related line of research concerns Bochner-type positivity statements. In the case
of ordinary Fourier(-Stiltjes) transform, the latter states that
a function on a real line is a transform of a positive measure iff it is continuous and positive-definite. \cite{BochnerBook}
Generalizations
to the context of Heckman-Opdam theory are formulated in terms of  positive convolution
structures.\cite{BergBook} They were addressed in \cite{Flensted-Jensen1973,Flensted-Jensen1979} for $\BC_1$,
in \cite{Trimeche2016} for $\BC_2$ and in \cite{ROSLER20102779} for $\BC_N$, but with very
specific choices of the multiplicities $k$. This mathematical framework provides the natural
context for Caron-Huot's Froissart-Gribov formula. In fact, the Lorentzian inversion formula
of \cite{Caron-Huot:2017vep} extracts positive operator product coefficients $c(\Delta,l)$ analytic in spin
through an integral transform of the discontinuity of the correlation function
$\mathcal{G}(z,\bar z)$. Instead of the usual Heckman-Opdam hypergeometric function $F$,
Caron-Huot obtains an integral transform defined by the block
\begin{equation} \label{eq:E}
E(z) = G_{l+d-1,\Delta+1-d} \ ,
\end{equation}
i.e. $E(z)$ is built as in eq.\ \eqref{eq:GDl}, but with $(\lambda_1,\lambda_2)$
replaced by $(-\lambda_1,\lambda_2)$. Similar integral transforms that are obtained
from $\Theta$-hypergeometric functions in the terminology of section \ref{subsec:Monodromy}
have also been studied in the mathematical literature \cite{Olafsson2004}.

In order to derive the conformal Froissart-Gribov formula, Caron-Huot departs
from the Euclidean inversion formula that goes back to the early harmonic analysis on
the conformal group \cite{Dobrev:1977qv}, see also \cite{Costa:2012cb}, and performs an analytic continuation to Lorentzian signature. The Euclidean
inversion formula involves a transform with the wave functions $F^E = F^E_{\Delta,l},
l \in \mathbb{Z}$ which we discussed at the end of section \ref{subsec:Monodromy}.
Following \cite{Costa:2012cb}, Caron-Huot denotes this function by the letter $F$, a nice
coincidence with our interpretation of it as a Euclidean hypergeometric function. In order
to pass to Lorentzian signature, one needs a good understanding of the continuation
of the functions $F = F^E$. In in context of Heckman-Opdam theory this is provided
my the monodromy matrices $M_i$. Let us note that $M_1$ and $\tilde M_2$ coincide
with the inverse of the monodromies which are encoded in eqs.\ (A.23) and (A.22)
of \cite{Caron-Huot:2017vep}. From our discussion in section \ref{subsec:Monodromy}
we know that they provide an eight-dimensional representation of an affine braid
group that satisfies Hecke relations \eqref{eq:Hecke}. This insight provides
significant algebraic control of the monodromy properties of Harish-Chandra (or
pure, in the terminology of \cite{Caron-Huot:2017vep}) functions that can be
exploited without ever using the explicit representation matrices. In order to
demonstrate how constraining these algebraic properties are, we want to rephrase
the main derivation of \cite{Caron-Huot:2017vep} (section 3.3) in our context.

\begin{figure}[htb]
\centering\includegraphics[scale=.4]{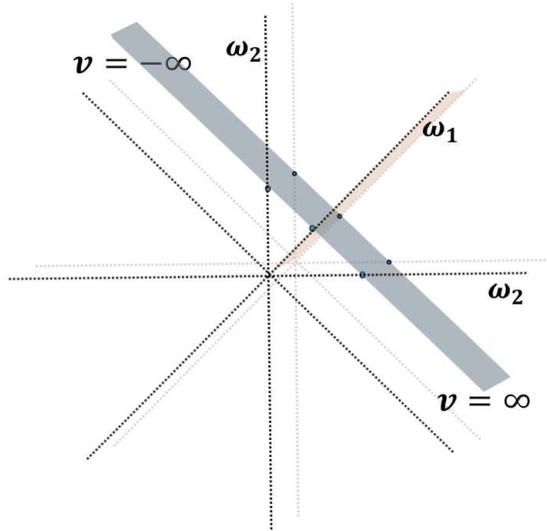}
\caption{Illustration of the upper $w$-half plane used by Caron-Huot inside the domain
of Calogero-Sutherland models. }
\label{figure5}
\end{figure}

Let us first review the general setup. It will be convenient to introduce the
following functions
\begin{eqnarray}
F_{(d/2-\lambda_1-\lambda_2,\lambda_1-\lambda_2-\epsilon/2)}(z,\bar z)
& = & 2^{4a+d-2\lambda_1 - 2\lambda_2}(z\bar z)^{-a} F^E_{\lambda_1,\lambda_2}(z,\bar z)
\ ,  \\[2mm]
g_{(d/2-\lambda_1-\lambda_2,\lambda_1-\lambda_2-\epsilon/2)}(z,\bar z) &  = &
2^{4a+d-2\lambda_1 - 2\lambda_2}(z\bar z)^{-a} \Phi(\lambda_1,\lambda_2,k;z,\bar z) \ .
\end{eqnarray}
that agree with the corresponding functions in \cite{Caron-Huot:2017vep} apart from the
normalization. Caron-Huot studies the function $F$ in the complex $w$-plane. In our
$u-$space, the latter is given by an infinite strip that is perpendicular to the space
$A^E_2$ in Figure \ref{fig:D2} and intersects it along the line $u_1 = \ln\sigma + i\theta$ where
$\theta \in [0,2\pi[$, see the grey strip in Figure \ref{figure5}. We can parametrize
this strip by a complex coordinate $v \equiv v + 2\pi i$ such that $v \mapsto (u_1 =
v + \ln\sigma,u_2 = -v +\ln\sigma)$. The parameter $v$ is related to the variable $w$
used by Caron-Huot through the exponential map, i.e. $w = \exp(v)$. The region in
which $\Re v \rightarrow - \infty$ is mapped to $w \sim 0$ while we can reach $w
\rightarrow \infty$ by sending $\Re v$ to positive infinity.

The strip we have just described intersects the walls in six different places. In fact,
as illustrated in Figure \ref{figure5}, thee are two intersections with the wall $\wall_1$
in $v = 0, i\pi$. Intersections with the walls $\wall_2$ and $\wall_0$, on the other hand,
are located at $v=\pm \ln\sigma$ and $v = \pm \ln \sigma +i\pi$, respectively. After
passing to the $w$-plane, the intersections with the wall $\wall_0$ become the
end-points $w = - \sigma,-1/\sigma$ of the t-channel cuts described by Caron-Huot.
The intersections with $\wall_2$ map to the end-points of the u-channel cuts in
the $w$-plane.
Caron-Huot now wants to continue the function $F$ to $w = 0$, i.e. $v = - \infty$. Along
the way, one has to pass by the intersection with the walls $\omega_1$ and $\omega_2$.
While $F$ is regular at the wall $\omega_1$, i.e. $M_1 F = M_1$, it picks up some
monodromy $M_2$ as it passes by $\omega_2$. So, we can take the limit $v = - \infty$ by
looking at the function $M_2 F$. According to our general characterization of $F$, the
function $M_2 F$ trivializes the monodromy $M_1$, i.e.
\begin{equation} \label{prop1}
 M_1 M_2 F = M_2 F \ .
\end{equation}
Let us restrict our discussion to generic dimension $d$, i.e.\ we assume that $\epsilon/2$
is not integer. As we explained at the end of the \hyperref[subsec:BlocksHCh]{previous section}, for such generic $d$ the
function $F$ is a linear combination of two Harish-Chandra functions,
$$ F(z,\bar z) =  2^{4a+d-2\lambda_1 - 2\lambda_2}(z\bar z)^{-a}
\left(\Phi(\lambda_1,\lambda_2,k;z_1,z_2) +
k \Phi(-\lambda_2,-\lambda_1,k;z_1,z_2) \right)\ . $$
When we act with $M_2$, two more Harish-Chandra functions appear, namely
$\Phi(\lambda_1,-\lambda_2)$ and $\Phi(-\lambda_2,\lambda_1)$. Because of the property
\eqref{prop1} the corresponding functions $g$ combine into a single block $G_{1-l,
1-\Delta}$. This new block that appeared while we continued $F$ through the wall
$\omega_2$ to the region of small $w$ does not vanish in the limit $w \sim 0$ and
thereby it prevents further evaluation of the integral in \cite{Caron-Huot:2017vep}.

The idea of Caron-Huot is to look for an appropriate combination of Harish-Chandra functions
that vanish at $w = \infty$ and can cancel the unwanted divergent terms at $w=0$. It is easy
to see that there are two Harish-Chandra functions that vanish at $w=\infty$, namely the
functions $\Phi(\lambda_2,-\lambda_1)$ and $\Phi(-\lambda_1,\lambda_2)$. From the
corresponding functions $g_{(l+d-1,1-\Delta)}$ and $g_{(l+d-1,\Delta+1-d)}$, see table
\ref{tab:table2}, we can construct the block $E(z) = G_{l+d-1,\Delta+1-d}$ we introduced
before in eq.\ \eqref{eq:E}. By construction, the function $E$ trivializes the monodromy
$M_1$, i.e.
$$ M_1 E(z) = E(z) \ . $$
In order to evaluate its behavior for small $w$, one has to continue $E$ from large absolute
values of $w$ to very small ones. Along the way, we pass the walls $\omega_2, \omega_1$ and
$\omega_2$ again. Cancellation of the singular terms then means that there exist constants
$\alpha$ and $\beta_i$ such that
$$ M_2 F + \alpha (M_1^{-1} M_2 M_1) M_1 M_2 E = \beta_1 g_{(\Delta,2-l-d)} +
\beta_2 g_{(d-\Delta,2-l-d)}\ .  $$
This is a precise formulation of Caron Huot's condition (3.13) in \cite{Caron-Huot:2017vep}
in the language we have set up.  In order to illustrate how to derive such statements using
algebra rather than the precise matrix elements the monodromy matrices, we want to prove, restricting to generic
$\epsilon/2  \not \in \mathbb{Z}$,
a closely related statement, namely
that there exist constants $\alpha$ and $\beta$ such that, the following equation holds
\begin{equation} \label{eq:CHvar}
M_2 F + \alpha M_2 M_1 M_2 E = \beta E\ .
\end{equation}
In order to derive this claim we shall consider eq.\ \eqref{eq:CHvar} as a formula that
allows us to construct the functions $F$ from $E$ as
\begin{equation} \label{eq:Fconstr}
F = \beta M_2^{-1} E - \alpha M_1 M_2 E \ \ .
\end{equation}
Through a short computation using braid and Hecke relations of the monodromy matrices, see
section \ref{subsec:Monodromy}, one may show that the function
$$ F' := M_1 M_2 E + \gamma_1 \gamma_2 E $$
which is proportional to the right hand side of the previous equation if $- \beta/\alpha
= \gamma_1 \gamma_2$, possesses the following properties
$$ M_1 F' = F' \quad \textrm{ and } \quad M_2^{-1} M_1 M_2 F' = F' \ . $$
On the other hand it is easy to see that $F'$ contains only six of the the eight
Harish-Chandra functions since we apply at most two monodromy matrices to $E$. Because
$F'$ is an eigenfunction of $M_1$ with unit eigenvalue we conclude that
$$ F' = \alpha_1 g_{(\Delta,l)} + \alpha_2 g_{(d-\Delta,l)} + \beta' E\ . $$
In addition $F'$ was shown to be an eigenfunction of $M_2^{-1} M_1 M_2$ with
unit eigenvalue. This fact imposes strong constraints on the coefficients $\alpha_i$
and $\beta'$. In fact, as in our discussion of $F^E$ at the end of section
\ref{subsec:Monodromy} we can conclude that $\alpha_2/\alpha_1$ is given by the
expression after eq.\ \eqref{eq:FEdef} and $\beta' = 0$. Hence, we have established
our formula eq.\ \eqref{eq:Fconstr} and thereby the equation \eqref{eq:CHvar} we
set out to prove.
In the argument we have only used braid and Hecke relations of the monodromy matrices
along with the fact that the Harish-Chandra functions $g_{(\Delta,l)}$ and
$g_{(\Delta,2-l-d)}$ are eigenfunctions of the monodromy $M_1$ with unit
eigenvalue, a very particular feature of representations with integer spin $l$
and generic $\epsilon/2 \not \in \mathbb{Z}$.

\section{Conclusions}
\label{sec:Conclusions}

In this work we began to elaborate on the very fruitful relation between conformal blocks and wave functions
of Calogero-Sutherland models that was uncovered in \cite{Isachenkov:2016gim}. In our review of Heckman-Opdam
theory we focused on features that may be understood from the quantum mechanical treatment of Calogero-Sutherland
scattering states. More advanced properties of the Calogero-Sutherland model that relate to its (super-)integrability
were not discussed. These will be briefly outlined below and then presented thoroughly in a forthcoming paper.\cite{alg-structures}

Even without tools from integrability we were able to recover essentially all important results about scalar
4-point blocks. The basic objects in Heckman-Opdam theory are the Harish-Chandra functions. Special cases
of these functions also appeared in the recent work \cite{Caron-Huot:2017vep} where they were called pure
functions. Here we obtained a new series expansion for these functions that is valid for non-integer spin
$l$ and we derived a number of important properties from it. In particular, we located all poles
in $\lambda$-space, which is the space of conformal weights and spins, and we computed their residues. In
addition, we reviewed the classical results of Heckman and Opdam on monodromy properties of Harish-Chandra
functions. Some of the monodromy matrices were also computed independently for $N=2$ in \cite{Caron-Huot:2017vep}.
Here, we learned in addition that they give rise to finite dimensional representation of an affine braid group
of reflection type that factors through the corresponding Hecke algebra. This algebraic characterization of the
monodromy properties of Harish-Chandra (or pure) functions led to a simple derivation of some central
observations in \cite{Caron-Huot:2017vep}. Once we understood how conformal blocks are built out of
Harish-Chandra functions, we were also able to rederive many important results on conformal blocks,
see e.g. \cite{Dolan:2003hv,Dolan:2011dv,Penedones:2015aga}, from our new series expansion and the residue
formulas for Harish-Chandra functions. In our discussion we focused on some of the most useful properties
of blocks. There exist many other consequences of Heckman-Opdam theory that lead to additional insights,
a few of which do not seem to be known in the conformal field theory literature.

We expect that the analytical toolkit we outlined in this paper will prove useful for various approaches to the analytical bootstrap, see in particular \cite{Alday:2016jfr, Mazac:2016qev, Simmons-Duffin:2016wlq, Caron-Huot:2017vep, Hogervorst:2017sfd, Li:2017lmh, Costa:2017twz, Jafferis:2017zna} and references therein for some recent work. In particular, our explicit understanding of the properties of conformal blocks should provide better control over large spin expansions, sum rules for operator product coefficients and multi-twist operators.

What makes the connection with Calogero-Sutherland models even more useful is that the techniques we have
described actually apply much beyond the theory of scalar 4-point blocks. The most immediate extension
concerns the study of defect blocks. In fact, scalar 4-point blocks are a very special case of a larger set
of conformal blocks that can be used to expand the correlation function of two conformal defects
of dimension $p,q < d$. Scalar 4-point blocks correspond to $p = q = 0$, i.e. a $0$-dimensional defect
should be considered as a pair of points. Correlations functions of two local fields in the presence of
a $p$-dimensional defect have received some attention for $p = d-1$ in \cite{Liendo:2012hy} and more
generally in \cite{Billo:2016cpy}. These involve one and two cross ratios, respectively, and the corresponding
blocks turn out to be identical to the ones we have studied here for $N=1$ and $N=2$. More recently, Gadde
also considered setups with $p \neq 0 \neq q$ which can involve a larger number $N > 2$ of cross ratios
\cite{Gadde:2016fbj}. As we will show in a forthcoming paper, the conformal blocks for such configurations
of two defects can all be built from the Harish-Chandra functions of \textit{BC}$_N$ Calogero-Sutherland
models. Hence, many of the results and techniques we have described in section
\ref{sec:Wavefunctions} and the appendices directly apply to such defect blocks.

Another direction concerns applications to spinning blocks. In \cite{Schomerus:2016epl} we have shown
that the Casimir equations for spinning blocks in dimension $d \leq 3$ also take the form of a
Calogero-Sutherland eigenvalue problem, though with a matrix valued potential. This observation
can be extended to arbitrary dimensions, see \cite{Schomerus:2017eny}. The theory of scattering states
for such matrix valued potentials is not really developed, at least for the case of \textit{BC}$_N$
root systems, but the general strategy we reviewed before should still apply. In the case of $A_N$
root system, a restricted set of matrix valued potentials is well understood \cite{CherednikBook} and some first
steps were taken for the most generic ones in \cite{Reshetikhin:2015rba}. It would be very interesting
to extend the results we have described above to matrix Calogero-Sutherland models for \textit{BC}$_2$
roots systems, i.e. to study their symmetries, the fundamental domain, monodromy representations, series
expansions etc.

Further extensions to superblocks are also feasible. As long as sufficiently many of the external
fields are in short (BPS) multiplets, the associated blocks are the same as for the bosonic conformal
symmetry. In general, however, the theory of superblocks with external fields in long multiplets is much
more involved. On the other hand, as was demonstrated in \cite{Cornagliotto:2017dup}, crossing symmetry
involving long multiplets is significantly more constraining when analysed in terms of superblocks
rather than their bosonic decomposition. Hence, it seems worthwhile to develop a systematic theory
of superblocks. We believe that the corresponding Casimir equations can still be rewritten as
eigenvalue equations for a supersymmetric version of Calogero-Sutherland models so that much of the
above could be extended to superblocks. This remains an interesting direction for future research.
\medskip

Let us stress however, that even in the case of scalar 4-point functions we did not even
come close to exploring all the existing features of the model. In fact, the most remarkable
property of the Calogero-Sutherland model is its (super-)integrability. It furnishes a wealth
of additional and very powerful algebraic structure. So far, the only algebra we have seen
above was the Hecke algebra that appeared in the context of the monodromy representation. It
acts in the $2^N N!$-dimensional spaces of Harish-Chandra functions $\Phi(w\lambda;z), w \in
W_N$, i.e.\ in finite dimensional subspaces of functions which all possess the same eigenvalue
of the Hamiltonian. This is just the tip of a true iceberg of algebraic structure.

In order to describe the contours of the remaining parts and thereby outline the content of
our forthcoming paper \cite{alg-structures}, it is useful to consider the example of
the very simplest quantum mechanical system, i.e.\ of a system of $N$ freely moving particles
on the real line. The standard Hamiltonian $H^x_\textit{free} = - \sum_{i}\partial_{i}^2$ of
such a system is certainly integrable. Actually, $H^x_\textit{free}$ can be built from a set
of $N$ commuting first order operators ${\mathcal Y}^\textit{free}_i = \partial_{x_i}=
\partial_i, i = 1, \dots, N,$ as $H^x_\textit{free} = - \sum_i {\mathcal Y}_i^2$. Although
Calogero-Sutherland particles are not free, they enjoy very similar properties. In fact, it
is well known that the Calogero-Sutherland Hamiltonian for \textit{BC}$_N$ roots systems can
be constructed out of a set of $N$ commuting first order operators ${\mathcal Y}^{\textit{CS}}_i$
that are known as \textit{Dunkl operators}. Of course, these Dunkl operators have to account
for the non-trivial potentials in the Calogero-Sutherland model and hence possess a non-trivial
dependence of the coordinates. Eigenfunctions of the Dunkl operators belong to a slighly larger
class of non-symmetric Harish-Chandra functions. These are similar to the Harish-Chandra functions
we discussed above except that they are no longer invariant under the action of the  Weyl group
$W_N$ on coordinates $x_i$. Within the space of such non-symmetric Harish-Chandra functions, the
Dunkl operators give rise to a systems of Knizhnik-Zamolodchikov-like first order differential
equations.\cite{cherednik1992} Once these are solved, symmetric Harish-Chandra functions and in particular the
conformal blocks can be obtained through an appropriate projection. The space of non-symmetric
Harish-Chandra functions is acted upon by an infinite dimensional algebraic structure that is
generated by elements of the Weyl group, Dunkl operators and multiplication with coordinates:
The degenerate double affine Hecke algebra (DAHA) or trigonometric Cherednik algebra.\cite{CherednikBook} In
addition to the monodromy algebra we have seen above, it contains generators that do not
commute with the Hamiltonian and hence can relate eigenfunctions with different eigenvalues.
In this sense, the degenerate DAHA is a spectrum generating symmetry.

This is not the end of the story. In order to prepare for its next chapter, we return to
example of $N$ freely moving particles. The eigenfunctions $\psi_\lambdan(x)$ in such a
system possess a remarkable property: Namely, their dependence on the eigenvalues
$\lambdan_i$ is the same as for the variables $x_i$. This implies that $\psi_\lambdan(x)$
are also eigenfunctions of a dual second order operator $H^y_\textit{free} =
- \sum_i \partial_{\lambdan_i}^2$. We have put a superscript $y$ in this Hamiltonian since
it controls dependence of the eigenfunctions $\psi$ on the eigenvalues $\lambdan$ of the
first order operators $\mathcal{Y}_i$. The dual Hamiltonian $H^y_\textit{free}$ has precisely
the same form as the original one. In this sense, the theory of a freely moving particles may
be considered as self-dual. Some of these facts remain true for our Calogero-Sutherland
model. As for free particles, the dependence of Calogero-Sutherland wave functions on the
eigenvalued $\lambdan$ is controlled by a dual Hamiltonian $H^y$. For the hyperbolic
Calogero-Sutherland model, the latter is a second order difference operator that
describes the so-called \textit{rational Ruijsenaars-Schneider model}. \cite{Ruijsenaars:1986vq, vanDiejen1994a, vanDiejen1994b} This dual
Hamiltonian $H^y_\textit{rRS}$ is also integrable, i.e. there is a set of first order
difference equations that determines the dependence of eigenfunctions on the
eigenvalues $\lambdan^i$. In the case of the free particle we saw that the
dual Hamiltonian $H^y$ had the same form as $H^x$. This is is sharp contrast
to the case of the hyperbolic Calogero-Sutherland system where the dual theory
is controlled by difference rather than differential equations. It turns out,
however, that the setup can be $q$-deformed such that the self-duality is
restored. In order to so so, the rational Ruijsenaars-Schneider model is
deformed into the trigonometric one. After this deformation, the dependence
of eigenfunctions on both sets of variables is controlled by the same type of
difference equations. Upon deformation the degenerate DAHA we briefly mentioned
at the end of the previous paragraph becomes a full fledged DAHA. All this will
be explained in detail in our forthcoming paper \cite{alg-structures} along
with a number of applications.

\section{Acknowledgements}
We wish to thank
T. Bargheer, S. Caron-Huot, A. Gadde, M. Hogervorst, M. Honda, Z. Komargodski, S. Komatsu, P. Kravchuk,
M. Lemos, P. Liendo, G. Mack, D. Mazac, C. Meneghelli,  V. Narovlansky, D. Poland, L. Rastelli, B. van Rees,
N. Reshetikhin, S. Shatashvili, D. Simmons-Duffin,  W. Skiba, Zh. Sobko, G. Torrents and R. Yacoby
for useful conversations and/or remarks and especially Ivan Cherednik for illuminating discussions. We are also
grateful to the the participants and organizers of the 'Back to bootstrap 2016', 'Bootstrap 2017' and 'Strings 2017' conferences, of the 2016 Les Houches Summer
School on Integrability and of the 2017 Cargese Summer School for the occasion to discuss our work. MI thanks the DESY Theory Group, the
Simons Center for Geometry and Physics, Yale University, the Institute for Advanced Study, Duke University and the Math Department of Chapel Hill University, North Carolina,
for a stimulating environment and hospitality during various stages of this work.
MI is supported in part by Israel Science Foundation (grant number 1989/14), by the ERC STG grant 335182 and by a Koshland Postdoctoral fellowship,
partially financed by the Koshland Foundation.

\appendix

\section{$z$- and $x$-expansions for $\BC_2$ Harish-Chandra function}
\label{app:zxexpansions}

As far as we know, the formulas for Harish-Chandra functions in this section are new.
Let us recall here that conformal blocks are obtained via formulas \eqref{eq:GDl} from
twisted Harish-Chandra functions, which are related to the Harish-Chandra function by a
simple phase shift as given in our eq. \eqref{twistedHCh-BC2}. According to \cite{Pasquale2004},
the resulting expansion is convergent for generic $\lambda$ on the entire shifted fundamental domain.

We start from the formula \eqref{BC2-zexpansion} in the main text. For the twisted Harish-Chandra
function $\tilde\Phi$ it takes the form
\begin{align}\label{tildeBC2-zexpansion}
&\tilde \Phi(\lambda_1,\lambda_2;k_i;z_1,z_2)=\frac{1}{4^{2a+1+\epsilon/2-\Lambda_{12}}}\sum_{n,m=0}^{\infty}
\frac{\left(1/2+a-\lambda_1,1/2+b-\lambda_1,\frac{\epsilon}{2}-\lambda_{12}\right)_{n}}{\left(1-2\lambda_1,1-\lambda_{12}\right)_n}\nonumber\\[2mm]
&\hspace*{2cm} \times\frac{\left(1/2+a-\lambda_2,1/2+b-\lambda_2, \epsilon/2+\lambda_{12}\right)_m}{\left(1-2\lambda_2,1+\lambda_{12}\right)_m} \frac{\left( 1-\epsilon/2-\lambda_{12}\right)_{n-m}}{\left(-\lambda_{12}\right)_{n-m}}\nonumber\\[2mm]
&\hspace*{2cm}\times \pFq{4}{3}{-n \pFcomma  -m \pFcomma 1-\epsilon/2 \pFcomma  1-\epsilon/2-\Lambda_{12}  }{1-\epsilon/2 +\lambda_{12} -n \pFcomma
1-\epsilon/2 -\lambda_{12} -m \pFcomma 1-\Lambda_{12} }{1}\\[2mm]
&\hspace*{2cm} \times\frac{1}{n!m!}\, z_1^{\frac{1+\epsilon}{2}+a-\lambda_1+n}z_2^{\frac{1}{2}+a-\lambda_2+m} \,\pFq{2}{1}{\epsilon/2-\lambda_{12}-m+n
\pFcomma \epsilon/2}{1-\lambda_{12}-m+n}{\frac{z_1}{z_2}}\nonumber\ .
\end{align}
Here an in the following we use the shorthand $\lambda_{12} = \lambda_1 - \lambda_2$ and
$\Lambda_{12} = \lambda_1 + \lambda_2$. Before we plunge into details, let us first outline
its derivation.

The key is an observation from \cite{Isachenkov:2016gim} where we directly related scalar
conformal blocks to $q\rightarrow 1^-$ degeneration of virtual Koornwinder polynomials \cite{Rains:2005}.
Due to the fast (and still on-going) progress in the end of 1990's, combinatorial aspects of Koornwinder
polynomials are quite well-studied, in particular the so-called binomial expansion of Koornwinder polynomials
that were obtained in \cite{Okounkov1998}. Now we will factor this connection through the connection to
Harish-Chandra functions of specialized parameters (for $\BC_2$, it means looking at Harish-Chandra functions
with $l=\lambda_1-\lambda_2-\epsilon/2$ integer) and then will analytically continue in these parameters. To
start with, the (non-symmetric $q-$) Harish-Chandra functions are related to Macdonald-Koornwinder polynomials
in \cite{vanMeer2010}.
One may write down a binomial formula for virtual Koornwinder polynomials \cite{Rains:2005}
and then use combinatorial expansions for the interpolating polynomials it contains (combinatorial expansions of
$\BC_N$ interpolation Macdonald polynomials can be found in \cite{Okounkov1998} and the degeneration to $\BC_N$
interpolation Jack polynomials in \cite{Stokman2005LimitTF}), see \cite{Koornwinder2014} for a nice review.
Next one has to perform an analytical continuation of the resulting formula to non-integer differences between
the row lengths of Young diagrams. By analysing the asymptotics, one may note that the resulting series expansion
is a linear combination of $N!$ asymptotically free solutions ($q$-Harish-Chandra functions), corresponding
to a $S_N\subset W_N$ subgroup of the Weyl group. It is then possible to single out a solution with the
correct asymptotics.

This way of obtaining explicit series expansions in $z$, which solves a rather complicated recurrence relation \eqref{Heckman-recursion}
for expansion coefficients (after one passes to $x$-variable), seems to work, provided one has sufficient control
of the analytical continuation of the constituents of the binomial expansion, for any $\BC_N$ Harish-Chandra function.
A degeneration $q\rightarrow 1^-$ is possible at any step, depending rather on convenience (purely formally, one
can take the limit at any point of the derivation, but sometimes this should be justified with more care).
Regardless of that, the complexity of explicit computation increases very fast with $N$, and we are not aware
of any closed form expansion of a generic $\BC_N$ Harish-Chandra function valid for any $N$.

Let us look at $N=2$ case in more detail. The summand of a generalized binomial formula for virtual
Koornwinder polynomials (see \cite{Rains:2005}, formula (7.32)) contains a generalized binomial coefficient
(which is proportional to a $\BC_2$ interpolation Macdonald polynomial), a (virtual) $\BC_2$-interpolation
Macdonald polynomial and some remaining ratios of $q$-Pochhammer symbols. To avoid heavy expressions, we
will send $q \to 1^-$ now. This degeneration has different consequences for the two interpolation polynomials:
the one entering the generalized binomial coefficient will go to a $\BC_2$ interpolation Jack polynomial, whereas
the virtual $\BC_2$-interpolation polynomial will go to a virtual Jack polynomial ('virtual' -- here just means having
a monomial of non-integer powers in front of Gegenbauer polynomial) \cite{Stokman2005LimitTF}.

Using explicit expressions from \cite{Rains:2005} and a combinatorial expansion for the $\BC_2$-interpolation
Jack polynomial from \cite{Okounkov1998} combined with \cite{Stokman2005LimitTF}, one can see that the
generalized binomial coefficient (undeformed) is proportional to
\begin{align*}
&\frac{\left(\frac{\epsilon}{2}-\lambda_{12}\right)_{n-m}}{\left(1-\lambda_{12}\right)_{n-m}}\pFq{4}{3}{\epsilon/2-\lambda_{12}-m+n \pFcomma 2\lambda_2-m
\pFcomma \epsilon/2 \pFcomma  -m }{\epsilon/2 -\lambda_{12} -m \pFcomma \epsilon/2 +\Lambda_{12}  -m \pFcomma 1-\lambda_{12}-m+n }{1}\\[2mm]
&\hspace*{2cm} =\frac{\left(\frac{\epsilon}{2}-\lambda_{12}\right)_{n}}{\left(1-\lambda_{12}\right)_{n}}\frac{\left(\frac{\epsilon}{2}+\lambda_{12},
1-\Lambda_{12}\right)_{m}}{\left(1-\frac{\epsilon}{2}\pm\lambda_1-\lambda_2\right)_{m}}\\[2mm]
&\hspace*{1cm}\times\pFq{4}{3}{-n \pFcomma  -m \pFcomma 1-\epsilon/2 \pFcomma  1-\epsilon/2-\Lambda_{12} }{1-\epsilon/2 +\lambda_{12}  -n
\pFcomma 1-\epsilon/2 -\lambda_{12} -m \pFcomma 1-\Lambda_{12}}{1}
\end{align*}
which we just related to a balanced $_4F_3$ appearing in eq.\ \eqref{BC2-zexpansion} via a Whipple transformation
\cite{AAR}, combined with inverting of summation index. Clearly, the analytical continuation in $l=\lambda_1-\lambda_2-\epsilon/2$
of the $_4F_3$ function on the right-hand side will not change anything (the series will stay terminating), unlike the continuation
of the one on the left (there the series is terminating for integer $l$ and becomes non-terminating for non-integer, so another $_4F_3$
term should appear, see \cite{BaileyBook} for a full account on the symmetries of $_4F_3$ hypergeometric function).

The 'virtual' Jack polynomial, which as we said for $N=2$, is just a product of a monomial and a Gegenbauer polynomial, upon
analytical continuation in $l$ also becomes a non-terminating series
\begin{align}
J(\lambda_1,\lambda_2;\epsilon;z_1,z_2) &=\left(z_1 z_2\right)^{\frac{\epsilon/2+1-\Lambda_{12}+n+m}{2}}\\[2mm]
&\times\pFq{2}{1}{\frac{\epsilon}{2}-\lambda_{12}-m+n \pFcomma \frac{\epsilon}{2} +\lambda_{12}-n+m}{\frac{\epsilon+1}{2}}{\frac{1-\frac{z_1+z_2}{2\sqrt{z_1 z_2}}}{2}}.\nonumber
\end{align}
By using a particular quadratic transformation \cite{AAR} (chapter 3.1) for $_2F_1$, this can also be written as
\begin{align}
c_{\alpha_1}(\lambda_1,\lambda_2;k)\frac{\left(\lambda_{12}\right)_{m-n}}{\left(\lambda_{12}+\frac{\epsilon}{2}\right)_{m-n}}&\frac{\left(z_1 z_2\right)^{n+\frac{1+\epsilon}{2}-\lambda_1}}{\left(z_1+z_2\right)^{n-m+\frac{\epsilon}{2}-\lambda_{12}}}\nonumber\\[2mm]
&\times\pFq{2}{1}{\frac{\epsilon/2-\lambda_{12}+n-m}{2} \pFcomma \frac{1+\epsilon/2-\lambda_{12}+n-m}{2}}{1-\lambda_{12}+n-m}
{\frac{4z_1z_2}{\left(z_1+z_2\right)^2}}\nonumber\\[2mm]
+\, c_{\alpha_1}(\lambda_2,\lambda_1;k)\frac{\left(-\lambda_{12}\right)_{n-m}}{\left(\frac{\epsilon}{2}-\lambda_{12}\right)_{n-m}}&\frac{\left(z_1 z_2\right)^{m+\frac{1+\epsilon}{2}-\lambda_2}}{\left(z_1+z_2\right)^{m-n+\frac{\epsilon}{2}+\lambda_{12}}}\\[2mm]&\times\pFq{2}{1}{\frac{\epsilon/2+\lambda_{12}+m-n}{2} \pFcomma \frac{1+\epsilon/2+\lambda_{12}+m-n}{2}}
{1+\lambda_{12}+m-n}{\frac{4z_1z_2}{\left(z_1+z_2\right)^2}}.\nonumber
\end{align}
Let us dress this back with the above binomial coefficient (and other Pochhammers from the generalized binomial formula
that we didn't display), take the first summand and strip off the $c_{\alpha_1}$ prefactor. This then gives us the
(twisted)\footnote{As the domain we work on is $\tilde A^+$.} Harish-Chandra series decomposition as
\begin{align*}
&\tilde \Phi(\lambda_1,\lambda_2;k_i;z_1,z_2)=\frac{1}{4^{2a+1+\epsilon/2-\Lambda_{12}}}\sum_{n,m=0}^{\infty}\frac{\left(1/2+a-\lambda_1,1/2+b-\lambda_1,
\frac{\epsilon}{2}-\lambda_{12}\right)_{n}}{\left(1-2\lambda_1,1-\lambda_{12}\right)_n}\nonumber\\[2mm]
&\hspace*{2cm}\times\frac{\left(1/2+a-\lambda_2,1/2+b-\lambda_2, \epsilon/2+\lambda_{12}\right)_m}{\left(1-2\lambda_2,1+\lambda_{12}\right)_m} \frac{\left( 1-\epsilon/2-\lambda_{12}\right)_{n-m}}{\left(-\lambda_{12}\right)_{n-m}}\nonumber\\[2mm]
&\hspace*{2cm}\times \pFq{4}{3}{-n \pFcomma  -m \pFcomma 1-\epsilon/2 \pFcomma  1-\epsilon/2-\Lambda_{12} }{1-\epsilon/2 +
\lambda_{12} -n \pFcomma 1-\epsilon/2 -\lambda_{12} -m \pFcomma 1-\Lambda_{12}}{1}\nonumber\\[2mm]
&\hspace*{2cm}\times\frac{1}{n!m!}\, \frac{\left(z_1 z_2\right)^{\frac{1+\epsilon}{2}+a-\lambda_1+n}}
{\left(z_1+z_2\right)^{\frac{\epsilon}{2}-\lambda_{12} +n-m}} \,\pFq{2}{1}{\frac{\epsilon/2-\lambda_{12}+n-m}{2} \pFcomma \frac{1+\epsilon/2-\lambda_{12}+n-m}{2}}{1-\lambda_{12}+n-m}{\frac{4z_1z_2}{\left(z_1+z_2\right)^2}}.\nonumber
\end{align*}
Notice that, in accord with the general definition of Harish-Chandra functions, this series development is explicitly
$W_2$-invariant, the non-trivial symmetry here being exactly $z_1 \leftrightarrow z_2$ since the symmetry with respect to
the subgroup $S_2 \subset W_2$ was already taken into account by our definition of $z$-variables in eq.\ \eqref{eq:zfromux}.
Of course, taking asymptotics at infinity breaks this symmetry explicitly.

To bring the last expansion to the form in eq.\ \eqref{tildeBC2-zexpansion}, one should apply another quadratic transformation
of the form
\begin{align}
\pFq{2}{1}{\frac{A}{2} \pFcomma\frac{A+1}{2}}{A-B+1}{\frac{4y}{\left(y+1\right)^2}}=(1+y)^A \pFq{2}{1}{A \pFcomma B}{A-B+1}{y}.
\end{align}
Here we run into a subtlety (first noticed by Gauss himself) which is sometimes encountered in working with quadratic transformations:
The absolute value of $y$ in the above formula should be less than one for it to be valid. Although one might imagine a wider
applicability range by analytical continuation, in fact it does not happen. Actually, a second term should appear in the
quadratic transformation to correct it in the latter case: Compare formula (3.1.3) in \cite{AAR} and ex.\ 6 for chapter 3 {\it ibid}.\footnote{When substituted into
the Harish-Chandra expansion, the emergence of a second term can be seen as appropriate to give a continuation of the
Harish-Chandra function past the wall $\omega_1$. In other words, it gives the action of monodromy matrix $\tilde M_1$, see
subsection \ref{subsec:Monodromy}.} So, we are forced to explicitly choose between two expansion domains related via $z_1 \leftrightarrow z_2$ and
to break the symmetry by setting $y=z_1/z_2$ ($|z_1|<|z_2|$ for $\tilde A^+$) along with $A=\epsilon/2-\lambda_1+\lambda_2+n-m,
B=\epsilon/2$, finally arriving at eq.\ \eqref{tildeBC2-zexpansion}.

The expansion \eqref{tildeBC2-zexpansion} is absolutely convergent when
\begin{align}
|z_1|<1, |z_2|<1, |z_1|<|z_2|,
\end{align}
as can be checked by using classical Horn's theorem \cite{Horn1889}. For the convergence analysis, it is
actually more convenient to pass to a different, resummed form
\begin{align}\label{conv-expansion-BC2}
&\tilde \Phi(\lambda_1,\lambda_2;k_i;z_1,z_2)=\frac{ z_1^{\frac{1+\epsilon}{2}+a-\lambda_1}z_2^{\frac{1}{2}+a-\lambda_2}}{4^{2a+1+\epsilon/2-\Lambda_{12}}}
\sum_{n,m,p,r=0}^{\infty} \frac{z_1^n z_2^m \left(z_1 z_2\right)^p\left(\frac{z_1}{z_2}\right)^r}{n!m!p!r!}
\frac{\left(\frac{\epsilon}{2}-\lambda_{12}\right)_{r+n-m}}{\left(1-\lambda_{12}\right)_{r+n-m}}\nonumber\\[2mm]
&\times \frac{\left(1-\lambda_{12}, 1-\frac{\epsilon}{2}-\lambda_{12}\right)_{n-m}}{\left(-\lambda_{12}, \frac{\epsilon}{2}-\lambda_{12}\right)_{n-m}}
\frac{\left(\frac{1}{2}+a-\lambda_1, \frac{1}{2}+b-\lambda_1\right)_{n+p}}{\left(1-2\lambda_1, 1-\lambda_{12}\right)_{n+p}}
\frac{\left(\frac{1}{2}+a-\lambda_2, \frac{1}{2}+b-\lambda_2\right)_{m+p}}{\left(1-2\lambda_2, 1+\lambda_{12}\right)_{m+p}}\nonumber\\[2mm]
&\times \left(\frac{\epsilon}{2}-\lambda_{12}\right)_{n} \left(\frac{\epsilon}{2}+\lambda_{12}\right)_{m} \frac{\left(1-\frac{\epsilon}{2},
1-\frac{\epsilon}{2}-\Lambda_{12}\right)_{p}}{\left(1-\Lambda_{12}\right)_{p}} \left(\frac{\epsilon}{2}\right)_{r}.
\end{align}
To obtain this formula from eq.\ \eqref{tildeBC2-zexpansion}, one can use an identity
\begin{align}\label{resum}
\sum_{n,m,p=0}^{\infty}F(n,m,p)\,\,\frac{y_1^{n+p}y_2^{m-p}}{n!m!p!}=
\sum_{n,m,p=0}^{\infty}F(p,m-p,n-p)\left(-n,-m\right)_p\frac{y_1^{n}y_2^{m-n}}{n!m!p!}
\end{align}
which follows from Bailey lemma \cite{AAR}. It is also instrumental to keep in mind another particular
presentation of the $z$-expansion which can be obtained by combining Whipple's identity and $S_6$ symmetry
of a terminating well-poised $_7F_6$ function \cite{BaileyBook}
\begin{align}
&\tilde\Phi(\lambda_1,\lambda_2;k_i;z_1,z_2)=\frac{1}{4^{2a+1+\epsilon/2-\Lambda_{12}}}\nonumber\\[2mm]&\times\sum_{n,m=0}^{\infty}\frac{\left(1/2+a-\lambda_1,1/2+b-\lambda_1\right)_n}{\left(1-2\lambda_1\right)_n}
\frac{\left(1/2+a-\lambda_2,1/2+b-\lambda_2\right)_m}{\left(1-2\lambda_2\right)_m}\nonumber\\[2mm]&\times\frac{\left(\epsilon/2-\lambda_{12}, 1-\epsilon/2-\lambda_{12},1+\epsilon/2\right)_{n-m}}{\left(-\lambda_{12}, 1-\lambda_{12}\right)_{n-m}\left(1+\epsilon/2\right)_n}\frac{\left(1-\epsilon/2-\Lambda_{12}\right)_m}{\left(1-\Lambda_{12}\right)_m} \\[2mm]&\times \pFq{7}{6}{\frac{\epsilon}{2}+n-m \pFcomma 1+\frac{\epsilon}{4}+\frac{n-m}{2} \pFcomma 1- \Lambda_{12}+n \pFcomma
\frac{\epsilon}{2}+\lambda_{12} \pFcomma \frac{\epsilon}{2}-\lambda_{12}-m+n \pFcomma \frac{\epsilon}{2} \pFcomma -m}
{\frac{\epsilon}{4}+\frac{n-m}{2} \pFcomma \frac{\epsilon}{2}+ \Lambda_{12}-m \pFcomma 1- \lambda_{12}-m+n
\pFcomma 1+\lambda_{12} \pFcomma 1+n-m  \pFcomma 1+\epsilon/2+n}{1}\nonumber\\[2mm]&\times\frac{(-1)^m (-n)_m}{n!m!}\, z_1^{\frac{1+\epsilon}{2}+a-\lambda_1+n}z_2^{\frac{1}{2}+a-\lambda_2+m}
\,\pFq{2}{1}{\epsilon/2-\lambda_{12}-m+n \pFcomma \epsilon/2}{1-\lambda_{12}-m+n}{\frac{z_1}{z_2}}.\nonumber
\end{align}
To complete our discussion of the $z$-expansions, let us finally spell out the one summing over the cone of
positive $\BC_2$ roots, which makes the partial ordering of terms in the series manifest. To achieve this goal,
we again use eq.\ \eqref{resum} and obtain
\begin{align}\label{BC2-zexpansion-ordered}
&\tilde\Phi(\lambda_1,\lambda_2;k_i;z_1,z_2)=\frac{1}{4^{2a+1+\epsilon/2-\Lambda_{12}}}\sum_{n,m=0}^{\infty}
\frac{1}{n!m!}\, z_1^{\frac{1+\epsilon}{2}+a-\lambda_1+n}z_2^{\frac{1}{2}+a-\lambda_2+m-n}\left(\frac{\epsilon}{2}\right)_n\nonumber\\[2mm]&\times\frac{\left(1/2+a-\lambda_2,1/2+b-\lambda_2, 1-\epsilon/2+\lambda_{12}\right)_m}{\left(1-2\lambda_2,\lambda_{12}\right)_m} \frac{\left( \epsilon/2-\lambda_{12}\right)_{n-m}}{\left(1-\lambda_{12}\right)_{n-m}}\nonumber\\[2mm]&\times \sum_{p=0}^{\infty}\frac{(-1)^p}{p!}\frac{\left(-\lambda_{12}-m,1+\frac{-\lambda_{12}-m}{2}, \frac{1-\frac{\epsilon}{2}-\lambda_{12}-m}{2}, \frac{2-\frac{\epsilon}{2}-\lambda_{12}-m}{2}\right)_p}{\left(\frac{-\lambda_{12}-m}{2}, \frac{1+\frac{\epsilon}{2}-\lambda_{12}-m}{2},
\frac{\frac{\epsilon}{2}-\lambda_{12}-m}{2}\right)_p}\\[2mm]&\times\frac{\left(\frac{\epsilon}{2}-\lambda_{12}+n-m,1/2+a-\lambda_1,1/2+b-\lambda_1,-n, -m\right)_p}{\left(1-\frac{\epsilon}{2}-n,
1/2-a+\lambda_2-m,1/2-b+\lambda_2-m, 1-\lambda_{12}+n-m, 1-\lambda_{12}\right)_p}\nonumber\\[2mm]&\times\frac{\left(\frac{\epsilon}{2}-\lambda_{12},2\lambda_2-m\right)_p}{\left(1-2\lambda_1,
1-\frac{\epsilon}{2}-\lambda_{12}-m\right)_p} \pFq{4}{3}{-p \pFcomma  p-m \pFcomma 1-\frac{\epsilon}{2}
\pFcomma  1-\frac{\epsilon}{2}-\Lambda_{12} }{1-\frac{\epsilon}{2} +\lambda_{12} -p
\pFcomma 1-\frac{\epsilon}{2} -\lambda_{12} -m+p \pFcomma 1-\Lambda_{12} }{1}.\nonumber
\end{align}
Notice that the Pochhammers of index $p$ in the first two lines of the inner double sum are "almost well-poised", i.e.
the upper $u_a$ and the lower $l_a$ parameters are related by $l_a=1+A-u_a$ with $A=-\lambda_1+\lambda_2-m$. This implies
that such a double sum should be a close two-variable analogue of a well-poised $_7F_6$ function. To see this somewhat more
clearly, the last line can be additionally rewritten through Whipple identity as
\begin{align*}
&\frac{\left( \frac{\epsilon/2-\lambda_{12}-m}{2}, \frac{1+\epsilon/2-\lambda_{12}-m}{2},
1-2\lambda_1-m, 2\lambda_2-m\right)_p}{\left(\frac{1-2\lambda_1-m}{2}, 1-\frac{2\lambda_1+m}{2}, \epsilon/2-\lambda_{12}-m, 1-\epsilon/2-\lambda_{12}-m\right)_p} \ \times \\[2mm]
& \pFq{7}{6}{-2\lambda_1+p-m \pFcomma 1+\frac{-2\lambda_1+p-m}{2} \pFcomma \frac{\epsilon}{2}-\Lambda_{12} \pFcomma 1-\frac{\epsilon}{2}-\Lambda_{12}
\pFcomma -\lambda_{12}+p-m   \pFcomma p-m \pFcomma -p}{\frac{p-m-2\lambda_1}{2} \pFcomma 1-\frac{\epsilon}{2}- \lambda_{12}+p-m \pFcomma
\frac{\epsilon}{2}- \lambda_{12}+p-m \pFcomma 1- \Lambda_{12}\pFcomma 1-2\lambda_1  \pFcomma 1-2\lambda_1+2p-m }{1}.
\end{align*}
In order to obtain expansions in the 'radial' coordinates ($x_i$ or $u_i$), we again start with eq.\
\eqref{tildeBC2-zexpansion}. For generic values of parameters, the $u$-expansion then can be derived in the same
way as in the $N=1$ example from the main text, i.e. through the use of a binomial theorem folowed by a change of
summation indices
\begin{align}
&\Phi(\lambda_1,\lambda_2;k_i;u_1,u_2)=\sum_{i,j,n,m=0}^{\infty}\frac{1}{i!j!n!m!}x_1^{\lambda_1-\frac{1+\epsilon}{2}-a-n-i}x_2^{\lambda_2-\frac{1}{2}-a-m-j}\nonumber\\[2mm]
&\times\frac{\left(1+\epsilon+2a-2\lambda_1\right)_{i+2n} \left(1+2a-2\lambda_2\right)_{j+2m} \left(\frac{1}{2}+a-\lambda_1,\frac{1}{2}+b-\lambda_1,\frac{\epsilon}{2}
-\lambda_{12}\right)_{n}}{\left(-\lambda_{12}\right)_{n-m}\left(\frac{\epsilon}{2}+\lambda_{12}\right)_{m-n}\left(1-2\lambda_1,1-\lambda_{12}, \frac{1+\epsilon}{2}+a-\lambda_1,
1+ \frac{\epsilon}{2}+a-\lambda_1 \right)_n}\nonumber\\[2mm]
&\times\frac{\left(\frac{1}{2}+a-\lambda_2,\frac{1}{2}+b-\lambda_2, \frac{\epsilon}{2}+\lambda_{12}\right)_m}{\left(1-2\lambda_2,1+\lambda_{12},
\frac{1}{2}+a-\lambda_2, 1+a-\lambda_2\right)_m}\\[2mm]
&\hspace*{4cm} \times \pFq{4}{3}{-n \pFcomma  -m \pFcomma 1-\frac{\epsilon}{2} \pFcomma  1-\frac{\epsilon}{2}-\Lambda_{12} }
{1-\frac{\epsilon}{2} +\lambda_{12} -n \pFcomma 1-\frac{\epsilon}{2} -\lambda_{12} -m \pFcomma 1-\Lambda_{12}}{1}\nonumber\\[2mm]
&\hspace{-16mm}\times \pFq{6}{5}{\frac{\epsilon}{2}-\lambda_{12}-m+n \pFcomma \frac{\epsilon}{2} \pFcomma -a+\lambda_2-m \pFcomma \frac{1}{2}-a+\lambda_2-m
\pFcomma \frac{1+\epsilon}{2}+a-\lambda_1+n+\frac{i}{2} \pFcomma 1+\frac{\epsilon}{2}+a-\lambda_1+n+\frac{i}{2} }{1-\lambda_{12}-m+n
\pFcomma \frac{1+\epsilon}{2}+a-\lambda_1+n \pFcomma 1+\frac{\epsilon}{2}+a-\lambda_1+n \pFcomma -a+\lambda_2-m-\frac{j}{2}
\pFcomma \frac{1}{2}-a+\lambda_2-m -\frac{j}{2} }{\frac{x_2}{x_1}}\nonumber.
\end{align}
This form seems to be most useful for getting an expansion of the blocks in Jack (Gegenbauer) polynomials of radial variables,
which explicitly solves the recursion relations given in \cite{Hogervorst:2013sma}. We will not present the result here, as
we do not use it. Instead, by changing summation indices and employing again the resummation identity \eqref{resum}, we obtain
an explicit, manifestly partially ordered (i.e., over the $\mathbb{Z}_{\geq 0}$-cone of positive roots) expansion of $\BC_2$
Harish-Chandra functions in the coordinates $u_1, u_2$  as\footnote{The coefficient of the powers in this expansion, i.e.
the inner triple sum of Pochhammer symbols times the balanced $_4F_3(1)$ function, is, of course, a finite sum. It can be
easily written as a product of Pochhammer quotients and a quadruple hypergeometric polynomial.}
\begin{align}\label{BC2-uexpansion}
&\Phi(\lambda_1,\lambda_2;k_i;u_1,u_2)=\sum_{i,j=0}^{\infty}e^{u_1\left(\lambda_1-\frac{1+\epsilon}{2}-a-i\right)+u_2\left(\lambda_2-\frac{1}{2}-a+i-j\right)}\frac{1}{i!j!}\nonumber\\[2mm]&\times\frac{\left(1+2a-2\lambda_2-2i\right)_{j} \left(\frac{\epsilon}{2}, \frac{\epsilon}{2}-\lambda_{12}\right)_{i} }{\left(1-\lambda_{12}\right)_{i}}\nonumber\\[2mm]&\times \sum_{n,m,p=0}^{\infty}\frac{(-1)^n}{n!m!p!} \,\, \pFq{4}{3}{-n \pFcomma  -m \pFcomma 1-\frac{\epsilon}{2} \pFcomma  1-\frac{\epsilon}{2}-\Lambda_{12} }
{1-\frac{\epsilon}{2} +\lambda_{12} -n \pFcomma 1-\frac{\epsilon}{2} -\lambda_{12} -m \pFcomma 1-\Lambda_{12} }{1}\nonumber\\[2mm]&\times\frac{\left(\lambda_{12}-i\right)_{m+p-n}}{\left(1-\frac{\epsilon}{2}+\lambda_{12}-i\right)_{m+p-n}}\frac{\left(-j,1+2a-2\lambda_2-2i+j \right)_{m+p}}
{\left(\frac{1}{2}+a-\lambda_2-i, 1+a-\lambda_2-i\right)_{m+p}}\\[2mm]&\times\frac{\left(-\frac{\epsilon}{2}-a+\lambda_1-i, \frac{1-\epsilon}{2}-a+\lambda_1-i\right)_{p-n}}{\left(-\epsilon-2a+2\lambda_1-2i\right)_{p-n}}
\frac{\left(1-\lambda_{12}, 1-\frac{\epsilon}{2}-\lambda_{12}\right)_{n-m}}{\left(-\lambda_{12}, \frac{\epsilon}{2}-\lambda_{12}\right)_{n-m}}\nonumber\\[2mm]& \times \frac{\left(-p, \frac{1}{2}+a-\lambda_1,\frac{1}{2}+b-\lambda_1,\frac{\epsilon}{2}-\lambda_{12}\right)_{n}}{\left(1-2\lambda_1,1-\lambda_{12}
\right)_n}\frac{\left(\frac{1}{2}+a-\lambda_2,\frac{1}{2}+b-\lambda_2, \frac{\epsilon}{2}+\lambda_{12}\right)_m}{\left(1-2\lambda_2,1+\lambda_{12}\right)_m}\frac{\left(-i\right)_{p}}{\left(1-\frac{\epsilon}{2}-i\right)_{p}}\nonumber.
\end{align}
This is the main result of this section. As the coefficient of the expansion term solves the recursion relation
\eqref{Heckman-recursion}, according to general theorems of Heckman and Opdam \cite{HeckmanBook} for generic values
of $\lambda$ the expansion is convergent on the whole domain $A^+$ and can be analytically continued to the entire (complex) fundamental domain.
The expansion \eqref{BC2-uexpansion} combined with Heckman-Opdam results gives a full control over the analytical properties of the $\BC_2$ Harish-Chandra functions.

The two expansions in $x$ coordinates above can be recast in many equivalent forms. All of them can in principle
be regarded as instances of sextuple Srivastava-Daoust hypergeometric functions \cite{Srivastava-Daoust}, a rather
general class of analogues of hypergeometric functions. Since information on the analytical behaviour of a generic
member of this family seems to be quite scarce, it is fortunate that we have the well-developed Heckman-Opdam theory
at our disposal (and a root symmetry at the root of it). We are not aware of any significant simplification of the
above formula for generic values of the parameters, it would be interesting to prove this rigorously.

To conclude this appendix, let us comment on the case of general $N$. We claim that in fact all the building blocks
required for writing down the expansions are already present here, although writing up a general $\BC_N$ formula
requires extra work along the lines described in the beginning of the appendix. It appears that such general
expansions of $\BC_N$ Harish-Chandra functions will involve $N(N+1)$ $(x-)$ and $N^2$ $(z-)$ summations, which
are numbers of roots of the $\BC_N$ root system and of its corresponding reduced root subsystem ($B_N$ or $C_N$),
especially when looking at such beautifully symmetric expansions as eq. \eqref{conv-expansion-BC2}. Clearly, for
$N=2$ we get back $6$ $x$-summations and $4$ $z$-summations of the above.

We conjecture that these numbers of summations are minimal possible for the generic values of parameters and,
moreover, that the same is true also for Harish-Chandra functions of other root systems, of which the ones
associated to exceptional root systems are probably mathematically most interesting, since they cannot be
obtained as special cases or limits of the $\BC$ case. E.g. for $E_8$ Harish-Chandra function this would
imply a minimum of $120$ $x-$/$z-$ summations. As such explicit general expansions are not yet available,
it is an interesting mathematical problem to solve.  As explained in \cite{Schomerus:2017talk}, the $\BC_N$ Harish-Chandra
functions describe conformal blocks of parallel defects of codimension $N$ (in ambient space of a
sufficiently high dimensionality). So, such expansions are of potential interest for the conformal
bootstrap programme, provided there is a corresponding analogue of crossing symmetry for such defects.

\section{Alternative expansions for integer spin}
\label{app:Integerspin}

The aim of this appendix is to specialize our general expansion \eqref{tildeBC2-zexpansion} to
integer spin $l = \lambda_1-\lambda_2 - \epsilon/2$ and to rewrite it as an expansion in 
the usual cross ratios $z$ and $\bar z$, or rather variables $u =z\bar z $ and $v= (1-z)(1-\bar z)$.
The latter can then be compared with existing expressions from other approaches, such as the
shadow or the embedding formalism.

Before we study Harish-Chandra series for such special values of the parameters, we want to make
some general comments. Actually, one has to be a bit careful when evaluating formulas that are
valid for generic values of the
parameters, like the ones for Harish-Chandra functions in the \hyperref[app:zxexpansions]{previous
appendix}, at special values of the parameters such as the momenta $\lambda_i$. In order to do so,
one should first take a limit for the multiplicities, and only then for the momenta (i.e. the
continued spin or conformal dimension). In other words, if we want to study blocks/Harish-Chandra
functions for particular spins (or twists) through the general series expansions, our prescription
is to fix the dimension $d$ and conformal weights $\Delta_i$ of external fields first and only
then, after carefully taking into account all implications on the summation range, to send the spin
(or twist) to some particular value. This recipe is directly related to the fact that an
(appropriately normalized) Harish-Chandra function is an entire function of its multiplicities, but
can have simple poles in $\lambda$ (and further subtleties related to its series expansion for
$\lambda$ along specific affine hyperplanes, see subsections \ref{subsec:HarishChandra}, \ref{subsec:HChPoles},
\ref{subsec:Poles}). The opposite  order of limits, provided it exists at all, also gives some
solution of the Calogero-Sutherland problem, but one that may be a non-trivial linear combination
of all the $|W_N|$ Harish-Chandra functions. The prescription we propose here is the same as usually used when
taking limits in formulas for the hypergeometric function $_2F_1$ (i.e.\ for $N=1$) to special values of
its parameters \cite{OlverBook}: when e.g. one upper and one lower parameter are negative
integers, one takes
$$z^{-\frac{k+m}{2}}\pFq{2}{1}{-m \pFcomma b-\frac{k+m}{2}}{-k-m}{z}=\lim_{h\to -(k+m)/2}\,\,
\lim_{a\to -h-m} z^h\pFq{2}{1}{h+a \pFcomma h+b}{2h}{z}$$
rather than
$$\lim_{a\to(k-m)/2} \pFq{2}{1}{a-\frac{k+m}{2} \pFcomma b-\frac{k+m}{2}}{-k-m}{z},$$
where $k,m$ are positive integers and $b$ is generic. One can show that the second function is a
linear combination of the first and its shadow.

After these introductory remarks we now want to start with our general formula
\eqref{tildeBC2-zexpansion} and set $l=0$, i.e. $\lambda_2=\lambda_1-\frac{\epsilon}{2}$. For the
moment we shall assume that the dimension $d$ is not even so that upon specializing $\lambda$, the
twisted Harish-Chandra function is actually just equal to a block, up to a normalizing prefactor $4^{2a+1+\epsilon+l-2\lambda_1}(z_1 z_2)^{-a}c_{\alpha_1}(\lambda_1,\lambda_2)$. By applying a
Whipple transformation \cite{AAR} to the inner $_4F_3$ function in eq.\ \eqref{tildeBC2-zexpansion},
we can perform a sum via Saalsch\"utz formula, yielding\footnote{The analysis by Horn's theorem \cite{Horn1889} shows that the absolute convergence for this series representation of the twisted Harish-Chandra function in fact takes place on a
the {\it extended} domain $\Re u_i > 0$ and $0< \Im u_i< 2\pi$. This is, of course, consistent with the fact that the single twisted Harish-Chandra function for generic dimension, being proportional to a block, is bound to be regular at the wall $z_1=z_2$ according to our discussion in subsection \ref{subsec:Poles}.}
\begin{align}
&\tilde \Phi(\lambda_1,\lambda_2; z_1,z_2)\biggl |_{l=0}=\frac{\left(z_1 z_2\right)^{\frac{1+\epsilon}{2}+a-\lambda_1}}{4^{2a+1+\epsilon-2\lambda_1}}\sum_{n,m,p=0}^{\infty}
\frac{\left(-z_1\right)^{n+p}z_2^{m-p}}{n!m!p!}\left(\frac{\epsilon}{2}\right)_{m-n-p}
\left(\frac{\epsilon}{2}\right)_{p}\left(-m\right)_{n+p}\nonumber\\[2mm]
&\quad \times \frac{\left(\frac{\epsilon}{2}+1\right)_{m-n}}{\left(\frac{\epsilon}{2}\right)_{m-n}}\frac{\left(\frac{1}{2}+a-\lambda_1, \frac{1}{2}+b-\lambda_1\right)_{n}}{\left(1+\frac{\epsilon}{2}-2\lambda_1\right)_{n}}\frac{\left(\frac{1+\epsilon}{2}+a-\lambda_1, \frac{1+\epsilon}{2}+b-\lambda_1\right)_{m}}{\left(1+\frac{\epsilon}{2}, 1+\epsilon-2\lambda_1 \right)_{m}}.
\end{align}
By changing the order of summation, we can rewrite this expression as
\begin{align}
&= \left(\frac{z_1z_2}{16}\right)^{\frac{1+\epsilon}{2}+a-\lambda_1}\hspace*{-3mm}\sum_{n,m=0}^{\infty}
\frac{\left(-z_1\right)^{n}z_2^{m-n}}{n!m!}\left(\frac{\epsilon}{2}\right)_{m-n}\!
\left(\frac{\epsilon}{2}\right)_{n}\!
\left(-m\right)_{n}\frac{\left(\frac{1+\epsilon}{2}+a-\lambda_1, \frac{1+\epsilon}{2}+b-\lambda_1\right)_{m}}{\left(\frac{\epsilon}{2}, 1+\epsilon-2\lambda_1 \right)_{m}}\times \nonumber\\[2mm]
&\pFq{7}{6}{-\frac{\epsilon}{2}-m \pFcomma 1+\frac{-\frac{\epsilon}{2}-m}{2} \pFcomma -n \pFcomma -m+n \pFcomma -\epsilon+2\lambda_1-m \pFcomma \frac{1}{2}+a-\lambda_1 \pFcomma \frac{1}{2}+b-\lambda_1}{\frac{-\frac{\epsilon}{2}-m}{2} \pFcomma 1-\frac{\epsilon}{2}-m+n \pFcomma 1-\frac{\epsilon}{2}-n \pFcomma 1+\frac{\epsilon}{2}-2\lambda_1 \pFcomma \frac{1-\epsilon}{2}-a+\lambda_1-m \pFcomma \frac{1-\epsilon}{2}-b+\lambda_1-m }{1}.\nonumber
\end{align}
The inner well-poised $_7F_6(1)$ is now to be transformed to another one using the $S_6$ symmetry of this function. Namely, we can apply a particular identity labeled $W(1;2)=W(3;4)$ by Bailey \cite{BaileyBook}, employ the series representation for the resulting inner $_7F_6(1)$ function and shift the summation indices. This leads to
\begin{align}
&= \left(\frac{z_1z_2}{16}\right)^{\frac{1+\epsilon}{2}+a-\lambda_1}\sum_{p=0}^{\infty}\frac{\left(z_1 z_2\right)^{p}}{p!}
\frac{\left(\frac{\epsilon}{2}, \epsilon-2\lambda_1, \frac{1+\epsilon}{2}\pm a-\lambda_1,  \frac{1+\epsilon}{2}\pm b-\lambda_1 \right)_{p}}{\left(1+\epsilon-2\lambda_1\right)_{2p}^2\left(1+\frac{\epsilon}{2}-2\lambda_1\right)_{p}}\\[2mm]
&\times \pFq{2}{1}{\frac{1+\epsilon}{2}+ a-\lambda_1+p \pFcomma \frac{1+\epsilon}{2}+b-\lambda_1+p}{1+\epsilon-2\lambda_1+2p}{z_1}\pFq{2}{1}{\frac{1+\epsilon}{2}+ a-\lambda_1+p \pFcomma \frac{1+\epsilon}{2}+b-\lambda_1+p}{1+\epsilon-2\lambda_1+2p}{z_2},\nonumber
\end{align}
where $n$ and $m$ summations were absorbed into the two $_2F_1$ hypergeometric functions. Using Chu-Vandermonde identity \cite{AAR} in the form
\begin{align*}
\frac{\left(\frac{\epsilon}{2}\right)_p}{\left(1+\frac{\epsilon}{2}-2\lambda_1\right)_p}=(-1)^p \sum_{r=0}^p\frac{\left(-p \pFcomma 1-2\lambda_1\right)_r}{\left(1+\frac{\epsilon}{2}-2\lambda_1\right)_r}\frac{1}{r!}
\end{align*}
and shifting $p,r$ summation indices, we can rewrite our expression further as
\begin{align}
&= \left(\frac{z_1z_2}{16}\right)^{\frac{1+\epsilon}{2}+a-\lambda_1}\sum_{r=0}^{\infty}\frac{\left(z_1 z_2\right)^{r}}{r!}\frac{\left(\frac{1+\epsilon}{2}\pm a-\lambda_1,  \frac{1+\epsilon}{2}\pm b-\lambda_1 \right)_{r}}{\left(1+\epsilon-2\lambda_1\right)_{2r}\left(1+\frac{\epsilon}{2}-2\lambda_1\right)_{r}}\nonumber\\[2mm]
&\quad \times \sum_{p=0}^{\infty}\frac{\left(-z_1 z_2\right)^{p}}{p!}
\frac{\left( \epsilon-2\lambda_1+2r, \frac{1+\epsilon}{2}\pm a-\lambda_1+r,  \frac{1+\epsilon}{2}\pm b-\lambda_1+r \right)_{p}}{\left(\epsilon-2\lambda_1+2r, 1+\epsilon-2\lambda_1+2r\right)_{2p}}\\[2mm]
&\quad \times \pFq{2}{1}{\frac{1+\epsilon}{2}+ a-\lambda_1+r+p \pFcomma \frac{1+\epsilon}{2}+b-\lambda_1+r+p}{1+\epsilon-2\lambda_1+2r+2p}{z_1}\nonumber\\[2mm]
&\quad\times\pFq{2}{1}{\frac{1+\epsilon}{2}+ a-\lambda_1+r+p \pFcomma \frac{1+\epsilon}{2}+b-\lambda_1+r+p}{1+\epsilon-2\lambda_1+2r+2p}{z_2},\nonumber
\end{align}
so that for the inner ($p-$) sum we can now use a beautiful identity of Burchnall-Chaundy (\cite{Burchnall1940}, formula (50))
\begin{align}\label{BCh}
\pFq{2}{1}{A \pFcomma B}{C}{z_1+z_2-z_1 z_2}&=\sum_{p=0}^{\infty}\frac{\left(-z_1 z_2\right)^p}{p!}\frac{\left(A,B,C-A,C-B\right)_p}{\left(C\right)_{2p}\left(C+p-1\right)_p}\\[2mm]
&\times \pFq{2}{1}{A+p \pFcomma B+p}{C+2p}{z_1} \pFq{2}{1}{A+p \pFcomma B+p}{C+2p}{z_2}\nonumber
\end{align}
(provided both sides converge or assuming analytical continuation), and finally arrive at\footnote{Note that, while previous manipulations just reshuffled summation order in an absolutely convergent series, the application of Burchnall-Chaundy formula changes the domain of (absolute) convergence which now becomes $\{u,v\in\mathbb{C} |\,\, |1-v|<1, \sqrt{|u|}-\sqrt{1-|1-v|}<1\}$.}
\begin{align}
\tilde \Phi(\lambda_1,\lambda_2; z_1,z_2)\biggl |_{l=0}=\left(\frac{u}{16}\right)^{\frac{1+\epsilon}{2}+a-\lambda_1}&\sum_{r,s=0}^{\infty}\frac{u^{r}\left(1-v\right)^s}{r!s!}\frac{\left(\frac{1+\epsilon}{2}+ a-\lambda_1,  \frac{1+\epsilon}{2}+ b-\lambda_1 \right)_{r+s}}{\left(1+\epsilon-2\lambda_1\right)_{2r+s}}\nonumber\\
&\times
\frac{\left(\frac{1+\epsilon}{2}- a-\lambda_1,  \frac{1+\epsilon}{2}- b-\lambda_1 \right)_{r}}{\left(1+\frac{\epsilon}{2}-2\lambda_1\right)_{r}}, \label{eq:resl0}
\end{align}
where $u, v$ are the usual cross ratios defined in eq.\ \eqref{zixirel}. As in section \ref{sec:Blocks}, we
identify $z_1=z$, $z_2=\bar{z}$. Once we pass to conformal blocks using our relation \eqref{eq:GDl}, we
recognize formula (3.32) in \cite{Dolan:2011dv} (see also \cite{Dolan:2000ut}). Dolan and Osborn obtained this formula from the shadow
formalism by calculating the corresponding integrals via Mellin transform, ignoring all pole terms form the
shadow contributions. As noticed e.g. in \cite{Exton1995, Dolan:2000ut}, this double hypergeometric expression
\eqref{eq:resl0} can be rewritten as a linear combination of two $F_4$ Appell functions.\footnote{Notice
that the convergence region after using this connection formula changes to $\{\sqrt{|u|}+\sqrt{|v|}<1\}$,
so a diagonal wall is in fact re-introduced.} It is clear that we can lift our restriction on $\epsilon$
and admit even values, literally repeating the above calculation for the scalar exchange block which for even $\epsilon
\in2\mathbb{Z}_{\geq 0}$ is composed from two twisted Harish-Chandra functions.

The type of analysis we have presented here and in particular the use of Burchnall-Chaundy type formulas,
generalizes much beyond the specific case analysed here and allows to relate expressions from our approach
to those obtained through the shadow formalism \cite{Dolan:2011dv} or results from the embedding formalism \cite{SimmonsDuffin:2012uy, Fortin:2016lmf, Fortin:2016dlj}. In particular, we can go through the same steps
for any spin $l>0$ (and 'push out' the finite summations associated to the spin being integer) to obtain
finite sums of double hypergeometric functions of Srivastava-Daoust (i.e. generalized Kamp\'e de F\'eriet) type \cite{Srivastava-Daoust}. While the resulting expressions can be used to compare with results from other
approaches, we want to stress that the expansion formulas we have derived in appendix \ref{app:zxexpansions} and their direct
specializations are typically simpler than those that are obtained by other techniques.

\bibliographystyle{JHEP}
\bibliography{literatureCalogero}

\end{document}